\documentclass[preprintnumbers,nofootinbib,showpacs,eqsecnum,pre,12pt]{revtex4-1}
\usepackage{amsmath}         
\usepackage{amssymb}          
\usepackage{bbold}          
\usepackage[pdftex]{graphicx}	
\usepackage[pdftex]{hyperref} 
\usepackage[utf8x]{inputenc}
\usepackage{xcolor}
\usepackage{ulem}
\usepackage{afterpage}
\usepackage{youngtab}
\usepackage{multirow}
\usepackage[capitalise]{cleveref}
\usepackage{slashed}
\usepackage{booktabs}
\usepackage{subcaption}

\newcommand{\ra}[1]{\renewcommand{\arraystretch}{#1}}
\newcommand{\adj}[1]{\overline{#1}}

\crefname{section}{Sec.}{Secs.}
\crefname{table}{Tab.}{Tabs.}
\crefname{figure}{Fig.}{Figs.}
\crefname{equation}{Eq.}{Eqs.}
\crefname{appendix}{Appendix}{Appendix}

\newcommand{\hc}{\text{h.c.}}
\newcommand{\SO}{\text{SO}}
\newcommand{\SU}{\text{SU}}
\newcommand{\U}{\text{U}}
\newcommand{\Sp}{\text{Sp}}

\newcommand{\Tr}{\text{Tr}}
\newcommand{\id}{\mathbb{1}}
\newcommand{\vev}{vev}

\def\gsim{\raise0.3ex\hbox{$\;>$\kern-0.75em\raise-1.1ex\hbox{$\sim\;$}}}
\def\lsim{\raise0.3ex\hbox{$\;<$\kern-0.75em\raise-1.1ex\hbox{$\sim\;$}}}
\def\ptmiss{p_{T}^{\rm miss}}

\begin{document}

\title{Phenomenology of unusual top partners in composite Higgs models}

\author{G.\ Cacciapaglia}
\email[E-mail: ]{g.cacciapaglia@ipnl.in2p3.fr}
\affiliation{
Institut de Physique des Deux Infinis de Lyon (IP2I), CNRS/IN2P3, 4 rue Enrico Fermi, 69622 Villeurbanne Cedex, France}
\affiliation{Universit\'e Claude-Bernard Lyon 1, Univ Lyon, Lyon, France}
\author{T.\ Flacke}
\email[E-mail: ]{flacke@kias.re.kr}
\affiliation{
Center for AI and Natural Sciences, KIAS, Seoul 02455, Korea}
\author{M. Kunkel}
\email[E-mail: ]{manuel.kunkel@physik.uni-wuerzburg.de}
\author{W. Porod}
\email[E-mail: ]{porod@physik.uni-wuerzburg.de}
\affiliation{
Institut f\"ur Theoretische Physik und Astrophysik, Uni W\"urzburg, Emil-Hilb-Weg 22,
D-97074 W\"urzburg, Germany
}

\begin{abstract}
We consider a particular composite Higgs model which contains SU(3) color octet top partners besides the
usually considered triplet representations. Moreover, color singlet top partners are present as well which can in principle serve as dark matter candidates. We investigate the LHC phenomenology of these unusual top partners. Some of these states could be confused with gluinos predicted in supersymmetric models at first glance.
\end{abstract}

\preprint{KIAS-A21002}

\maketitle

\tableofcontents

\newpage

\section{Introduction}

The discovery of a Higgs-like scalar resonance at the LHC \cite{ATLAS:2012yve,CMS:2012qbp} has materialized long-standing questions on the Standard Model (SM) as the ultimate theory of particle interactions. Why is the Higgs boson mass insensitive to the scale of new physics (Planck mass)? Is there a dynamical origin for the spontaneous breaking of the electroweak (EW) symmetry?
Do elementary scalar particles really exist?

The two time-honored avenues addressing the above questions are supersymmetry (SUSY) and compositeness. In the former, scalars are associated with fermions via a new symmetry extending Poincar\'e invariance of particle interactions. In the latter, scalars emerge as resonances of underlying bound fermions. In both cases, the solution involves tying the properties of the Higgs-like scalar to those of fundamental fermionic states, which do not suffer from the quantum sensitivity to large scales. In this work we will focus on the composite avenue, which was proposed as an alternative to the SM Higgs mechanism shortly after the SM itself was established \cite{Weinberg:1975gm,Susskind:1978ms}. While the first incarnations, inspired by Quantum Chromo Dynamics (QCD), were essentially Higgsless \cite{Yamawaki:1985zg,Sannino:2004qp,Hong:2004td}, it was later realized that a SM-like limit could be achieved by extending the global symmetry of the condensing theory to allow for a misaligned vacuum \cite{Kaplan:1983fs}. At the price of a moderate tuning, this class of models features a limit where a light Higgs-like state emerges as a pseudo-Nambu-Goldstone boson (pNGB). This idea was revved up in the early 2000s, thanks to the holographic principle \cite{Contino:2003ve}, which links near-conformal theories in 4 dimensions to gauge/gravity theories in 5 dimensions on a warped background. A `minimal' model emerged \cite{Agashe:2004rs,Agashe:2005dk}, based on the symmetry breaking pattern $\SO(5)/\SO(4)$, where the only pNGBs match the 4 degrees of freedom of the SM Higgs doublet field. The phenomenology of this model has been widely explored, and we refer the interested reader to the excellent reviews \cite{Contino:2010rs,Bellazzini:2014yua,Panico:2015jxa}. 

While the minimal model, and variations thereof, can be considered as a useful template to understand the phenomenology of a composite pNGB Higgs, models in this class  are not UV complete and its UV embeddings contain additional BSM states which are relevant at colliders, like the LHC. From the holographic point of view, a 5-dimensional model is consistent only if it includes all the operators of the corresponding 4-dimensional conformal field theory: in models with top partial compositeness \cite{Kaplan:1991dc}, for instance, this implies the inevitable presence of QCD-charged bosonic operators, corresponding to two-point functions of the spin-1/2 top partners \cite{Cacciapaglia:2020kgq}.
From the point of view of an underlying gauge-fermion theory, \`a la QCD, the symmetry breaking pattern $\SO(5)/\SO(4)$ cannot be obtained as the global symmetry group of the underlying fermions is unitary \cite{Cacciapaglia:2014uja,Cacciapaglia:2020kgq}. The minimal breaking patterns for a composite Higgs which arises from models with underlying fermions are $\SU(4)/\Sp(4)$, $\SU(5)/\SO(5)$ or $\SU(4)\times \SU(4) / \SU(4)$ \cite{Dugan:1984hq,Galloway:2010bp,Cacciapaglia:2014uja,Ma:2015gra}. The models thus contain additional pNGBs besides the Higgs doublet. Furthermore, in models with top partial compositeness \cite{Barnard:2013zea,Ferretti:2013kya,Vecchi:2015fma}, additional QCD-charged underlying fermions are required in order to allow for fermionic bound states with the same quantum numbers as the SM top. Like their electroweak counter parts, the underlying colored fermions condense, and the spontaneous breaking of the color sector global symmetry yields additional colored pNGBs in the low energy effective theory \cite{Ferretti:2016upr}. 

Regarding the LHC signatures, the presence of the additional pNGBs generates new decay channels for the top partners \cite{Bizot:2018tds}, besides the ones usually considered in direct searches at the LHC. Some of those have been studied in detail \cite{Chala:2017xgc,Aguilar-Saavedra:2017giu,Han:2018hcu,Kim:2018mks,Alhazmi:2018whk,Xie:2019gya,Cacciapaglia:2019zmj,Benbrik:2019zdp,Aguilar-Saavedra:2019ghg,Wang:2020ips,Corcella:2021mdl}. The QCD-charged pNGBs can also be directly pair-produced at the LHC via their QCD interactions, leading to relatively strong bounds \cite{Cacciapaglia:2015eqa,Belyaev:2016ftv,Cacciapaglia:2019bqz,Cacciapaglia:2020vyf}. Besides a ubiquitous color-octet scalar, the spectrum may also contain exotically charged states, like color sextets \cite{Cacciapaglia:2015eqa}.

In this work we will focus on one specific model, first proposed in Ref.~\cite{Ferretti:2013kya} and dubbed M5 in Ref.~\cite{Belyaev:2016ftv}. It belongs to a class of models, comprising 12 candidates, where the top partners arise as chimera baryons made of two different species of confining fermions. The peculiarity of the model M5 compared to the others is that the baryon spectrum contains color-octet fermionic states. They are predicted to be among the lightest top partners, hence they play the leading role in the LHC phenomenology of this model. Furthermore, the pNGB spectrum contains a color-triplet with the charge of a right-handed stop. As we will see, together with color-neutral baryons and the ubiquitous color-octet pNGB, the spectrum and phenomenology of the model M5 shows similarities to SUSY models. 
Finally, the properties of the confining gauge dynamics, based on Sp(4), is being studied on the Lattice with promising results~\cite{Bennett:2017kga,Lee:2018ztv,Bennett:2019jzz,Lee:2019pwp,Bennett:2019cxd}.
Complementary information on the mass spectrum and decay constants of the composite states can also be obtained by use of Nambu-Jona-Lasinio models \cite{Bizot:2016zyu} or holographic techniques \cite{Erdmenger:2020lvq,Erdmenger:2020flu}.

The article is organized as follows: After introducing the basic properties of the model M5 in Section~\ref{sec:modelaspects}, we characterize the phenomenology at colliders in Section~\ref{sec:LHCpheno}. 
In Section~\ref{sec:numerics} we present the current bounds on scenarios with a dark matter candidate, before offering our conclusions and outlook in Section~\ref{sec:outlook}.

\section{Model aspects} \label{sec:modelaspects}

We consider an $\Sp(2N_c)$ hyper-color gauge theory with 5 Weyl fermions $\psi_i$ in the antisymmetric and 6 Weyl fermions $\chi_j$ in the fundamental representation as an underlying model of the composite sector. The fermion sector exhibits an $\SU(5)\times \SU(6) \times \U(1)$ global symmetry. It has been named M5 in \cite{Belyaev:2016ftv}
and its EW sector has been investigated in \cite{Agugliaro:2018vsu}.
The chiral condensates $\langle \psi\psi \rangle$ and $\langle \chi\chi \rangle$ spontaneously break the global symmetry to the stability group $\SO(5)\times \Sp(6)$. The SM color group $\SU(3)_c$ is realized as a gauged $\SU(3)$ subgroup of $\Sp(6)$, while the weak gauge group $\SU(2)_L$ is a gauged subgroup of $\SO(5)$, which also contains a custodial subgroup $\SO(4) \sim \SU(2)_L \times \SU(2)_R$. The
$\U(1)_Y$ hypercharge $Y = T^3_R + X$ is a gauged linear combination of the diagonal generator of $\SU(2)_R\subset \SO(5)$ and $\U(1)_X\subset \Sp(6)$. In addition, the model
contains two global abelian symmetries $\U(1)_\chi$ and $\U(1)_\psi$, acting independently on the two hyper-fermion species. One linear combination of these $\U(1)$ factors is $\Sp(2N_c)$ anomaly free, and the spontaneous breaking by the condensates yields a pNGB, while the would-be pNGB associated to the orthogonal $\U(1)$ combination is expected to receive a mass through the $\Sp(2N_c)$ anomaly. 
We summarize the microscopic field content that replaces the Higgs sector of the SM in \cref{tab:micro}.

\begin{table}[htb]
	\begin{center}
		\begin{tabular}{|c|c|c|c|c||c|c|c|}
			\hline
			& $\Sp(2N_c)$&${\SU(3)}_c$&${\SU(2)}_L$&${\U(1)}_Y$ & SU(5) & SU(6) & U(1) \\
			\hline
			$\psi_{1,2}$&${\tiny{\yng(1,1)}}$&$\bf 1$&$\bf 2$&$1/2$& \multirow{3}{*}{\bf 5} & \multirow{3}{*}{\bf 1} & \multirow{3}{*}{$-\frac{3 q_\chi}{5 (N_c-1)}$}\\
			\cline{1-5}
			$\psi_{3,4}$&${\tiny{\yng(1,1)}}$&$\bf 1$&$\bf 2$&$-1/2$ & & &\\
			\cline{1-5}
		    $\psi_{5}$&${\tiny{\yng(1,1)}}$&$\bf 1$&$\bf 1$&$0$ & & & \\
			\hline
			$ \begin{array}{c} \chi_1 \\ \chi_2 \\ \chi_3\end{array} $&${\tiny{\yng(1)}}$&$\bf 3$&$\bf 1$&$-x$ & \multirow{4}{*}{\bf 1} & \multirow{4}{*}{\bf 6} & \multirow{4}{*}{$q_\chi$}\\
			\cline{1-5}
			$ \begin{array}{c} \chi_4 \\ \chi_5 \\ \chi_6\end{array} $&${\tiny{\yng(1)}}$&$\bf \bar{3}$&$\bf 1$&$x$ & & &\\
			\hline
		\end{tabular} 
	\end{center}
		\caption{Field content of the microscopic fundamental theory and transformation properties under the gauged symmetry group $\Sp(2N_c)\times\SU(3)_c \times \SU(2)_L \times  \U(1)_Y$, and under the global symmetries $\SU(5)\times\SU(6)\times\U(1)$.}
		\label{tab:micro}
	
\end{table}

\subsection{pNGBs}

The condensates $\langle \psi \psi\rangle$ and 
$\langle \chi \chi\rangle$ break the global
group $\SU(5) \times \SU(6) \times \U(1)$ to
\mbox{$\SO(5) \times \Sp(6)$}. This will give rise to three classes of pNGBs in this model:
\begin{enumerate}
 \item A SM singlet pNGB $a$ from the $\Sp(2N_c)$ anomaly-free spontaneously broken $\U(1)$. It is expected to be light and with couplings of an axion-like particle (see \cite{Belyaev:2016ftv,Cacciapaglia:2019bqz,BuarqueFranzosi:2021kky} for studies of collider bounds and projections for the pNGB $a$ of the model considered here, as well as for other composite axion-like particles).
  \item 14 pNGBs in the EW sector in the $\bf{14}$ of $\SO(5)$, which decomposes into $\bf{3}_1+\bf{3}_0+\bf{3}_{-1}+\bf{2}_{1/2}+\bf{2}_{-1/2}+\bf{1}_0$ under $\SU(2)_L\times \U(1)_Y$, with the 4 degrees of freedom in $\bf{2}_{1/2} + \bf{2}_{-1/2}$ identified as the composite Higgs doublet.
  The EW pNGB sector has been studied in \cite{Agugliaro:2018vsu}, to which we refer for further details. The main aspect relevant for the present work is that, after the EW symmetry
  breaking, the bi-triplet $\bf{3}_1+\bf{3}_0+\bf{3}_{-1}$ of $\SU(2)_L\times \SU(2)_R$ decomposes into a singlet, a triplet and
  a fiveplet of the custodial diagonal $\SU(2)$ denoted as
  $\eta_1$, $\eta_3$ and $\eta_5$, respectively.
  Further details are given in \cref{app:ew_pngbs}.

\item 14 pNGBs in the color sector in the $\bf{14}$ of $\Sp(6)$, which decomposes into $\bf{8_0}+\bf{3_{-2x}}+\bf{\bar{3}_{2x}}$ under $\SU(3)_c\times \U(1)_Y$. They play a central role in the phenomenology we discuss here, hence we provide more details below as they have not been discussed elsewhere in the literature.

\end{enumerate}

The pNGB sector of $\SU(6)/\Sp(6)$ can be parameterized by a scalar field $\Sigma_\chi$ in the anti-symmetric 2-tensor representation $\bf{15}$ of $\SU(6)$, transforming like $g \Sigma_\chi g^T$ where $g\in \SU(6)$.\footnote{We structurally follow Ref. \cite{Cheng:2020dum} which however discusses $\SU(6)/\Sp(6)$ breaking in the EW sector.}
The vacuum, which respects the stability group $\Sp(6)$, can be written as
\begin{equation} \label{eq:vacuumchi}
\Sigma_{\chi,0}=  \begin{pmatrix} 0 & - \id_3 \\ \id_3 & 0 \end{pmatrix}\,,
\end{equation}
where $\id_3$ is a $3\times 3$ identity matrix.
Hence, within $\Sp(6)$, the QCD $\SU(3)_c$ subgroup is defined by the following 8 generators
\begin{align}
\label{eq:SU3_embedding}
 \frac{1}{2 \sqrt{2} } \begin{pmatrix} \lambda^a & 0 \\ 0 & - (\lambda^a)^* \end{pmatrix}\,,
\end{align} 
where $\lambda^a$ are the Gell-Mann matrices corresponding to the usual $\SU(3)_c$ generators,
normalized as tr$(\lambda^a \lambda^b)= 2 \delta^{ab}$.
In addition, there is a $\U(1)_X$ subgroup of $\Sp(6)$ 
\begin{align}
X =  {\bf x} \, \begin{pmatrix} \id_3 & 0 \\ 0 & - \id_3  \end{pmatrix}\,,
\label{eq:Ygen}
\end{align}
needed to assign the correct hypercharge to the composite colored states: the SM hypercharge is defined as $Y = T^3_R + X$, where $T_R^3$ is the diagonal generator of a global $\SU(2)_R$ subgroup of $\SO(5)$ in the EW coset. 
Choosing ${\bf x}=-1/3$ yields the correct hypercharge for the top partners, as we will show below.

The pNGB matrix can be written as
\begin{equation}
U_\chi = e^{2i\Pi^I_\chi X^I/f_\chi}\,,
\end{equation}
where $X^I$ are the 14 generators broken by the vacuum in \cref{eq:vacuumchi}.
It transforms as $U_\chi \rightarrow g U_\chi h^\dagger$ with $g\in \SU(6)$, $h\in \Sp(6)$.
In terms of $\SU(3)_c$, the 14 degrees of freedom decompose into a color octet $\pi_8$ and a complex color triplet $\pi_3$, where
$\pi_8$ has charge 0 with respect to the $\U(1)_X$ subgroup and the triplet $2/3$. As these
are $\SU(2)_L$ singlets, these numbers correspond automatically to their hypercharges
and consequently also to their electric charges. The explicit decomposition of $\Pi_\chi$ 
in terms of these fields reads 
\begin{align}
\label{eq:pi_chi_expl}
\Pi_\chi &= \Pi_\mathbf{8} + \Pi_{\mathbf{3}+\bar{\mathbf{3}}}\,, \\
\Pi_\mathbf{8} &=
\frac{1}{2 \sqrt{2}} \sum_{a=1}^8 \begin{pmatrix} \pi_8^a\ \lambda^a & 0 \\ 0 & \pi_8^a\ (\lambda^a)^T \end{pmatrix} \quad \text{ and } \quad
\Pi_{\mathbf{3}+\bar{\mathbf{3}}} =  \begin{pmatrix} 0 & \kappa^\dagger \\ \kappa & 0\end{pmatrix}
\text{ with } \kappa_{ij} = \frac 12 \epsilon_{ijk} \pi_3^k\,.
\end{align}
In the following we will refer to $\pi_8$ and $\pi_3$ as octet and triplet pNGBs, respectively.
 
The $\pi_8$ couples via
 the Wess-Zumino-Witten term to two gluons. The corresponding coupling is proportional to $f^{-1}_\chi$ and, thus,
 one could get in principle  contraints
 on the decay constant $f_\chi$ from existing LHC data \cite{Cacciapaglia:2020vyf}. These bounds depend on the $\pi_8$ mass $m_{\pi_8}$ and vanish for  $m_{\pi_8}\gsim 1.1$~TeV.
 However, we
note for completeness that one has in addition a decay constant  $f_\psi$ in the electroweak sector beside $f_\chi$. One expects that both have the same size. It is well known, that
 electroweak precision data give constraints on $f_\psi$, see e.g.~\cite{Cacciapaglia:2020kgq} and references therein. One finds that
 the ratio $v/f \lsim 0.25$, where $v$ is electroweak vaccum expection value. This implies  $f_\chi \simeq f_\psi \gsim 1$~TeV.

\subsection{Chimera hyper-baryons}
\label{sec:hyperbaryons}

In the confined phase, the model contains fermionic resonances (chimera hyper-baryons) corresponding to composite operators made of one $\psi$ and two $\chi$ hyper-fermions. They can be classified in terms of their transformation properties under the stability group $\SO(5) \times \Sp(6)$, as shown in \cref{tab:bdstates}, where we also show the corresponding three-fermion operators and their transformation under the global symmetry $\SU(5) \times \SU(6)$. 

\begin{table}[htb]
	\begin{center}
		\begin{tabular}{|c|c|c|c|}
			\hline
			 & SU(5)$\times$SU(6) & SO(5)$\times$Sp(6) & names \\
			\hline \hline
			$\psi \chi \chi$  & $(\bf{5}, \bf{15})$ & $(\bf{5}, \bf{14})$ & $\mathcal{B}^1_{14}$ \\
			                         &                             & $+ (\bf{5}, \bf{1})$ & $\mathcal{B}^1_{1}$\\			
			                         & $(\bf{5}, \bf{21})$ & $(\bf{5}, \bf{21})$ & $\mathcal{B}^1_{21}$\\
			\hline
			$\psi \bar{\chi} \bar{\chi}$  & $(\bf{5}, \bf{\overline{15}})$ & $(\bf{5}, \bf{14})$ &$\mathcal{B}^2_{14}$  \\
                                     &                                & $+ (\bf{5}, \bf{1})$ & $\mathcal{B}^2_{1}$ \\
                                    & $(\bf{5}, \bf{\overline{21}})$ & $(\bf{5}, \bf{21})$ & $\mathcal{B}^2_{21}$\\
			\hline
			$\bar{\psi} \bar{\chi} \chi$  & $(\bf{\bar 5}, \bf{35})$ & $(\bf{5}, \bf{14})$ & $\mathcal{B}^3_{14}$  \\
			                                    &                           & $+ (\bf{5}, \bf{21})$ & $\mathcal{B}^3_{21}$\\
			                                 &  $(\bf{\bar 5}, \bf{1})$  & $(\bf{5}, \bf{1})$ & $\mathcal{B}^3_{1}$\\
			
			\hline
			\hline
		\end{tabular} 
		\caption{Three-fermion bound states of the model and their quantum numbers with respect to the global flavor group and the unbroken subgroups.}
		\label{tab:bdstates}
	\end{center}
\end{table}

Under $\SU(3)_c\times \U(1)_X$, the $\Sp(6)$ representations decompose as:
\begin{align}
\bf{14}& \rightarrow \bf{8_0}+\bf{3_{-2x}}+\bf{\bar{3}_{2x}} \,, \\
\bf{21}&\rightarrow \bf{8_0}+\bf{6_{2x}}+\bf{\bar{6}_{-2x}}+\bf{1_0}
\,.
\end{align}
Hence, it is the $\bf 14$ that contains color-triplets that can mix with the SM elementary top fields to generate partial compositeness for the top quark mass origin. In terms of $\Sp(6)$, the components of the $\bf 14$ are embedded in the anti-symmetric matrix
\begin{equation} 
\Psi_{\bf 14} =  \begin{pmatrix}  -Q_3^{c} & - \frac{1}{2\sqrt{2}} Q_8^{a}\ \lambda^a  \\  \frac{1}{2\sqrt{2}} Q^{a}_8\ (\lambda^a)^T  & -Q_3  \end{pmatrix}\,,
\label{eq:tom_a17}
\end{equation}
where $Q^{(c)}_{3,ij}=\frac{1}{2} \epsilon_{ijk}Q^{(c)}_{3,k}$  gives correct transformations for the $\SU(3)_c$ generator embedding (i.e. the diagonals  transform like $\bf{3}\times \bf{3} \supset \bf{\bar{3}}$ and $\bf{\bar{3}}\times \bf{\bar{3}} \supset \bf{3}$ while the off-diagonals transform like octets). Each component $Q_3$ and $Q_8$ also transform as a fundamental of $\SO(5)$:
\begin{subequations}
\begin{align}
Q_3 &=  (X_{5/3}, X_{2/3},T_L,B_L,iT_R)^T\, , \label{eq:q3compo}\\
Q^{c}_{3} &=  ( B_L^c,-T_L^c, -X^c_{2/3}, X^c_{5/3},-i T_R^c)^T\, , \label{eq:q3ccompo}\\
Q_8 &= (\tilde G^+_u, \tilde G^0_u, \tilde G^0_d, \tilde G^-_d, i \tilde g)^T\,, \label{eq:q8compo}
\end{align}
\end{subequations}
where we have fixed ${\bf x}=-1/3$ in order to have the correct hypercharges for the color triplets that will mix with the top quark. The first four components transform as two doublets of $\SU(2)_L$ (and a bi-doublet of $\SU(2)_L \times \SU(2)_R$), while the fifth is a singlet. 

The couplings of the top fields to the hyper-baryons depend on the choice of three-hyper-fermion operator. In fact, the partial compositeness couplings for the left-handed top (in the doublet $q_{L,3}$) and right-handed top $t_R^c$ are assumed to originate from a four-fermion interaction. The simplest possibility is
\begin{equation} \label{eq:topPC4F}
    \frac{\xi_L}{\Lambda_t^2} \psi \chi \chi q_{L,3} + \frac{\xi_R}{\Lambda_t^2} \psi \chi \chi t_{R}^c 
\end{equation}
with the appropriate components of the hyper-fermions. Here we have chosen to couple them to the operator $\psi \chi \chi$ in the channel $({\bf 5}, {\bf 15})$, see \cref{tab:bdstates}. This implies the presence of the singlet baryons as well:
\begin{equation}
Q_1 = (\tilde h^+_u, \tilde h^0_u, \tilde h^0_d, \tilde h^-_d, i \tilde B)^T\,. \label{eq:q1compo}
\end{equation}
Note that all the hyper-baryon fields introduced so-far are 2-component Weyl spinors.  
Henceforth, the top partners in this model include all the components of the $\bf 14$ and the singlet, as all their components couple with the top, as we will show in the next subsection. In the following we will not consider the hyper-baryons in the $(\bf{5}, \bf{21})$ as they do not participate in the couplings generated by \cref{eq:topPC4F}.

\begin{figure}[tb]
	\centering
	\includegraphics[width=0.9\textwidth]{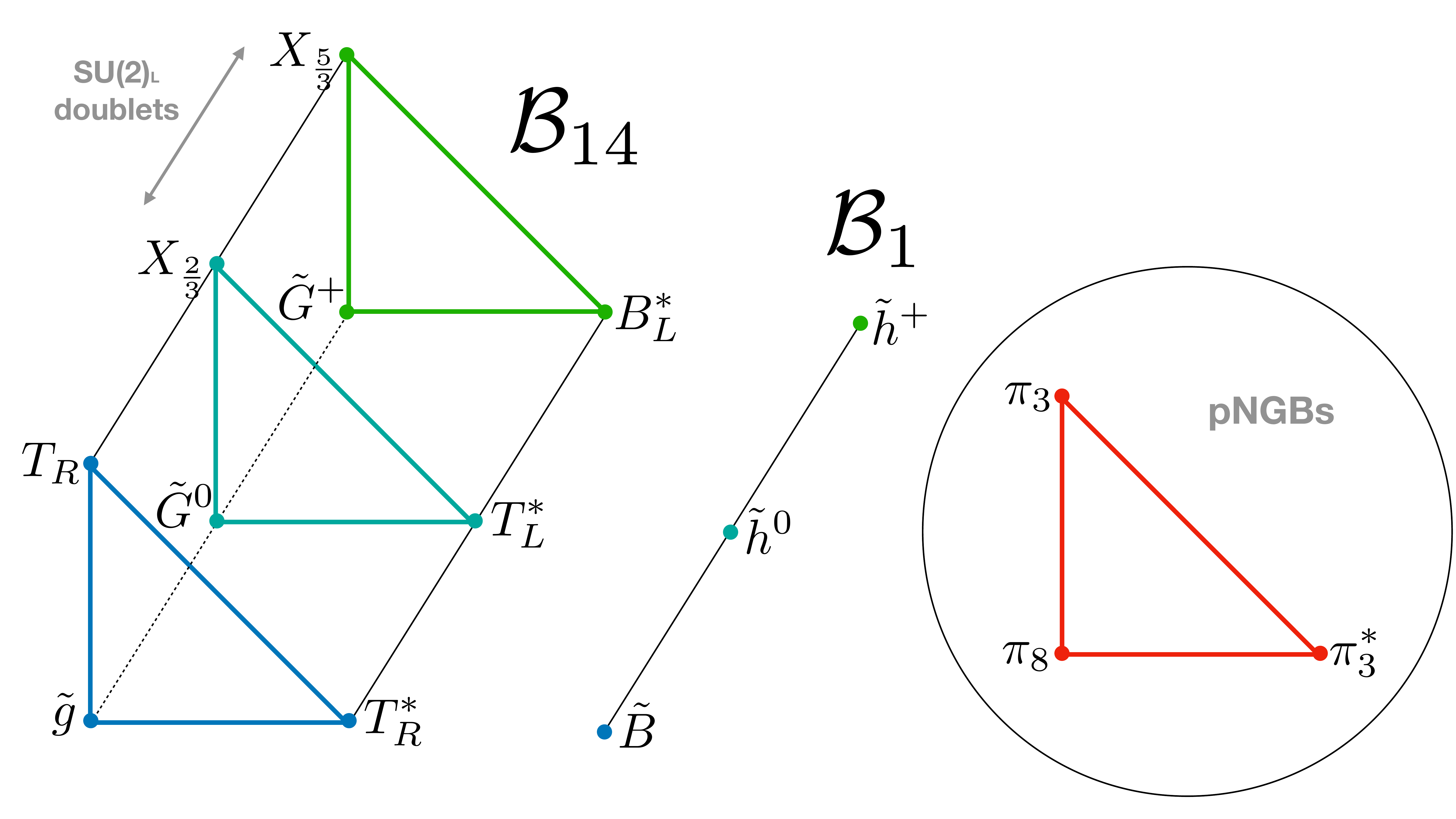}
	\caption{Schematic representation of the $\mathcal{B}_{14}$ and $\mathcal{B}_1$ chimera baryons  in terms of the global symmetries $\Sp(6)$ and $\SO(5)$. The triangles represent the $\bf 14$ of $\Sp(6)$, with the color octet at the square vertex, the triplet at the top and the anti-triplet at the right. The diagonal direction corresponds to the $\bf 5$ of $\SO(5)$, with the points in the foreground (blue) corresponding to singlets, and the other two composing complex doublets. In the circle, we also show the $\SU(6)/\Sp(6)$ pNGBs, also transforming as a $\bf 14$ of $\Sp(6)$.}
	\label{fig:diagrams}
\end{figure}

Besides the usual top partners in \cref{eq:q3compo,eq:q3ccompo}, the low energy spectrum of this model contains unusual top partners, transforming as color octets and color singlets. They correspond to the following states, written in terms of 4-components spinors:
\footnote{The indices $u$ and $d$ indicate that the corresponding $\SU(2)_L$ doublets have isospin $1/2$ and
$-1/2$, respectively.} 
\begin{subequations}\label{eq:hyperbaryonnames}
\begin{align}
\mbox{Octoni (Dirac): \hspace{10pt}} \tilde G^+ &= \begin{pmatrix} \tilde G^+_u \\ \bar{\tilde {G}}^-_d \end{pmatrix} \,\,,\,\,
\tilde G^0 = \begin{pmatrix} \tilde G^0_u \\ \bar{\tilde {G}}^0_d \end{pmatrix}\,\,, \\
\mbox{Gluoni (Majorana): \hspace{10pt}} \tilde g &= \begin{pmatrix} \tilde g \\ \bar{\tilde{g}} \end{pmatrix} \,\,; \\
\mbox{Higgsoni (Dirac): \hspace{10pt}}\tilde h^+ &= \begin{pmatrix} \tilde h^+_u \\ \bar{\tilde{h}}^-_d \end{pmatrix} \,\,,\,\,
\tilde h^0 = \begin{pmatrix} \tilde h^0_u \\ \bar{\tilde{h}}^0_d \end{pmatrix} \,\,, \\
\mbox{Boni (Majorana): \hspace{10pt}}\tilde B &= \begin{pmatrix} {\tilde B} \\ \bar{\tilde{B}} \end{pmatrix} \,.
\end{align}
\end{subequations}
The choice of the names is motivated by the fact that the states $\tilde h$, $\tilde B$ and $\tilde g$ have the same
quantum numbers as the higgsino, bino, and gluino in supersymmetric extensions of the SM.
Their embedding in the global symmetries are schematically shown in Fig.~\ref{fig:diagrams}, together with that of the colored pNGBs.

\subsection{Partial compositeness couplings and baryon number}

Following the Coleman-Callan-Wess-Zumino (CCWZ) prescription \cite{Coleman:1969sm,Callan:1969sn}, the low energy Lagrangian for the hyper-baryons contains three types of couplings \cite{Marzocca:2012zn}:
\begin{equation} \label{eq:LagB}
    \mathcal{L}_{\mathcal{B}} \supset  \overline{\mathcal{B}} i \bar{\sigma}^\mu D_\mu \mathcal{B} - M_\mathcal{B} \mathcal{B} \mathcal{B} + y_{L/R}\ \mathcal{F}_{L/R} (\Pi_\chi, \Pi_\psi)\, \mathcal{B}\ t_{L/R}^{(c)} + \frac{c}{f_\chi} \mathcal{F}' (\Pi_\chi)\ \partial_\mu \Pi_\chi\, \overline{\mathcal{B}} \bar{\sigma}^\mu \mathcal{B}\,,
\end{equation}
where $\mathcal{F}_{L/R}$ and $\mathcal{F}'$ are functions of the pNGB matrices $\Pi_\chi$ and $\Pi_\psi$. The kinetic term encodes gauge interactions with the SM via the covariant derivative $D_\mu$. The second term is generated by the operators in Eq.~\eqref{eq:topPC4F} and encodes the mixing between the SM top fields and the hyper-baryons, hence generating a mass for the top quark proportional to the product $y_L y_R$. 
We note that the three-hyper-fermion operator $\psi\chi\chi$ must have conformal dimension $d\simeq 5/2$ in order to obtain a top quark mass, since $y_{L/R} \sim (\Lambda_\mathrm{HC}/\Lambda_t)^{d-5/2}$ and $\Lambda_\mathrm{HC} \ll \Lambda_t$ \cite{Panico:2015jxa}.
Finally, we include derivative couplings of the colored pNGBs in the third term, as they comprise couplings between the octet and singlet baryons, which are of phenomenological relevance for our purposes. These interactions are generated by the strong dynamics itself, and they are suppressed by the compositeness scale $f_\chi$.

We first need to embed the SM fields $q_{L,3}$ and $t_R^c$ in a representation of $\SU(5) \times \SU(6)$ 
to write down the couplings to the top, second term in Eq.~\eqref{eq:LagB}. First we define two unique embeddings as $\bf \bar{5}$ of $\SU(5)$ as follows:
\begin{subequations}
\begin{eqnarray} 
\zeta_L &=&  (b_L,-t_L,0,0,0)\, , \\
\zeta_R &=&  (0,0,0,0, -i t^c_R)\, .
\end{eqnarray}
\end{subequations}
Then, we embed them in the $\bf \overline{15}$ ($\bar{A}$) representation of $\SU(6)$ as follows:
 \begin{eqnarray}
 \zeta_{L,\bar{A}} = \begin{pmatrix}   \frac 12 \epsilon_{ijk} \zeta_{L,k}  & 0 \\ 0  & 0 \end{pmatrix} \,, \qquad
 \zeta_{R,\bar{A}} = \begin{pmatrix}  0  & 0 \\ 0  & \frac 12 \epsilon^{ijk} \zeta_R^k  \end{pmatrix} \,,
\end{eqnarray}
where $ijk$ are QCD indices showing explicitly the embedding of the triplet and anti-triplet in an anti-symmetric matrix.
They form incomplete representations of the global symmetry, and as such they can be understood as spurions that explicitly break $\SU(5) \times \SU(6)$. 
The second ingredient we need are baryonic operators, also transforming as representations of the global symmetry $\SU(5) \times \SU(6)$. As the hyper-baryons form representations of the stability group $\SO(5) \times \Sp(6)$, they need to be dressed by pion matrices as follows:
\begin{equation}\label{eq:dressedoperators}
    \mathcal{O}_{A,14} = U_\chi \left( U_\psi \cdot \Psi_{14} \right) U_\chi^T\,, \qquad \mathcal{O}_{A,1} = \left( U_\psi \cdot Q_1 \right)\; U_\chi \cdot \Sigma_{\chi,0} \cdot U_\chi^T\,,
\end{equation}
such that the two operators transform as $({\bf 5},{\bf 15})$ of $\SU(5) \times \SU(6)$.

Finally, the couplings with the top fields can be written as follows. For the left-handed top field:
\begin{subequations} \label{eq:mixingAL0}
\begin{align}
y_{L}\, \mathcal{O}_{A,14}\ \zeta_{L,\bar A} &= y_{L} \left( \frac{1}{2}\ \zeta_{L,i} \cdot Q^{c}_{3,i}  -\frac{\sqrt 2 i}{4 f_\chi}\ \pi^*_{3,i} \lambda^a_{ij}\ \zeta_{L,j} \cdot Q^a_8 -\frac{\sqrt 2 i}{4 f_\chi}\  \pi^a_8 \lambda^a_{ij}\  \zeta_{L,j}\cdot Q^{c}_{3,i} \right. \nonumber \\  & \left. \phantom{\frac{1}{2}}+\mathcal{O}(\Pi^2_\chi) + \mathcal{O} (\Pi_\psi) \right)\, ,\label{eq:mixingAL}\\
y'_{L}\, \mathcal{O}_{A,1}\ \zeta_{L,\bar A} &= y'_{L} \left(\frac{i}{f_\chi}\ \pi^*_{3,i} \zeta_{L,i}\cdot Q_{1}+\mathcal{O}(\Pi^2_\chi) + \mathcal{O} (\Pi_\psi)\right)\,,
\end{align}
\end{subequations}
where $i,j$ are QCD color indices and the $\SU(5)$ contractions lead to
\begin{equation} \label{eq:zetaLQ}
     \zeta_{L} \cdot Q^c_3 = B_L^c b_L + T_L^c t_L\,, \qquad \zeta_{L} \cdot Q_8 = \tilde{G}^+_u b_L - \tilde{G}^0_u t_L\,, \qquad \zeta_{L} \cdot Q_1 = \tilde{h}^+_u b_L - \tilde{h}^0_u t_L\,,
\end{equation}
where, for simplicity, we have set to zero the misalignment angle that breaks the EW symmetry (see \cref{app:ew_pngbs} for more details).
We only display couplings involving one colored pNGB.
Similarly, for the right-handed top singlet:
\begin{subequations}
\label{eq:mixingAR0} \begin{align}
y_{R}\, \mathcal{O}_{A,14}\ \zeta_{R,\bar A} &= y_{R} \left( \frac{1}{2}\ \zeta_{R,i} \cdot Q_{3,i}  -\frac{\sqrt 2 i}{4 f_\chi}\ \pi_{3,i} \lambda^a_{ij}\ \zeta_{R,j} \cdot Q^a_8 -\frac{\sqrt 2 i}{4 f_\chi}\  \pi^a_8 \lambda^a_{ij}\  \zeta_{R,j}\cdot Q_{3,i} \right. \nonumber \\  & \left. \phantom{\frac{1}{2}}+\mathcal{O}(\Pi^2_\chi) + \mathcal{O} (\Pi_\psi) \right)\, ,\label{eq:mixingAR}\\
y'_{R}\, \mathcal{O}_{A,1}\ \zeta_{R,\bar A} &= y'_{R} \left(\frac{i}{f_\chi}\ \pi_{3,i} \zeta_{R,i}\cdot Q_{1}+\mathcal{O}(\Pi^2_\chi) + \mathcal{O} (\Pi_\psi)\right)\,,
\end{align}
\end{subequations}
with (neglecting the EW misalignment)
\begin{equation} \label{eq:zetaRQ}
     \zeta_{R} \cdot Q_3 = t_R^c T_R\,, \qquad \zeta_{R} \cdot Q_8 = t_R^c \tilde{g}\,, \qquad \zeta_{R} \cdot Q_1 = t_R^c \tilde{B}\,.
\end{equation}
Alternative top quark embeddings in the global symmetries and couplings to other hyper-baryons, see \cref{tab:bdstates}, are reported for completeness in \cref{app:altebb}. The case of the $({\bf 5}, {\bf \overline{15}})$ leads to similar couplings to the one presented here, up to signs. The $({\bf \bar 5}, {\bf 35})$, instead, is very different, as it involves the $\bf 21$ of $\Sp(6)$ and the partial compositeness couplings always involve a minimum of 2 colored pNGBs. We leave the phenomenology of this case to future work.

The last term in Eq.~\eqref{eq:LagB} generates derivative couplings between the hyper-baryons and the colored pNGBs. In our case, this only involves the $\bf 14$ and the singlet of $\Sp(6)$, and it can be written as
\begin{equation}
    c\ \mbox{Tr} \left[\bar{\Psi}_1\cdot  \bar \sigma^\mu d_\mu\cdot  \Psi_{14}\right] + \text{h.c.}\,,
\end{equation}
where the trace acts on the $\Sp(6)$ indices and
\begin{equation}
    \Psi_1 = Q_1 \Sigma_{\chi,0}\,, \qquad d_\mu = \mbox{Tr} \left[U_\chi^\dagger\cdot  \partial_\mu U_\chi\cdot  X^I \right]\ X^I\,.
\end{equation}
Here, $\Psi_1$ contains the singlet top partners and $d_\mu$ is a Maurer-Cartan form written in terms of the colored pNGBs.
Expanding up to linear order in the pNGB fields, we find
\begin{align}\label{eq:Lder141}
		& c\ \Tr \left[\bar{\Psi}_1\cdot  \bar{\sigma}^\mu d_\mu\cdot  \Psi_{14}\right] \sim i \frac{c}{f_\chi}\  \bar{Q}_1 \Tr\left[\Sigma_{\chi,0} \cdot \bar{\sigma}^\mu \partial_\mu \Pi_\chi \cdot \Psi_{14}\right] \nonumber\\
		&=-i \frac{c}{2 f_\chi}\ \left( \bar{Q}_1 \bar{\sigma}^\mu Q_{3,i}^c \, \partial_\mu \pi_{3,i} +\bar{Q}_1 \bar{\sigma}^\mu Q_{3,i} \, \partial_\mu \pi_{3,i}^* + \bar{Q}_1 \bar{\sigma}^\mu Q_8^a \ \partial_\mu \pi_8^a \right)\,,
\end{align}
where the $\SO(5)$ contractions are left understood.

All the couplings stemming from \cref{eq:LagB} allow to assign SM baryon number charges B to all the top partners, so that B is still preserved in the presence of top partial compositeness. This results in all the color triplets to carry $\text{B}=1/3$, like quarks, while color octets and singlets remain neutral. This can be achieved by assigning appropriate baryon number to the hyper-fermions in the underlying theory, see \cref{tab:micro}: $\text{B} = 1/6$ to $\chi_{4,5,6}$ transforming as an anti-triplet and $\text{B} = -1/6$ for $\chi_{1,2,3}$ transforming as a triplet. Hence,
\begin{equation}
    \text{B} = 1/3\;\; \mbox{for}\;\; Q_3\,,\, \pi_3\,; \qquad \text{B} = 0\;\; \mbox{for}\;\; Q_8\,,\, Q_1\,,\, \pi_8\,. 
\end{equation}
This implies that the lightest state among the components of $Q_8,\ Q_1$ and $\pi_3$ must be stable in absence of either baryon or lepton number violation, as there is no matching SM final state.

\section{Phenomenology at hadron colliders}
\label{sec:LHCpheno}

\begin{figure}
	\centering
	\includegraphics[width=0.5\textwidth]{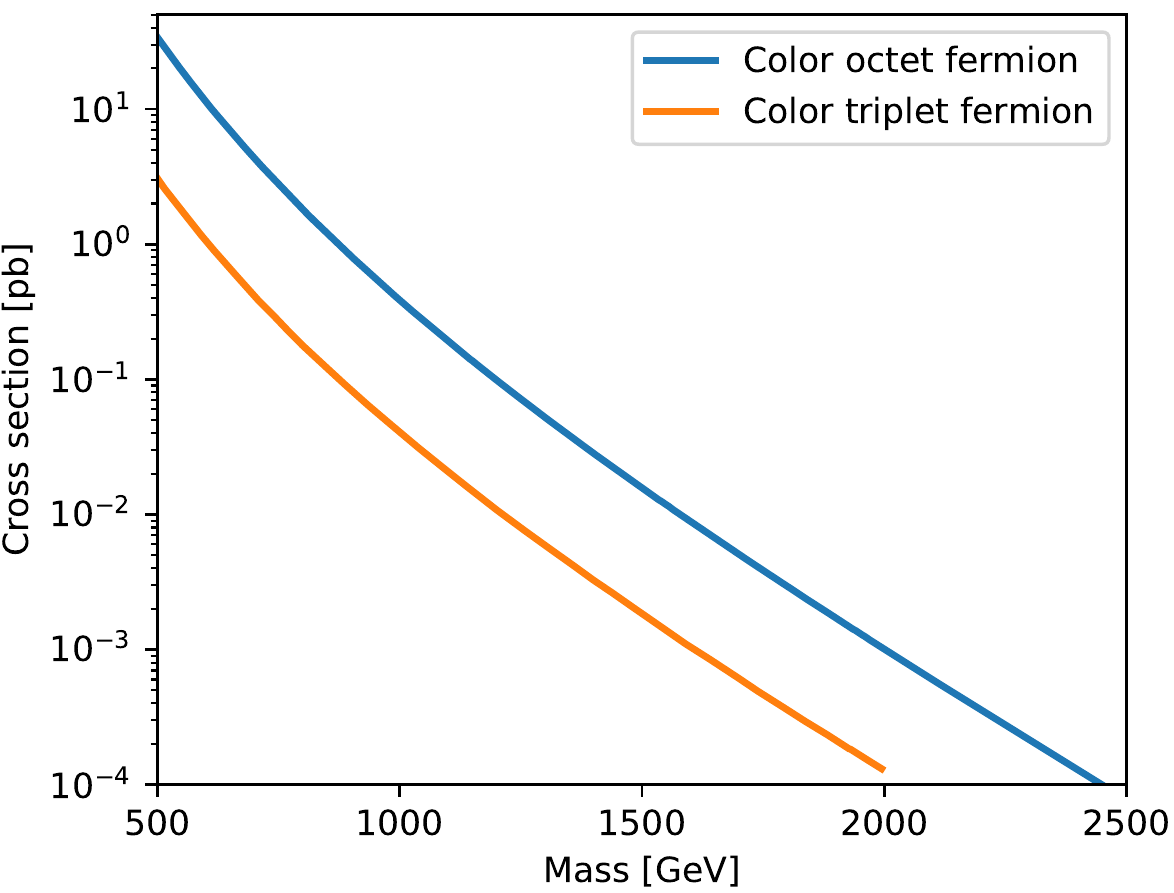}
	\caption{Comparison of the production cross sections of a color triplet top partner (at QCD NLO, from \cite{Fuks:2016ftf}) and a color octet Majorana top partner (at NNLO$_\text{approx}$+NNLL from \cite{Beenakker:2016lwe}).}
	\label{fig:q3q8xs}
\end{figure}

The phenomenology of M5 at hadron colliders depends crucially on the mass hierarchy among the QCD-colored baryons, which enjoy the largest production rates due to their QCD interactions. As they all belong to the same baryon multiplet -- the $\bf{14}$ of $\Sp(6)$ -- they have a common mass $M_{14}$ and the mass differences are only due to the SM gauge interactions and the top couplings.
The octet top partners, however, enjoy the largest pair production cross section thanks to their QCD quantum numbers, as shown in \cref{fig:q3q8xs}. The plot shows that the octets have a significantly larger cross section if the baryons have approximately the same mass. 

\begin{figure}[tb]
	\centering
	\includegraphics[width=0.9\textwidth]{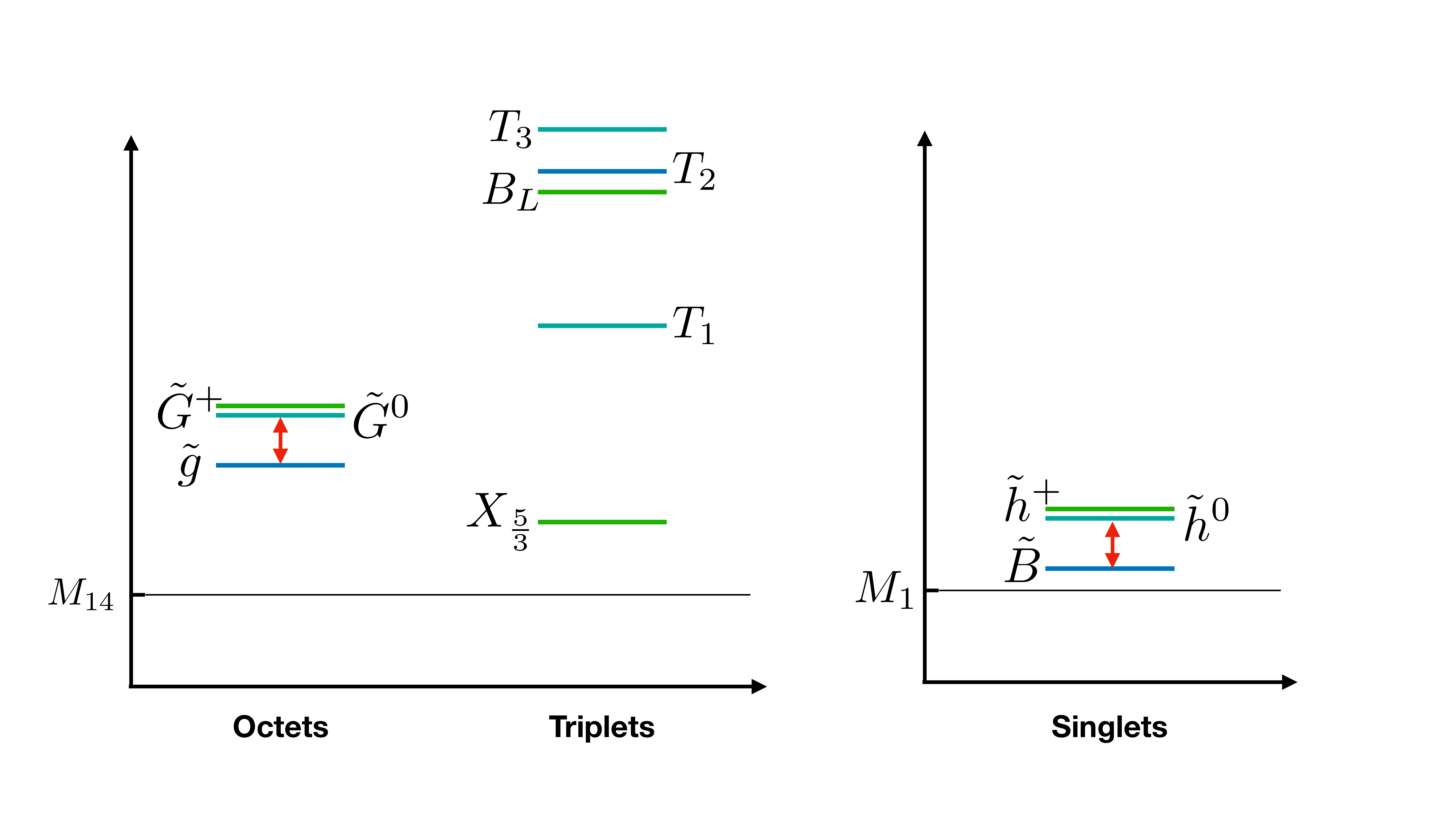}
	\caption{Template spectrum of the chimera baryon states, where $T_{1,2,3}$ are the mass eigenstates with the same charge as the top quark. Together with $B_L$, they receive a positive mass shift due to the mixing with the top. The mass splitting highlighted by the red arrow also receives significant contribution from the different masses in the $\psi$ sector.}
	\label{fig:spectrum}
\end{figure}

A typical spectrum for the colored top partners is illustrated in Fig.~\ref{fig:spectrum} (left). The triplets that have charges matching the top and bottom quarks, i.e. $2/3$ and $-1/3$, receive large positive corrections due to the mixing with the top fields, which makes them heavier than the octets and of $X_{5/3}$. On the other hand, the mass difference between the octets and $X_{5/3}$ is due to QCD corrections.
A crude estimate can be obtained starting from the electromagnetic contribution to the mass split between proton and neutron in QCD. We take as a starting point the results of \cite{Gasser:2020mzy} on the electromagnetic contribution to the proton neutron mass difference
$\Delta m^{\rm em} \simeq 0.58$~MeV, which is also confirmed by lattice studies \cite{Borsanyi:2014jba}.
This corresponds roughly to a relative mass difference
$r= 2 \Delta m^{\rm em}/(m_P+m_N) \simeq 6 \cdot 10^{-4}$.
We then re-scale this by the gauge couplings and the difference of Casimir factors for the two QCD representations, to obtain
\begin{align}
\frac{2 (m_{\tilde{g}} - m_{X_{5/3}})}{m_{\tilde{g}} + m_{X_{5/3}}} \approx \frac{\alpha_S (\text{TeV})}{\alpha_{em} (\text{GeV})} \left(3   -\frac{4}{3}\right)\ r \sim 1.4\%\,. 
\end{align}
This estimate confirms that the octet top partners are dominantly produced at the LHC compared to the usual triplet top partners. 
It is also interesting to note that the mass difference between the octonis and the gluoni is generated by EW corrections, which we estimate to be
\begin{align} \label{eq:weak_splitting}
\frac{2 (m_{\tilde{G}} - m_{\tilde{g}})}{m_{\tilde{G}} + m_{\tilde{g}}} \approx \frac{3}{4} \frac{1}{\sin^2 \theta_W} \ r \sim 0.2\%\,,
\end{align}
where we included the dominant contribution of $\SU(2)_L$. 
Note, however, that a contribution to this mass difference is also generated by $\SU(5)$-breaking mass differences in the $\psi$ sector, which could go in either direction.
Finally, we expect the charged octoni to be slightly heavier than the neutral one due to EW symmetry breaking effects.

The final states resulting from octet top partner decays depend on the mass hierarchy with the singlet top partners and the QCD-colored pNGBs. We reasonably expect the pNGBs to be lighter than the colored hyper-baryons. The singlet top partners, however,  receive a mass $M_1 \neq M_{14}$ from the confining strong dynamics, and the precise values can only be obtained from lattice studies. We will explore two different scenarios for $M_1$. As discussed in the last section, the lightest state among $\tilde{B}, \tilde{h}^{0,\pm}, \pi_3$ is stable, unless additional baryon or lepton number violating interactions are added to the model. If the lightest singlet top partner $\tilde{B}$ or $\tilde{h}^0$ is lighter than $\pi_3$, a stable $\tilde{B}/\tilde{h}^0$ provides potential Dark Matter (DM) candidates -- case A. If $\pi_3$ is lighter than $\tilde{B}$, a stable $\pi_3$ is not viable, and new interactions need to be present, which open baryon and/or lepton number violating $\pi_3$ decay channels -- case B.  Note that case B could also occur if the singlet top partners are the lightest. Both cases carry resemblance with SUSY signatures: gluinos decaying into tops plus missing transverse energy in case A, and R-parity violating decays in case B.

\subsection{Scenarios with a DM candidate}
\label{sec:LHCphenoDM}

This scenario can occur if the singlet top partners,  boni or higgsonis, are lighter than the colored pNGBs, and cannot decay into any SM final state being the lightest hyper-baryons. We also assume throughout this work that the colored top partners are heavier than the pNGBs. Henceforth, the lightest color-singlet baryon can be a DM candidate or, at least, be detector stable if it decays via higher order operators.

The top partners $\tilde h^0$, $\tilde h^+$ and $\tilde B$ and the pNGB $\pi_3$ have quantum numbers resembling the higgsino-bino and right-handed stop sectors of SUSY models, and their phenomenology depends on the mass mixing among them.
Considering only the EW interactions, one finds
\begin{align}
m_{\tilde h^+} \gtrsim m_{\tilde h^0} >  m_{\tilde B}\,, 
\end{align}
as shown in \cref{fig:spectrum} (right), with a mass splitting estimated to range around or less than $0.2$\% of the mass, i.e. below a GeV for masses in the TeV range, see \cref{eq:weak_splitting}.  A small mixing is also generated by the EW symmetry breaking. However, the mass difference between the boni and the higgsonis also receives a sizable contribution from the $\SU(5)$-breaking mass differences between the singlet and bi-doublet hyper-fermions $\psi$, which can go in either direction. The most natural expectation is that the spectrum remains fairly compressed, hence
we would expect $\tilde h^+$ and $\tilde h^0$ to decay into soft leptons and mesons plus $\tilde B$.
Thus, all three particles would effectively contribute to the missing transverse momentum as
the soft decay products are hardly registered in the detectors. Such a scenario
might, therefore, be easily confused with a SUSY model at first glance. 

The QCD-colored pNGBs $\pi_3$ and $\pi_8$ are heavier than the EW ones as their masses receive contributions from QCD loops \cite{Cacciapaglia:2015eqa}. As $\pi_3$ carries baryon number, its decay modes are strongly constrained: In the scenario under consideration, the only available decay channels are
\begin{align} \label{eq:pi3decayA}
\pi_3 \to t  \tilde B,\ t \tilde{h}^0,\ b \tilde{h}^+\,.
\end{align}
The first one dominates if $y_R' \gg y_L'$, while the decays in the higgsonis dominate for $y_L' \gg y_R'$.
In principle, decays into lighter families, like $c \, \tilde B$ and $u \, \tilde B$, are also possible, but in the spirit of composite Higgs models we expect those to be strongly suppressed. Hence, $\pi_3$ behaves exactly like a right-handed stop in supersymmetry, and LHC bounds from scalar top quark searches can be directly applied \cite{CMS:2019zmd,ATLAS:2020xzu,CMS:2021eha}. We will come back to this in \cref{sec:numerics}.

The color octet $\pi_8$ can decay directly into a pair of SM states:
\begin{align} \label{eq:pi8decays}
    \pi_8 \to t\bar{t},\ gg,\ g\gamma,\ gZ\,,
\end{align}
with decays into a pair of light jets as subleading channels. We will discuss bounds on $\pi_8$ further in \cref{sec:caseC}.

The octet top partners $\tilde G^{0,+}$ and $\tilde g$ feature the largest pair-production cross sections at hadron colliders such as the LHC or a prospective 100 TeV pp-collider. 
In the scenario considered here, their decays lead to final states similar to those of gluinos
in SUSY models. This applies, in particular, to the Majorana gluoni $\tilde g$ that decays via the following channels
\begin{subequations}
\begin{align}
\tilde g &\to \pi_3 \,\bar{t}\,,\,   \pi_3^* t \to t\,\bar{t} \, \tilde B \ / \ t\,\bar{t} \, \tilde{h}^0 + t\,\bar{t} \, \overline{\tilde{h}^0} \ / \ b\,\bar{t} \, \tilde{h}^+ +  t\,\bar{b} \, \tilde{h}^-  \,\,\, \mbox{and/or} \label{eq:gpi3dec}\\
\tilde g   & \to \pi_8 \tilde B \to t\,\bar{t} \, \tilde B\ / \  g\,g \, \tilde B \, ,\label{eq:gpi8dec}
\end{align}\end{subequations}
depending on the mass spectrum.
In all cases the final states contain large missing transverse momentum as we assume the $\tilde B$ to be collider stable. This implies that gluino searches at the LHC can be used to constrain these scenarios. 
A similar comment applies in case of the Dirac states $\tilde G^0$ and $\tilde G^+$, which -- depending on the mass spectrum -- decay into the following channels
\begin{subequations}\begin{align}
\tilde G^0  &\to \pi_3 \,\bar{t}  \to t\,\bar{t} \, \tilde B \ / \ t\,\bar{t} \, \tilde{h}^0 \ / \ b\,\bar{t} \, \tilde{h}^+ \,\,\, \mbox{and/or} \label{eq:G0pi3dec}\\
\tilde G^0  &\to \pi_8 \, \tilde h^0   \to t\,\bar{t} \, \tilde h^0 \ /\ g \, g \, \tilde h^0\,;  \label{eq:G0pi8dec}\\
\tilde G^+  &\to \pi_3 \,\bar{b}  \to t\,\bar{b} \, \tilde B \ / \ t\,\bar{b} \, \tilde{h}^0 \ / \ b\,\bar{b} \, \tilde{h}^+ \,\,\, \mbox{and/or}\label{eq:G+pi3dec}\\
\tilde G^+  &\to \pi_8 \, \tilde h^+   \to t\,\bar{t} \, \tilde h^+ \ / \ g \, g \, \tilde h^+ \,.\label{eq:G+pi8dec}
\end{align} \end{subequations}
Note that $\tilde{G}^0$ resembles a Dirac gluino in extended SUSY models \cite{Choi:2008pi}, while the charge one gluoni $\tilde{G}^+$ is a novel state from composite models without a SUSY analog.
For completeness, we note that for a fixed mass the QCD production cross sections fulfill the relation
\begin{align}
\sigma(p\,p\to \tilde G^0 \, \overline{\tilde G^0}) = \sigma(p\,p\to \tilde G^+ \, \tilde G^-)
= 2\, \sigma(p\,p\to \tilde g \, \tilde g)\,.
\end{align}
In addition, there are  subdominant EW production cross sections for the $\SU(2)_L$ doublet such as
\begin{align}
p \, p \to \overline{\tilde G^0} G^+ \,, \; \tilde G^0 G^-\,. 
\end{align} 

In Section \ref{sec:numerics} we will focus on the final states discussed above, and present numerical studies of the bounds coming from current LHC searches. For simplicity, we will focus on the case $y'_R \gg y'_L$, so that only $\tilde{B}$ appears in the $\pi_3$ decays.

\subsection{Scenarios without a DM candidate} 

If $\pi_3$ is lighter than $\tilde{B}$, lepton or baryon number violating interactions need to be included in order to avoid a stable $\pi_3$. As we have seen in the previous section, these interactions are not allowed by the symmetries of the low energy Lagrangian (including the top partial compositeness couplings), hence they must be generated by new couplings in the UV completion. The corresponding operators can then be added to the low energy effective Lagrangian via appropriate spurions. Also, $\tilde{B}$ will be allowed to decay via the inverse process in \cref{eq:pi3decayA}. 
The simplest possibilities are 
\begin{align} \label{eq:pi3decayBb}
\pi_3 \to \bar{d}_i \, \bar{d}_j \quad \text{ with } d_i = d,s,b \text{ and } i\ne j 
\end{align}  
or
\begin{align} \label{eq:pi3decayBl}
\pi_3 \to u_i \, \nu_{l_j} \,,\, d_i \, l_j \quad \text{ with } u_i = u,c,t \text{ and } 
l_j = e,\mu,\tau\,. 
\end{align}  
The former violates baryon number whereas the latter violates lepton number.
This implies that only one of the two interaction types can be present as otherwise there would be the danger of proton decays at a rate incompatible with experiment. This scenario corresponds to typical R-parity violating SUSY models for the stop decays, although the origin of the couplings is very different from the SUSY case and no R-parity analog exists in the composite model.

The QCD-singlet top partners $\tilde h^0$, $\tilde h^+$ and $\tilde B$ can decay according
to 
\begin{subequations}\begin{align}
\tilde B &\to \pi^*_3 \, t \,,\, \pi_3 \, \bar{t}\,, \\
\tilde h^0 & \to \pi_3  \, \bar{t}\,, \\
\tilde h^+ & \to \pi_3  \, \bar{b} \,.
\end{align}\end{subequations}
Furthermore, there could be mixing of $\tilde h^0$ and $\tilde h^-$ with the left-handed leptons,
extending the spirit of partial compositeness to the leptonic sector. In such a case one can well
imagine that $\tilde B$ plays the role of a heavy right-handed neutrinos. Additional decay channels into EW gauge bosons and pNGBs would be present, such as
\begin{subequations}\begin{align}
\tilde h^0 & \to Z \, \nu \,,\, W^+ \, l^- \,,\, h\, \nu \, ,\, \eta^+_{3,5} l^- \\
\tilde h^+ & \to W^+ \, \nu \,,\, Z \, l^+ \,,\, h\, l^+ \, ,\, \eta^+_{3,5} \nu \\
\tilde B & \to h\, \nu \, ,\, \eta^+_{3,5} l^- \, ,\, \eta^-_{3,5} l^+ \,,
\end{align}\end{subequations}
to name a few. Note that this possibility is only compatible with the $\pi_3$ decay in \cref{eq:pi3decayBl}, as it involves lepton number violation, and it also holds if $\pi_3$ is heavier than these states.

The final states from the decays of the color-octet baryon will contain additional jets and leptons from the new decays of $\pi_3$ and the singlet baryons, and reduced missing transverse momentum. We leave a detailed study of these signatures to future investigations.

\subsection{Phenomenology of other composite states} \label{sec:caseC}

The model contains other composite states, whose phenomenology does not depend on the cases A and B, and we summarize them here.

Firstly, the EW coset $\SU(5)/\SO(5)$ entails the presence of additional pNGBs besides the Higgs boson. In terms of the EW interactions, they resemble the scalar sector of the Georgi-Machacek (GM) model~\cite{Georgi:1985nv,Chanowitz:1985ug} with an additional pseudo-scalar singlet. However, there are important phenomenological differences: firstly, the composite model contains topological interactions with the EW gauge bosons \cite{Dugan:1984hq} that are only loop-induced in the GM model. Furthermore, due to the absence of a vacuum expectation value, no large couplings to a pair of EW gauge bosons is generated.  The couplings to fermions, instead, depend crucially on the operators that generate masses for the light SM fermions \cite{Agugliaro:2018vsu}. The EW pNGBs can be produced in Drell-Yann, pair produced via their EW gauge interactions, or appear in the final states of the decays of top partners, as we will discuss below.

The decays and the resulting LHC phenomenology of the octet pNGB $\pi_8$ are also independent on the spectrum, as listed in \cref{eq:pi8decays}. Their phenomenology has been widely studied, both in the case of composite models \cite{Cacciapaglia:2015eqa,Belyaev:2016ftv,Cacciapaglia:2020vyf} and in effective and supersymmetric  set-ups (sgluons) \cite{Plehn:2008ae,Choi:2008ub,Chen:2014haa,SekharChivukula:2014uoe,Darme:2018dvz,Carpenter:2020hyz,Darme:2021gtt,Carpenter:2021vga}. 
In the following we will assume that $\pi_8$
can either decay into $t\,\bar{t}$ and/or $g\,g$.~\footnote{From the experimental point of view, decays into $g\,g$ and into two quarks of the first two generations lead to two jets which are
not distinguished at the LHC. Thus, our investigation in \cref{sec:numerics} also covers
the case of decays into light quarks.}  Current bounds on the $\pi_8$ mass from LHC searches for pair production with dominant decays into $t\bar{t}$ and $gg$ lie at $\sim 1.05$~TeV and $\sim 0.85$~TeV, respectively, while bounds from single production are strongly model dependent \cite{Cacciapaglia:2020vyf}.

In this model under consideration, the usual color triplet top partners, $B$, $T_i$  and $X_{5/3}$, have additional exotic decay modes \cite{Bizot:2018tds} beside the ones used by the ATLAS and CMS for the searches. In absence of the exotic decays, top partner pair production searches ATLAS and CMS established bounds on the top partner mass of the order of $1.3$ - $1.6$ TeV (depending on the top partner branching ratios) \cite{Aaboud:2017zfn,Aaboud:2017qpr,Aaboud:2018xuw,Aaboud:2018saj,Aaboud:2018xpj,Aaboud:2018wxv,Aaboud:2018pii,Sirunyan:2017pks,Sirunyan:2018qau,Sirunyan:2018omb,Sirunyan:2019sza,Sirunyan:2018yun,Sirunyan:2020qvb,Aaboud:2018ifs,Sirunyan:2017ynj,Sirunyan:2018fjh,Sirunyan:2018ncp,Sirunyan:2019xeh,ATLAS:2021ibc}, but the presence of exotic decay modes can alter these mass bounds.
We list the possible decays for the color triplet top partners below, starting from the usually considered decays into third generation quarks and EW bosons, followed by the exotic decays with other EW and QCD-colored pNGBs:
\begin{subequations}\begin{align}
B &\to t \, W^- \,,\, Z \, b \,,\, \eta^-_{3} \,t \,, \eta^-_5 \,t \,, \eta \, b  \,, \eta^0_1 \, b  \,, \eta^0_5 \, b \,,\, \pi_8 \, b  \,,\, \pi_3 \, \tilde h^-\,, \\
T_i &\to  b \, W^+ \,,\,Z \, t \,, h \, t \,,\,  \eta \, t \,,\, \eta^0_{1,3,5} \, t\,,\, \eta^+_{3,5} \, b\,,\,  \pi_8 \, t  \,,\, \pi_3 \, \tilde h^0 \,,\, \pi_3 \, \tilde B\,, \\
X_{5/3} &\to  t \, W^+ \,,\,\  \eta^+_{3,5} \, t \,, \eta^{++}_5 \, b \,,\, \pi_3 \, \tilde h^+ \,.
\end{align}\end{subequations}
The spectrum in Fig.~\ref{fig:spectrum} also allows decays into the color octets $\tilde g$, $\tilde G^0$ or $\tilde G^+$ plus a colored pNGB, however due to the large mass of $\pi_3$ and $\pi_8$ we assume here that they are kinematically forbidden.
The signatures from top partner pair production will consist of at least two
jets (stemming either directly from a $b$ and/or from a $b$ resulting from a $t$ decay)
in combination with jets and leptons stemming from the decays of the electroweak bosons. Moreover, there will be substantial missing transverse momentum in final states with decays into $\tilde B$ and $\tilde h$. 
If decays into the strongly interacting pNGBs dominate one gets signatures alike those of color octet baryons discussed above. 
In case of single production of the color triplet top partners one has on the one hand the commonly considered signatures arising from the decay into a third generation quark and a SM boson. On the other hand, one has also final states consisting of a top quark and two vector bosons or a single top quark plus missing transverse momentum.
Several, but by far not all of the exotic decay channels have been discussed in the literature \cite{Chala:2017xgc,Aguilar-Saavedra:2017giu,Bizot:2018tds,Han:2018hcu,Kim:2018mks,Alhazmi:2018whk,Xie:2019gya,Cacciapaglia:2019zmj,Benbrik:2019zdp,Aguilar-Saavedra:2019ghg,Wang:2020ips,Corcella:2021mdl}.

\section{LHC bounds on fermionic color octets}
\label{sec:numerics}

In this section, we provide a first phenomenological study of the fermionic color octets. In the model M5, they are part of the same $\Sp(6)$ multiplet as the color triplet top partners, and -- as discussed in \cref{sec:LHCphenoDM} -- they are expected to have masses comparable to the usual top partners. Due to their QCD representation, the production cross section for the octet top partners will be significantly
larger than the one for the usual top partners, as is shown in \cref{fig:q3q8xs}. Henceforth, the octet top partner signatures might be the first sign of the model, if realized in nature. We focus on the scenario with DM candidate described in \cref{sec:LHCphenoDM}.

Based on the discussion in \cref{sec:LHCphenoDM}, we make a number of simplifying assumptions for our phenomenological study. We assume all the octet top partners $Q_8=(\tilde{g}, \tilde{G}^{+,0})$ to be mass degenerate. The boni $\tilde{B}$ is assumed to be the lightest hyper-baryon which is taken to be (at least collider scale) stable.
Furthermore, we assume the  higgsonis $\tilde{h}^{+,0}$ to be nearly mass degenerate with the boni and that they promptly decay to the boni plus soft leptons. Therefore all (color) singlet top partners $Q_1=(\tilde{B},\tilde{h}^{+,0})$ have a common mass scale. Moreover, we assume that $\pi_3$ decays dominantly into $t\tilde B$ for simplicity.

We study three scenarios which cover the different possible octet top partner decay channels based on the spectra and interactions discussed in \cref{sec:LHCphenoDM}:
\begin{itemize}
    \item Scenario 1: Octet top partners decay dominantly to $\pi_3$ and 3rd generation quarks.
    \item Scenario 2: Octet top partners decay dominantly to $\pi_8$ and singlet top partners, assuming decays $\pi_8\to t \bar{t}$ or/and $\pi_8\to j j$.
    \item Scenario 3: Octet top partners decay through both the above channels (with comparable branching ratios).
\end{itemize}
For each scenario, we determine bounds from current LHC searches for gluino pair production and decays in the various channels. In all cases we constrain ourselves to kinematical configurations where two-body decay channels are open. The case where two-body decays are kinematically forbidden will be left for a future study.
\bigskip

\emph{Simulation setup and bound setting:} We implement effective models for the octet top partners $Q_8=(\tilde{g}, \tilde{G}^{+,0}$) (gluoni and octonis), the color singlet top partners $Q_1=(\tilde{B}, \tilde{h}^{+,0}$) (boni and higgsonis), the scalar pNGB color triplet $\pi_3$, the scalar color octet $\pi_8$, the EW pNGBs and their interactions with SM particles in \textsc{FeynRules} \cite{Alloul:2013bka,Christensen:2009jx,Degrande:2011ua} to generate a leading-order \textsc{LO UFO} model. Details about the implementation can be found in \cref{app:FRimp}. We use \textsc{MADGRAPH5\_AMC@NLO} \cite{Alwall:2014hca} with the LO set of \textsc{NNPDF 3.0} of parton densities \cite{NNPDF:2014otw,Buckley:2014ana} in conjunction with \textsc{PYTHIA 8} \cite{Sjostrand:2014zea} to produce hadron-level events of pair-produced octet top partners with various decay modes. We simulate events at LO and rescale production cross sections for the Majorana gluoni to the NNLO$_\text{approx}$+NNLL result for gluinos with the corresponding mass of Ref.\cite{Beenakker:2016lwe}. For octonis, we rescale to twice the gluino cross section, correspondingly.

To determine bounds from LHC searches, we pass the generated signal events to \textsc{MADANALYSIS 5} version 1.8.44 \cite{Conte:2012fm,Conte:2014zja,Dumont:2014tja,Conte:2018vmg} for detector simulation, event reconstruction based on \textsc{DELPHES 3} \cite{deFavereau:2013fsa} and the \textsc{FASTJET} \cite{Cacciari:2011ma} implementation of the anti-$k_T$ algorithm \cite{Cacciari:2008gp}, and the extractions of $CL_s$ exclusions relative to the ATLAS and CMS searches at the LHC with $\sqrt{s}=13$~TeV which are publicly available in the \textsc{MADANALYSIS 5 PAD}. Analogously, we determine bounds by passing the signal events to \textsc{CheckMATE} \cite{Drees:2013wra,Dercks:2016npn} to test the model against the publicly available LHC search implementations in \textsc{CheckMATE} version 2.0.29. In this section, all bounds presented are the maximal $95\%$ CL bound obtained from an individual search available from \textsc{MADANALYSIS 5} or \textsc{CheckMATE}. We do not attempt any combination. As could be expected, we find that the most sensitive available searches for the final states under consideration are ATLAS and CMS searches for stops and gluinos \cite{CMS:2019zmd,ATLAS:2019vcq,ATLAS:2018yhd}. In \cref{app:tools_comparison}, we provide more details on the exclusion power of various existing searches in a few sample scenarios.

\subsection{Scenario I: $Q_8\to\pi_3+t/b$}

\begin{figure}[t]
{	\centering
		\begin{subfigure}[]{0.31\linewidth}
		\centering
		\includegraphics[width=\textwidth]{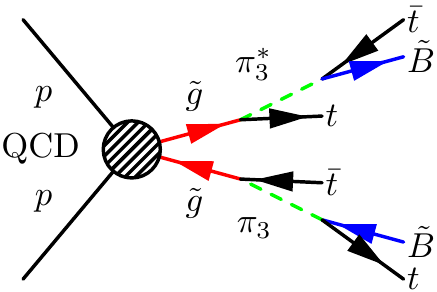} 
		\caption{$\tilde g\to \bar t\pi_3, t\pi_3^*\to \bar t t\tilde B$}
		\end{subfigure}
				\begin{subfigure}[]{0.31\linewidth}
		\centering
		\includegraphics[width=\textwidth]{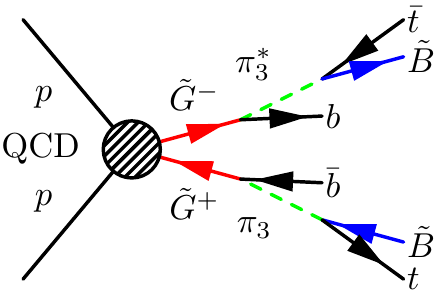} 
		\caption{$\tilde G^+\to \bar b\pi_3\to \bar bt\tilde B$}
		\end{subfigure}
				\begin{subfigure}[]{0.31\linewidth}
		\centering
		\includegraphics[width=\textwidth]{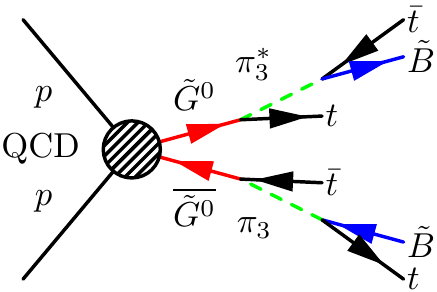} 
		\caption{$\tilde G^0\to \bar t\pi_3\to \bar tt\tilde B$}
		\end{subfigure}
		}
	\caption{Feynman diagrams of QCD pair production of octet top partners $Q_8=(\tilde g,\tilde G^+,\tilde G^0)$ with dominant decays  $Q_8\to\pi_3+t/b$.}	
	\label{fig:S1graphs}
\end{figure}

We first investigate the case in which the color octet top partners dominantly decay to color triplet pNGBs. The corresponding decay modes for the octet top partners $\tilde g,\tilde G^+,\tilde G^0$ are given in \cref{eq:gpi3dec,eq:G0pi3dec,eq:G+pi3dec}. The final states from QCD pair production are $4 t + \ptmiss$ in the cases of $\tilde g$ or $\tilde G^0$ pair production and
$2b + 2t + \ptmiss$ in case of $\tilde G^+\tilde G^-$ production. 
\cref{fig:S1graphs} shows the corresponding Feynman diagrams.

In the M5 model, all three states are present, and their decays populate the signal regions. It is nevertheless instructive to also determine bounds for the individual processes. Process (a) in \cref{fig:S1graphs} resulting from Majorana gluoni pair production is \emph{identical} to gluino pair production with a gluino to stop-top decay and the stop decay to top and bino in SUSY models. The bounds determined from solely this process thus provide a valuable cross check of the implementation and simulation chain. Process (c) from the neutral octoni pair production yields an identical final state with identical kinematics, but the octoni pair production cross section is twice as large. Process (b) in \cref{fig:S1graphs} resulting from charged octoni pair production yields a different final state and, to our knowledge, it has not been discussed in the literature yet. Thus, we also provide results assuming the sole presence of a charged octoni as these results could be of interest for other models containing a charged color octet fermion.

The pair production cross section is solely a function of the octet top partner mass, but the kinematics of the processes depend in addition on the boni mass $m_{\tilde B}$ and the mass of the color triplet pNGB $m_{\pi_3}$, leaving us with 3 relevant mass parameters. To present results, we chose two kinematically different setups: For setup (i) the mass difference $m_{Q_8}-m_{\pi_3}=200$~GeV
is fixed, and we scan over the octet and singlet top partner masses. In this case the $t/b$ of the $Q_8$ decay has little momentum in the $Q_8$ rest-frame. For setup (ii) we fix $m_{\pi_3}=1.4$~TeV and again scan over $m_{Q_8}$ and $m_{Q_1}$.

\begin{figure}[t]
	\centering
	\begin{subfigure}[]{0.31\linewidth}
		\centering
		\includegraphics[width=\textwidth]{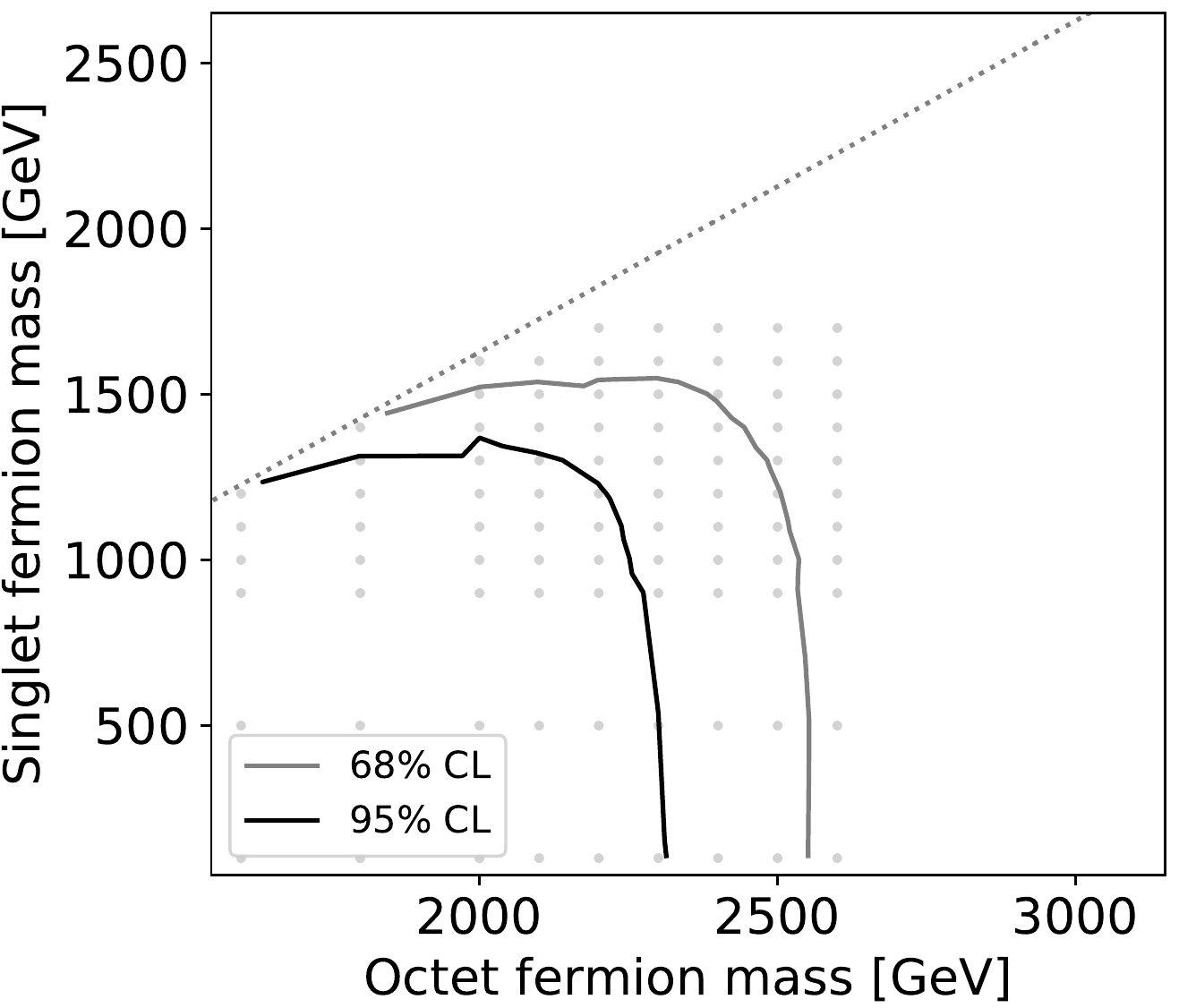} 
		\caption{$\tilde g\to \bar t\pi_3, t\pi_3^*\to \bar t t\tilde B$,\\ $m_{\tilde g} - m_{\pi_3} = 200$~GeV}\label{fig:fig1_gluoni_stop_varyingmass_MA}
	\end{subfigure}
	\begin{subfigure}[]{0.31\linewidth}
		\centering 
		\includegraphics[width=\textwidth]{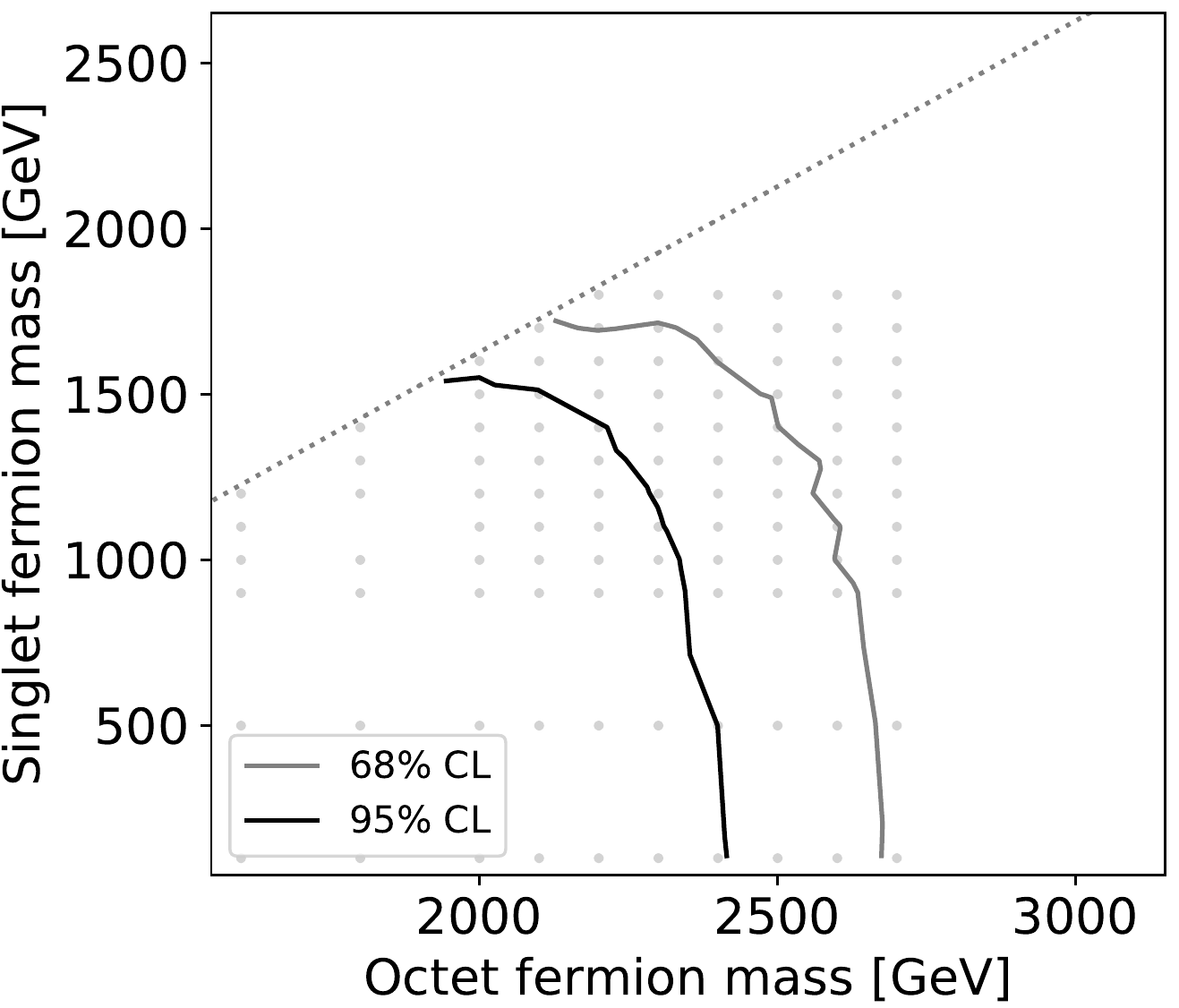} 
		\caption{$\tilde G^+\to \bar b\pi_3\to \bar bt\tilde B$,\\ $m_{\tilde G^+}-m_{\pi_3} = 200$~GeV}\label{fig:fig3_Gplus_stop_varyingmass_MA}
	\end{subfigure}
	\begin{subfigure}[]{0.31\linewidth}
		\centering
		\includegraphics[width=\textwidth]{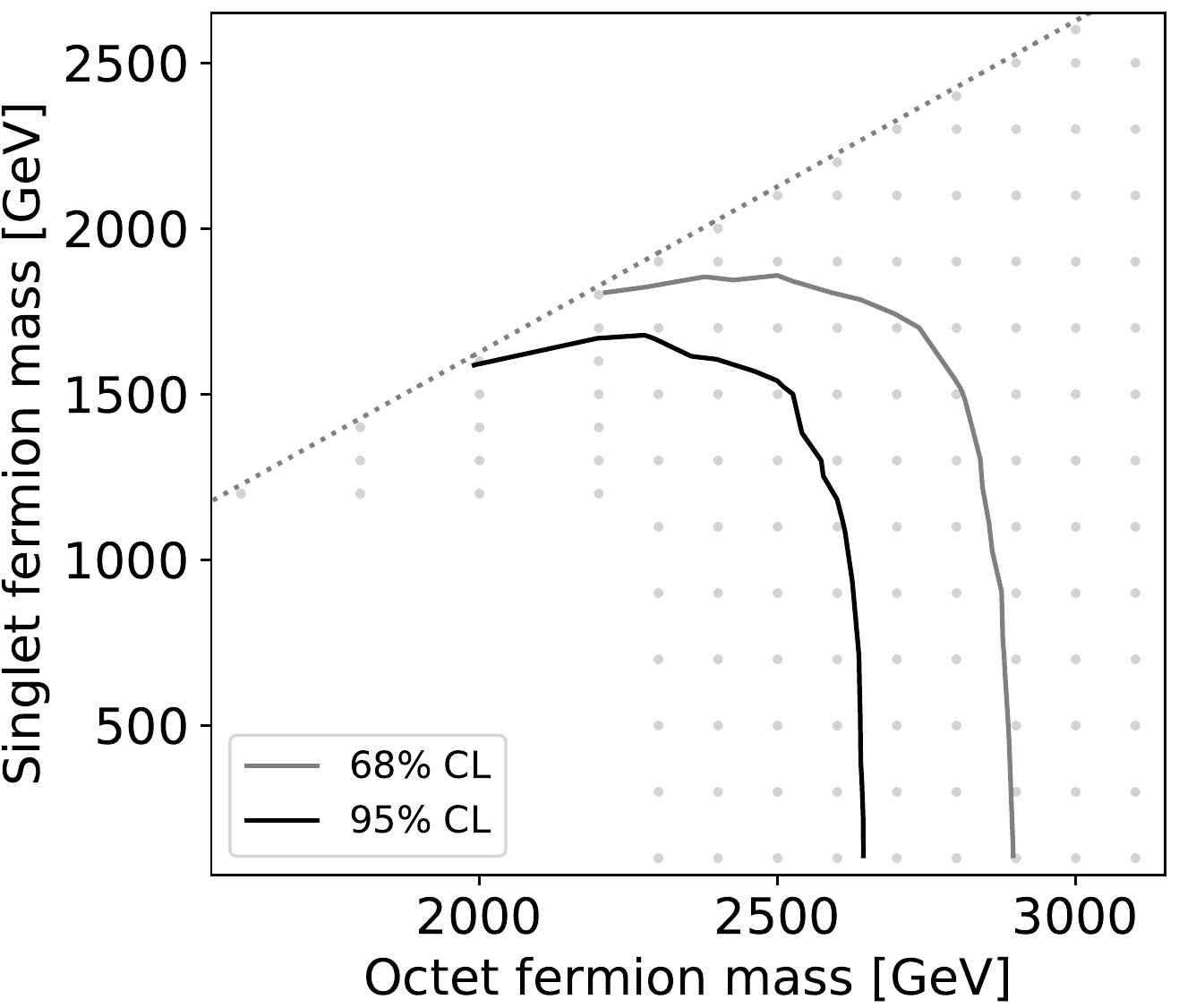} 
		\caption{$Q_8\to \bar q\pi_3\to \bar qt\tilde B$,\\ $m_{Q_8} - m_{\pi_3} = 200$~GeV}\label{fig:fig5_multiplet_stop_varyingmass_MA}
	\end{subfigure}
	\vspace{1.5ex}

	\begin{subfigure}[]{0.31\linewidth}
		\centering
		\includegraphics[width=\textwidth]{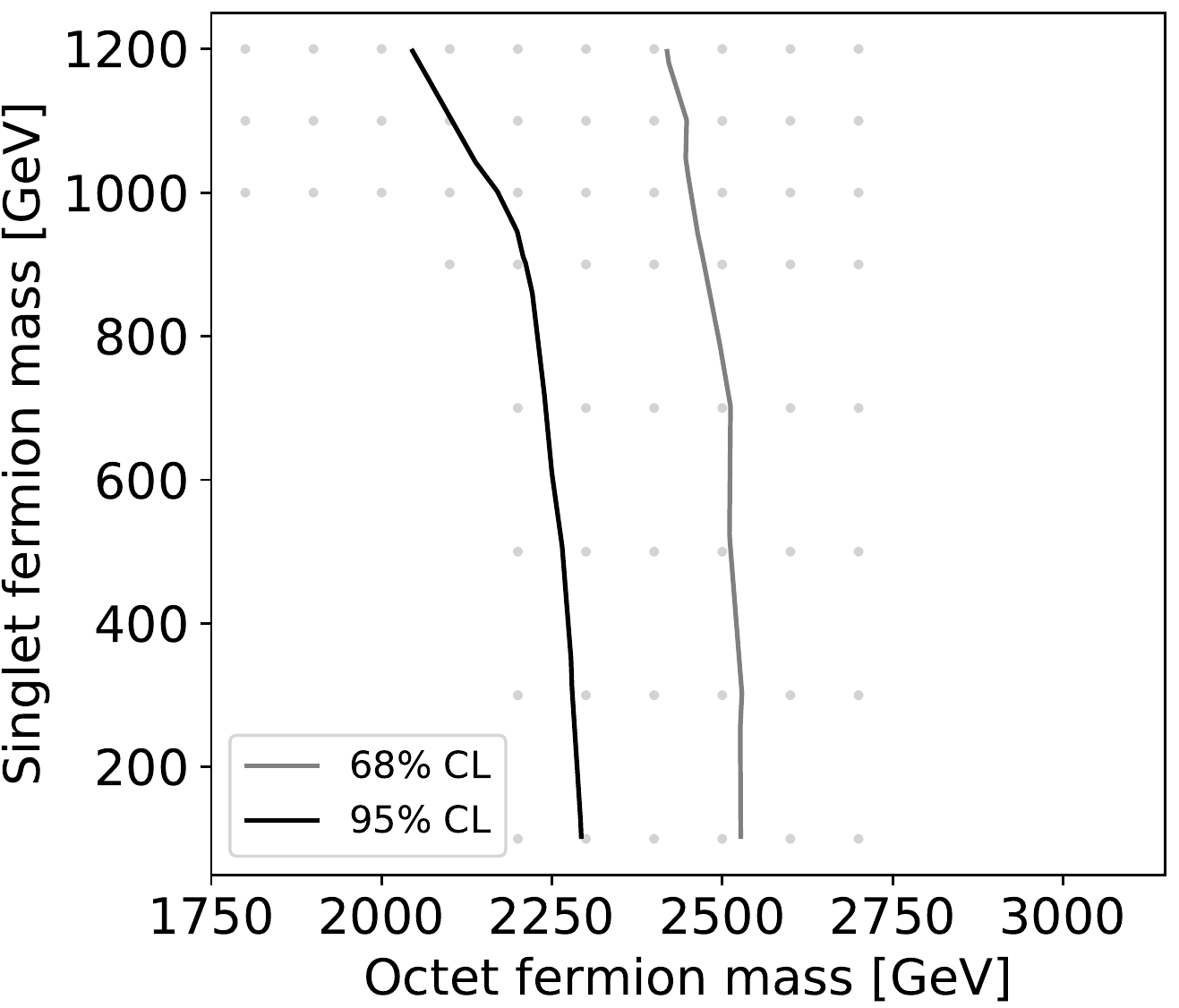} 
		\caption{$\tilde g\to \bar t\pi_3, t\pi_3^*\to \bar t t\tilde B$,\\ $m_{\pi_3}=1.4$~TeV}\label{fig:fig2_gluoni_stop_fixedmass_MA}
	\end{subfigure}
	\begin{subfigure}[]{0.31\linewidth}
		\centering
		\includegraphics[width=\textwidth]{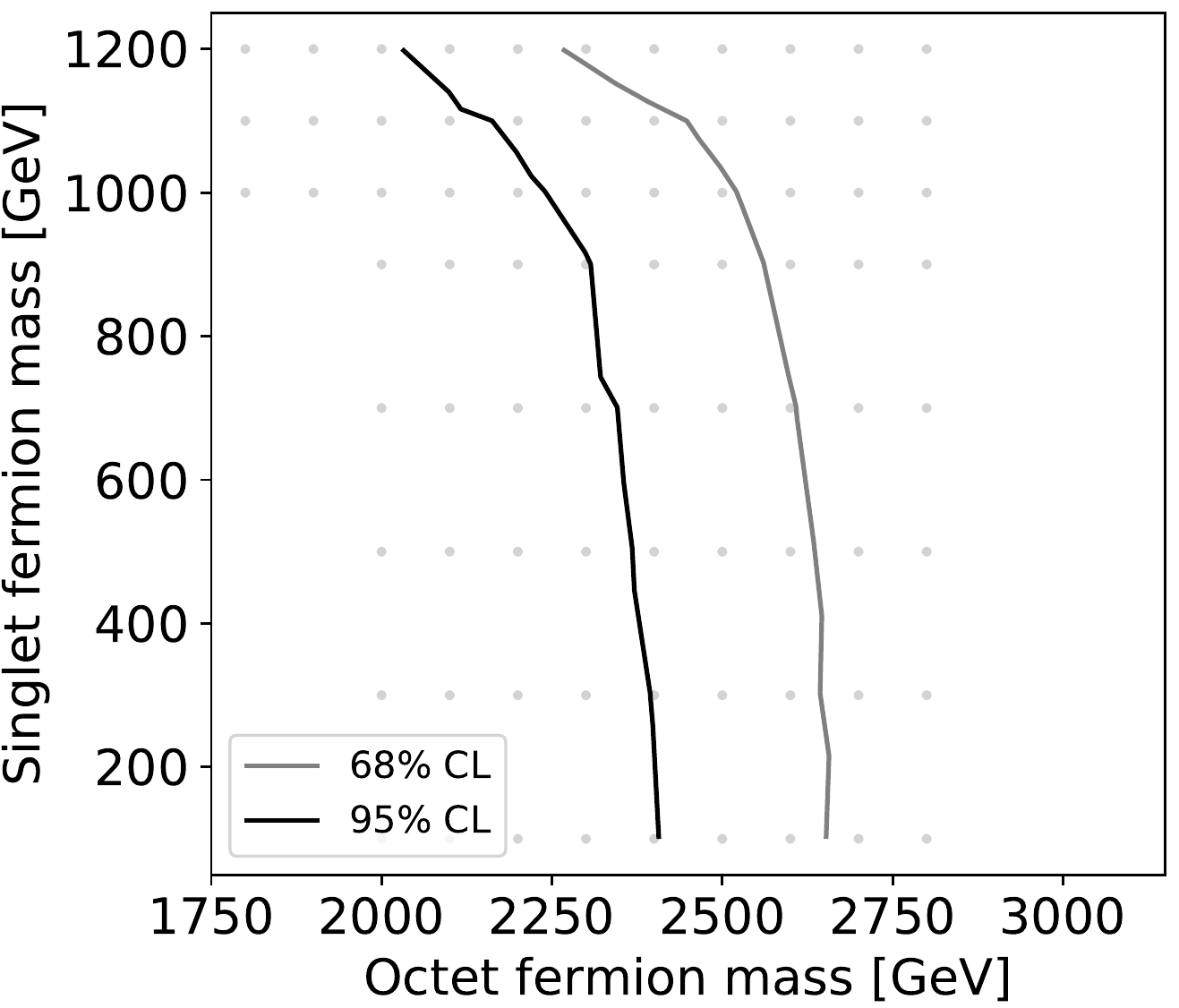} 
		\caption{$\tilde G^+\to \bar b\pi_3\to \bar bt\tilde B$,\\ $m_{\pi_3}=1.4$~TeV}\label{fig:fig4_Gplus_stop_fixedmass_MA}
	\end{subfigure}
	\begin{subfigure}[]{0.31\linewidth}
		\centering
		\includegraphics[width=\textwidth]{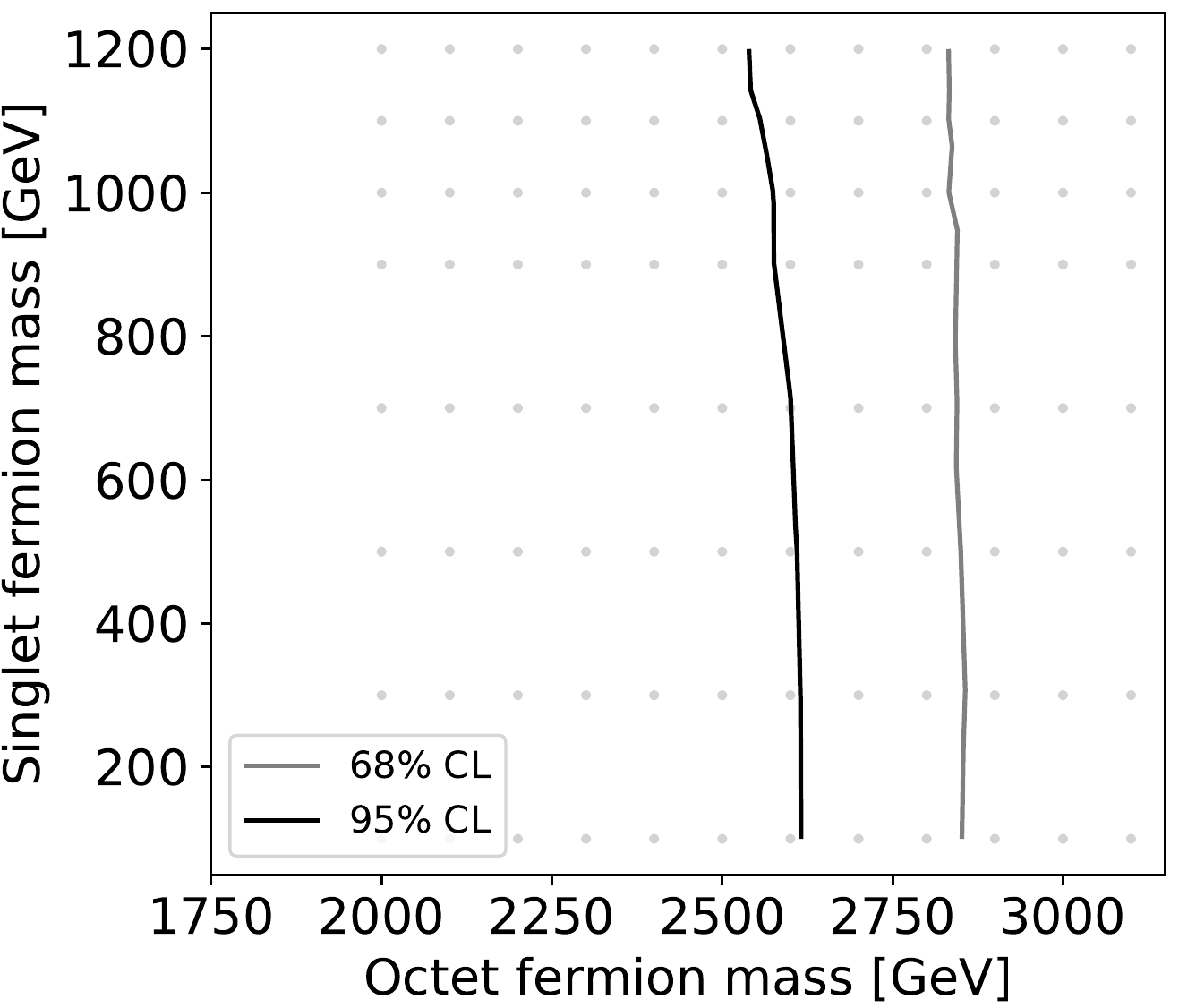} 
		\caption{$Q_8\to \bar q\pi_3\to \bar qt \tilde B$,\\ $m_{\pi_3}=1.4$~TeV}\label{fig:fig6_multiplet_stop_fixedmass_MA}
	\end{subfigure}
	\caption{Bounds on the fermion masses for QCD pair production of an octet fermion $Q_8$ with subsequent decay to a SM third generation quark $t/b$ and  $\pi_3$. 
	The first row shows the case of a constant difference between the octet mass and the $\pi_3$ mass of 200~GeV below the octet mass. In the second row we have fixed $m_{\pi_3}=1.4$~TeV.
	We include only the $\tilde g$ in the first and $\tilde G^+$ in the second column.
	The third column shows the bounds for the full octet multiplet $Q_8=(\tilde g,\tilde G^+,\tilde G^0)$, assuming it is mass degenerate.}
	\label{fig:bounds_stop}
\end{figure}

Figure \ref{fig:bounds_stop} shows the resulting bounds assuming only pair production of $\tilde{g}$ (left column), only production of $\tilde G^+\tilde{G}^-$ (middle column) and pair production of all color octet top partners (right column), for $m_{Q_8}-m_{\pi_3} = 200$~GeV (top row) and for fixed $m_{\pi_3}=1.4$~TeV (bottom row). Grey dots mark the scan points. For each scan point, we generated $10^5$  $\tilde{g}\tilde{g}$, $\tilde{G^+}\tilde{G^-}$, and $\tilde{G^0} \overline{\tilde{G^0}}$ events which were analyzed with LHC searches available in \textsc{MADANALYSIS 5} and \textsc{CheckMATE}. In most of the parameter space, the leading bounds arise from the CMS search focusing on multi jets plus missing transverse momentum  \cite{CMS:2019zmd,Mrowietz:2020ztq}.\footnote{See \cref{app:tools_comparison} for a comparison of bounds from various searches.} 
The black and grey  lines show the obtained 95\% and 68\% CL exclusion boundaries, below which singlet and octet fermion masses are excluded. In the case of $m_{Q_8}-m_{\pi_3} = 200$~GeV (top row), top left regions of the parameter space do not allow for on-shell color octet top partner decays.

As can be seen from \cref{fig:bounds_stop} (a) and (d), for $\tilde{g}$ pair production alone we obtain bounds on $m_{\tilde g}$ of up to $\sim 2.3$~TeV in case of light $\tilde B$ while the bound is weaker for a more compressed mass spectrum as the tops resulting from the decay are less energetic, and searches are less sensitive. These bounds are consistent with bounds on the gluino mass found in \cite{CMS:2019zmd} on which our analysis is based, hence providing a validation of the simulation setup we used.

Figure \ref{fig:bounds_stop} (b) and (e) show the bounds obtained for sole $\tilde{G}^+\tilde{G}^-$ production which results in the $2b + 2t + \ptmiss$ final state. Again, we obtain the leading bounds from the search \cite{CMS:2019zmd}. For light $\tilde B$, the bound on $m_{\tilde G^+}$ extends to $2.4$~TeV. This bound is higher than for $\tilde g$ pair production which is owed to the fact that the $\tilde G^+G^-$ production cross section is twice as large. In \cref{fig:bounds_stop} (b) for nearly degenerate mass spectrum, the bound on $m_{\tilde G^+}$ appears less reduced than for $m_{\tilde g}$. The reason for this apparent difference is that for \cref{fig:bounds_stop} (a) and (b) we chose $m_{Q_8}-m_{\pi_3} = 200$~GeV which for the $\tilde{g}$ decay results in a low $p_T$ top while for the $\tilde G^+$ decay it results in a moderate $p_T$ $b$-quark which evades cuts more easily.

Figure \ref{fig:bounds_stop} (c) and (f) show the bounds obtained when all color octet top partners $\tilde{g},\tilde{G}^+,\tilde{G}^0$ are taken into account. For mass degenerate color octets (as is assumed here), the summed pair production cross section of all states is five times larger than the $\tilde{g}$ pair production cross section. As a consequence we obtain a much higher bound of about 2.65~TeV on the color octet top partner scale $m_{Q_8}$ for light
$\tilde B$.

\subsection{Scenario II: $Q_8\to \pi_8+Q_1$ }

\begin{figure}
	\centering
		\begin{subfigure}[]{0.31\linewidth}
		\centering
		\includegraphics[width=\textwidth]{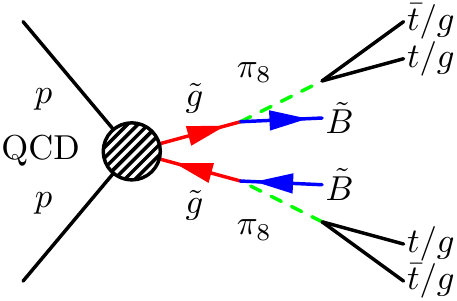} 
		\caption{$\tilde g\to \tilde B \pi_8$, $\pi_8\to \bar t t/g g$.}
		\end{subfigure}
				\begin{subfigure}[]{0.31\linewidth}
		\centering
		\includegraphics[width=\textwidth]{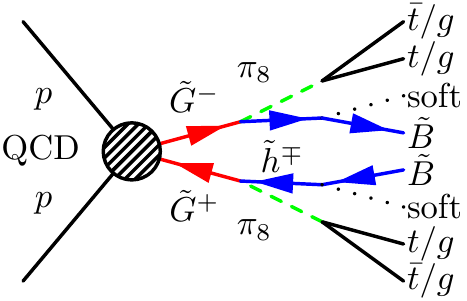} 
		\caption{$\tilde G^+\to \tilde h^+ \pi_8$, $\pi_8\to \bar t t/g g$.}
		\end{subfigure}
				\begin{subfigure}[]{0.31\linewidth}
		\centering
		\includegraphics[width=\textwidth]{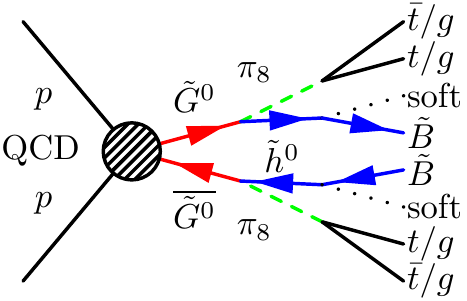} 
		\caption{$\tilde G^0\to \tilde h^0\pi_8$, $\pi_8\to \bar t t/g g$.}
		\end{subfigure}
	\caption{Feynman diagrams of QCD pair production of octet top partners $Q_8=(\tilde g,\tilde G^+,\tilde G^0)$ with dominant decays  $Q_8\to Q_1\pi_8$.}	
	\label{fig:S2graphs}
\end{figure}

Next, we investigate the case in which the color octet top partners dominantly decay to color octet pNGBs and color singlet top partners. The corresponding decay modes for the octet top partners $\tilde g,\tilde G^+,\tilde G^0$ are given in \cref{eq:gpi8dec,eq:G0pi8dec,eq:G+pi8dec}.  
We parameterize the relevant part of the effective interaction Lagrangian of the color octet pNGB following Ref.~\cite{Belyaev:2016ftv} as
\begin{align}
	\mathcal L_{\pi_8,int} = i \,C_{t,8} \frac{m_t}{f_\chi} \pi_8^a\, \overline t \gamma_5 \frac{\lambda^a}{2} t + \frac{\alpha_s \kappa_g}{8\pi f_\chi}\, \pi_8^a \, \epsilon^{\mu\nu\rho\sigma}\, \frac 12 d^{abc}\, G^b_{\mu\nu} G^c_{\rho\sigma} \,,
\end{align}
where $\kappa_g=2d_\chi$ and $d_\chi= \mathrm{dim}(\chi)=2N_c =4$ is the dimension of the $\chi$ representation.\footnote{We neglect  $\pi_8$ decays into $g\,\gamma$ and
$g\, Z$ (see \cite{Cacciapaglia:2020vyf}) as well as into a pair of light quarks here as they do not dominate in M5.} 
The ratio of $\pi_8$ decays into $t\,\bar t$ and $g\,g$ is given by
\begin{equation}
	\frac{\mathrm{Br}(\pi_8\to t\,\bar t)}{\mathrm{Br}(\pi_8\to g\,g)} = \frac{3\pi^2}{20\alpha_s^2} \, C_{t,8}^2 \, \frac{m_t^2}{m^2_{\pi_8}} \left( 1-4\frac{m_t^2 }{m_{\pi_8}^2} \right)^{1/2} \,.
\end{equation}
The value of $C_{t,8}$ depends on the details of the mixing of the top partners with the top, see Ref.~\cite{Cacciapaglia:2020vyf} for a discussion. Therefore we consider three cases:
(i) exclusive decay $\pi_8\to g\, g$, (ii)  decay  $\pi_8\to g\,g, t\, \bar t$ with equal branching ratio, and (iii) exclusive decay $\pi_8\to t\,\bar t$.
In each case, we fix $m_{\pi_8}=1.1$~TeV which is at the level of current experimental constraints on $m_{\pi_8}$ \cite{Cacciapaglia:2020vyf}.

Figure \ref{fig:S2graphs} shows the Feynman diagrams for QCD pair production of $\tilde g$ or $\tilde G^0$ and production of $\tilde G^+\tilde G^-$ with their subsequent cascade decays.
The $\SU(2)$ singlet $\tilde{g}$ decays directly into a boni and an octet pNGB. The $\SU(2)$ doublet $\tilde{G}^{+,0}$ decays into a higgsoni and an octet pNGB instead. The higgsoni decays to soft leptons or hadrons and a boni, as discussed in \cref{sec:LHCphenoDM}. Depending on the mass splitting, higgsoni decays could yield displaced vertices, but the LHC searches which we use to determine bounds, here, are not sensitive to the displacement, and the higgsoni effectively yields $\ptmiss$ like the boni. \footnote{On a technical level, for our event simulation, the higgsonis are chosen 5 GeV heavier than the boni and to decay promptly into boni and 1st generation leptons.} 
Thus the effective final states are $4j + \ptmiss$ for case (i), $4t + \ptmiss$ for case (iii) and both of these plus $2t + 2j + \ptmiss$ for case (ii). Note that the $4t + \ptmiss$ final state resembles the $\tilde{g}$ pair production final state considered in the last subsection, but it has completely different kinematics, as here, the two $t\bar{t}$ pairs form the $\pi_8$ resonances. Efficiencies and bounds are thus expected to be altered as compared to the bounds presented in the last subsection.

\begin{figure}
		\centering
	\begin{subfigure}[]{0.31\linewidth}
		\centering
		\includegraphics[width=\textwidth]{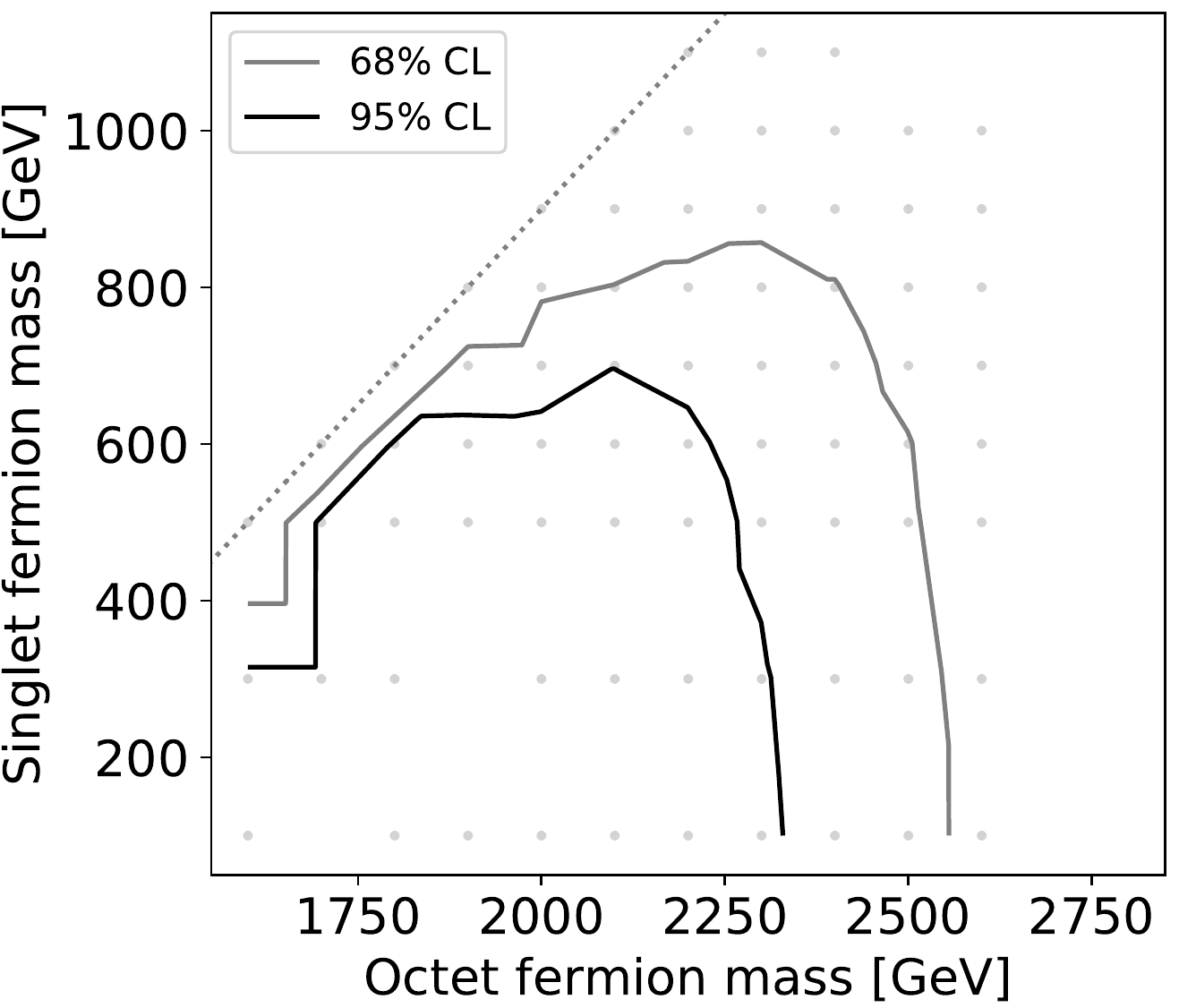} 
		\caption{$\tilde g\to \pi_8\tilde B,$ $\pi_8\to gg$}\label{fig:fig7_gluoni_pi8_gg_fixedmass_MA}
	\end{subfigure}
	\begin{subfigure}[]{0.31\linewidth}
		\centering
		\includegraphics[width=\textwidth]{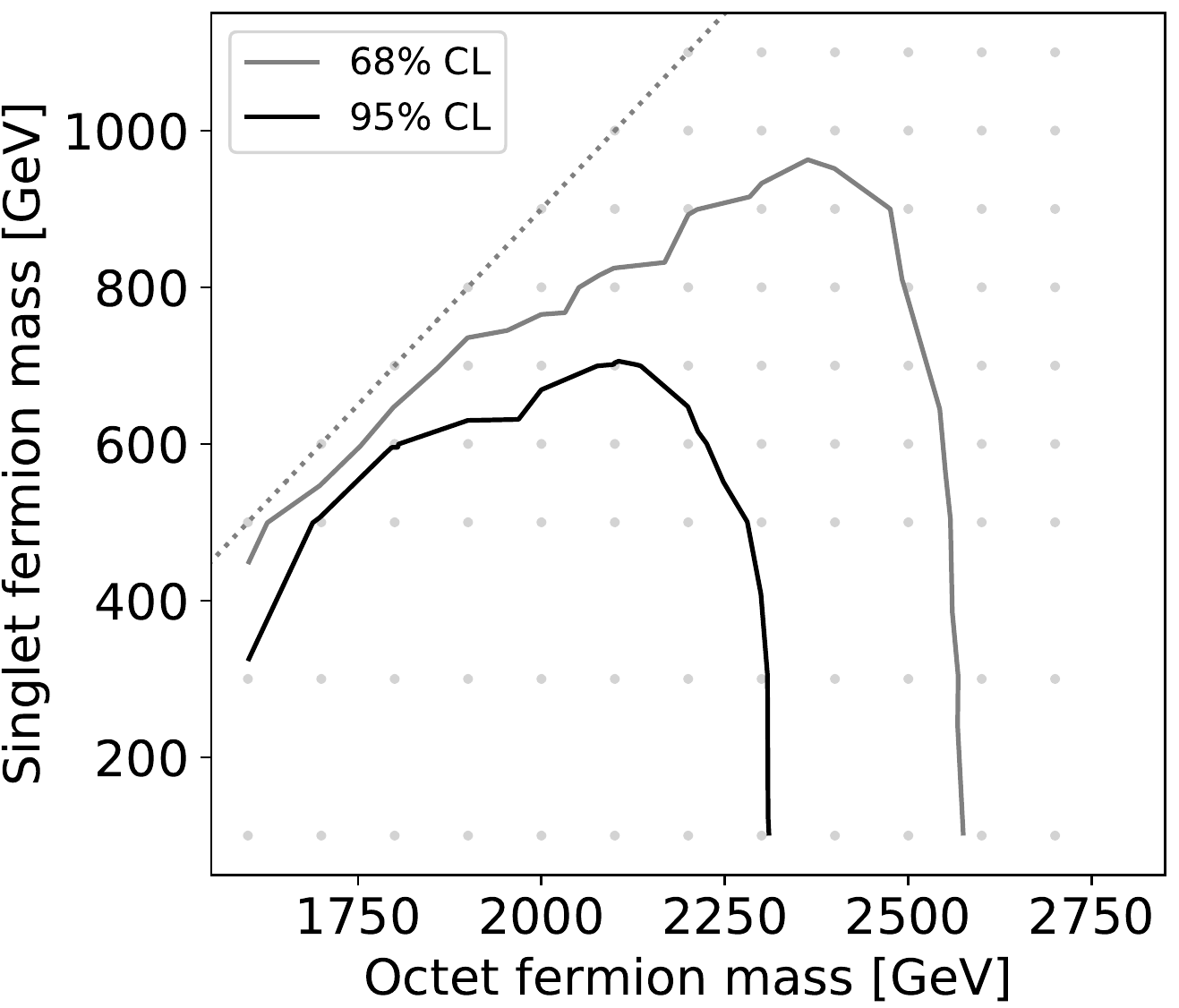} 
		\caption{$\tilde g\to \pi_8\tilde B,$ $\pi_8\to gg,t\bar t$}\label{fig:fig8_gluoni_pi8_ggtt_fixedmass_MA}
	\end{subfigure}
		\begin{subfigure}[]{0.31\linewidth}
		\centering
		\includegraphics[width=\textwidth]{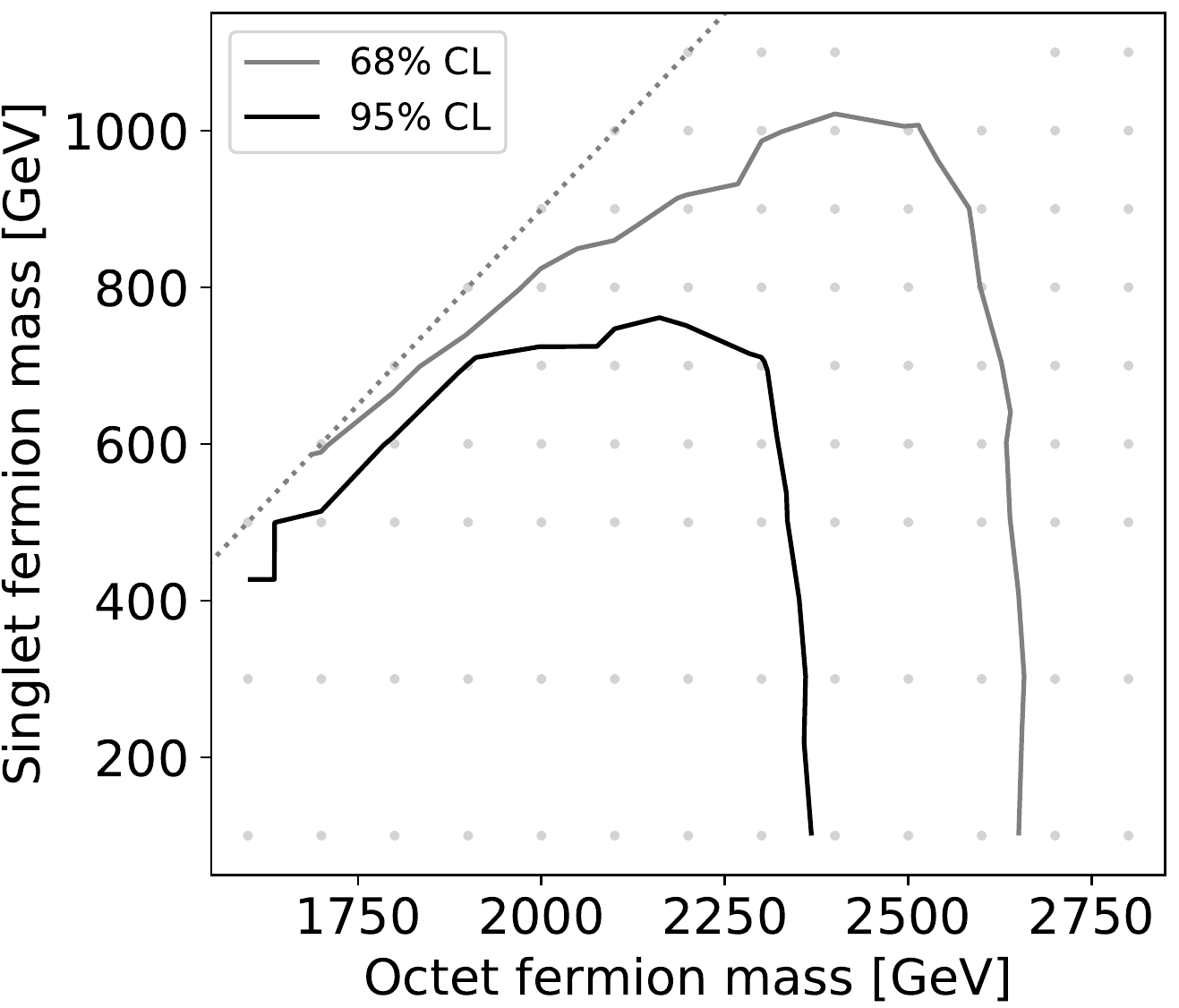} 
		\caption{$\tilde g\to \pi_8\tilde B,$ $\pi_8\to t \bar t$}\label{fig:fig9_gluoni_pi8_tt_fixedmass_MA}
	\end{subfigure}
	\vspace{1.5ex}
	
	\begin{subfigure}[]{0.31\linewidth}
		\centering
		\includegraphics[width=\textwidth]{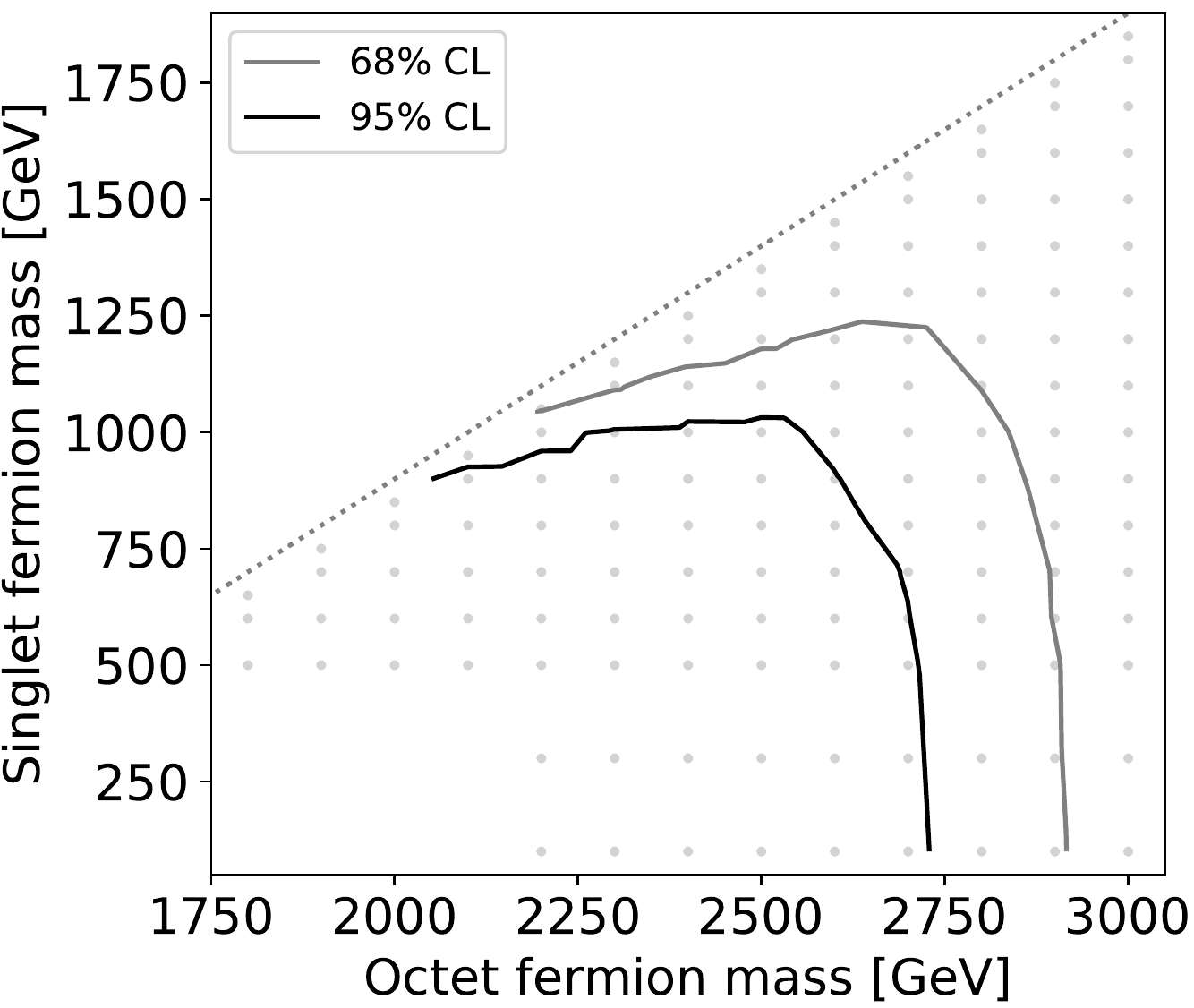} 
		\caption{$Q_8\to \pi_8 Q_1,$ $\pi_8\to gg$}\label{fig:fig10_multiplet_pi8_gg_fixedmass_MA}
	\end{subfigure}
	\begin{subfigure}[]{0.31\linewidth}
		\centering
		\includegraphics[width=\textwidth]{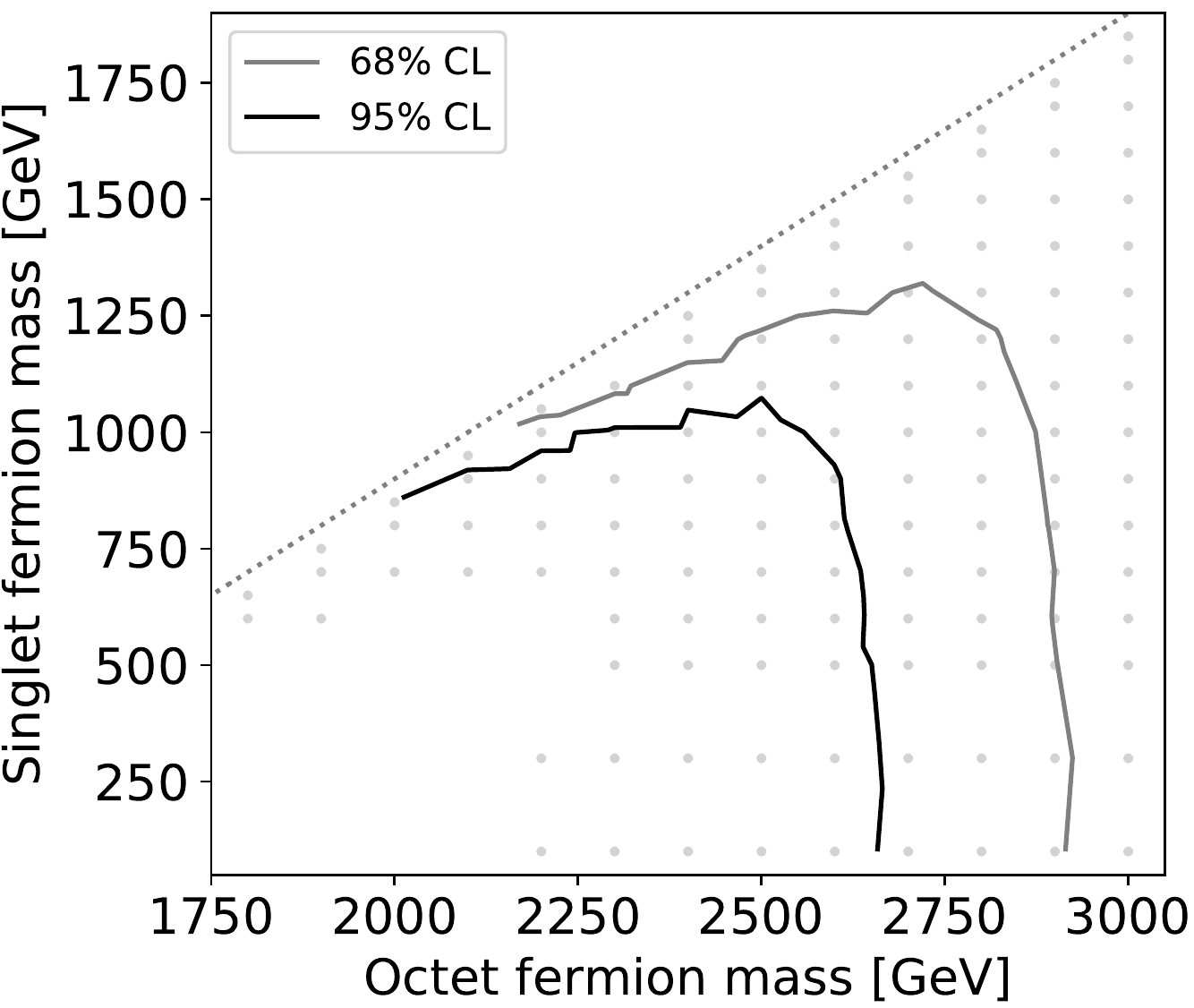} 
		\caption{$Q_8\to \pi_8 Q_1,$ $\pi_8\to gg,t\bar t$}\label{fig:fig11_multiplet_pi8_ggtt_fixedmass_MA}
	\end{subfigure}
	\begin{subfigure}[]{0.31\linewidth}
		\centering
		\includegraphics[width=\textwidth]{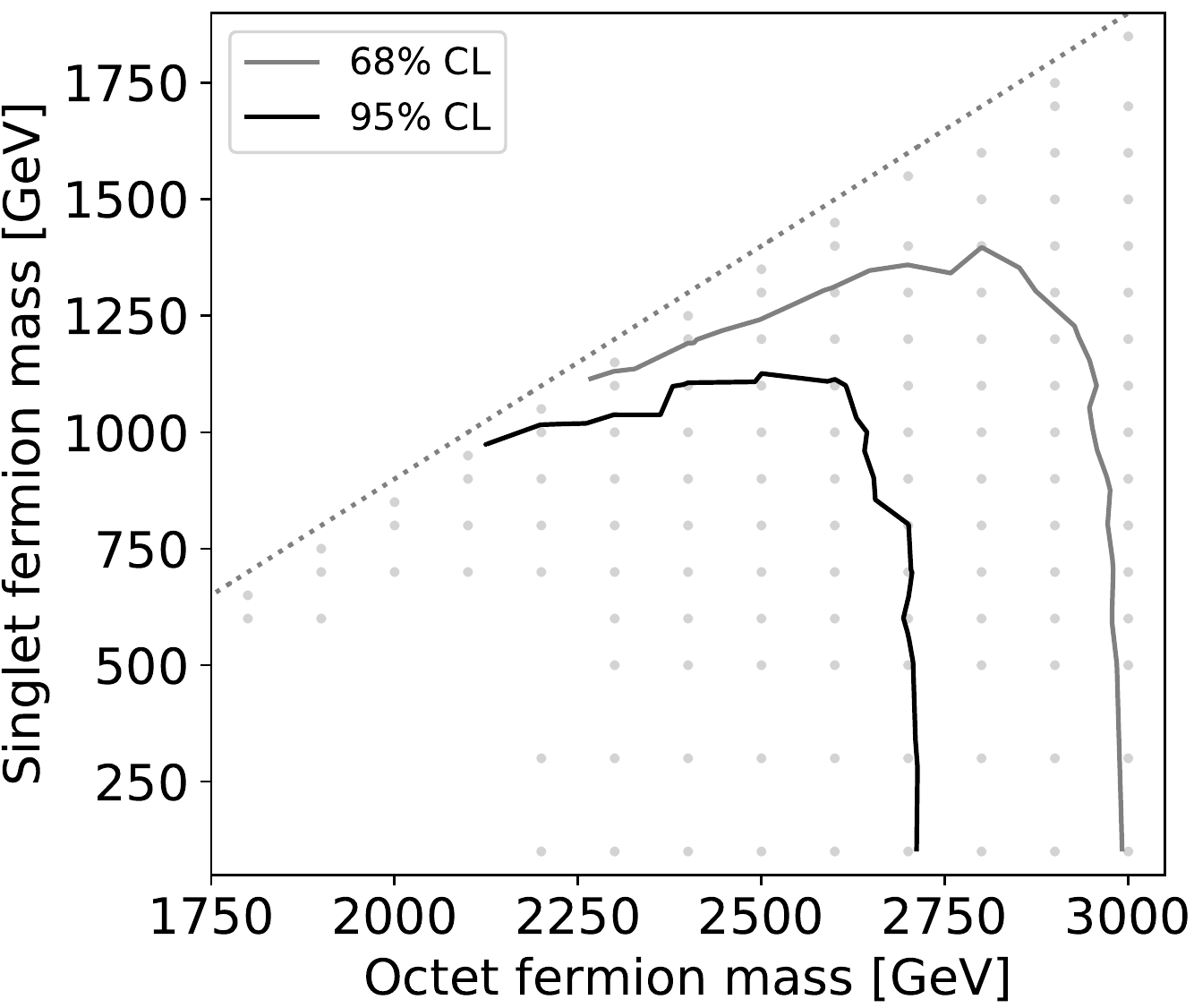} 
		\caption{$Q_8\to \pi_8 Q_1,$ $\pi_8\to t\bar t$}\label{fig:fig12_multiplet_pi8_tt_fixedmass_MA}
	\end{subfigure}
	\caption{Bounds on the fermion masses for QCD pair production of an octet fermion $Q_8$ with subsequent decay to a singlet fermion $Q_1$ and a $\pi_8$. The octet pNGB mass is fixed to $m_{\pi_8}=1.1$~TeV.
	In the first row only the gluoni is considered, which decays to a $\pi_8$ and a boni.
	In the second row the complete multiplets are taken into account, $Q_8=(\tilde G^+,\tilde G^0,\tilde g)$ and $Q_1=(\tilde h^+, \tilde h^0, \tilde B)$.
	The multiplets are assumed to be almost mass degenerate.
	The boni is stable, for the $\pi_8$ we consider the decays to $gg$ (left column), to $t\bar t$ (right column) or to either with equal branching ratio of 50\% (middle column).}
	\label{fig:bounds_sgluon}
\end{figure}

Figure \ref{fig:bounds_sgluon} shows the obtained bounds for various cases. Again, we find the implemented search \cite{CMS:2019zmd} to dominate the bound over most of the parameter space and refer to \cref{app:tools_comparison} for more information on bounds from other searches.

In the first row we consider a scenario where
only $\tilde g$ is present. We find that
in case of light $\tilde B$ the bounds are about 
2.3-2.4 TeV depending weakly on the assumed $\pi_8$ decay channel. For the $4t + \ptmiss$ final state, the bound obtained here is marginally stronger than the bound for the $4t + \ptmiss$ final state from the decay through $\pi_3$ in \cref{fig:fig2_gluoni_stop_fixedmass_MA}.
The dependence on the $\pi_8$ final states gets somewhat more pronounced for smaller differences
$m_{\tilde g}-m_{\tilde B}$. 
In the second row we show the bounds on $m_{Q_8}$ when pair production of the entire $Q_8$ multiplet is taken into account. For mass degenerate $Q_8$, the sum of the pair production cross sections is five times larger than for Majorana gluoni pair production alone, and the bound on $m_{Q_8}$ increases to about 2.7~TeV for small $\tilde B$ masses. Note, that these bounds are similar to the previous case where the $Q_8$ particles decay exclusively into final states containing a $\pi_3$. This finding is non-trivial as the kinematics of the decays are very different.

\subsection{Scenario III: $Q_8$ decays to both $\pi_8$ and $\pi_3$} 

\begin{figure}
	\centering
	\includegraphics[width=0.5\textwidth]{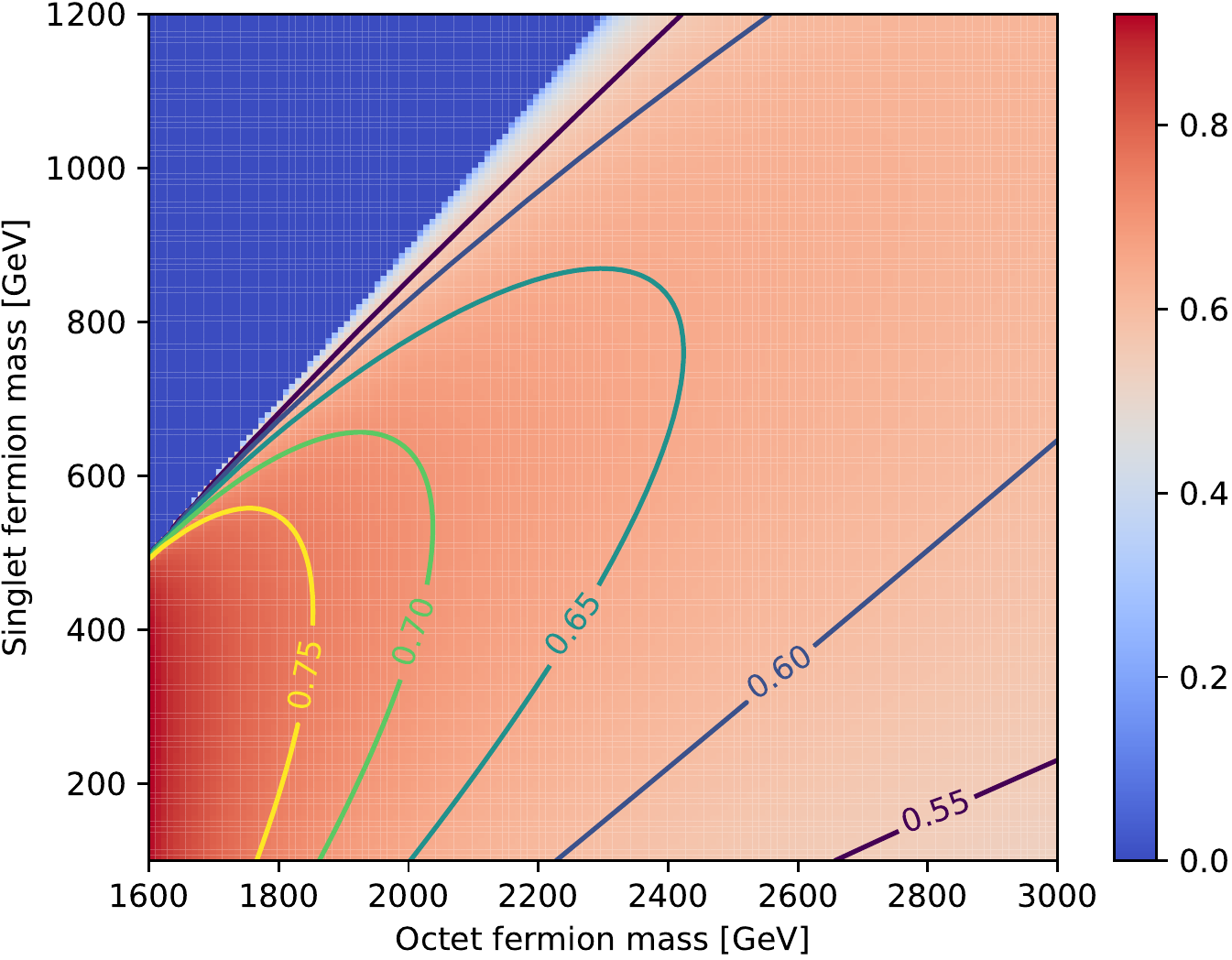}
	\caption{Branching ratio of $\tilde g\to \pi_8 \tilde B$ for the mixed decays. 
	The scalar masses are fixed to $m_{\pi_8}=1.1$~TeV and $m_{\pi_3}=1.4$~TeV. The ratio of the couplings to $\pi_8$ and $\pi_3$ is set to 1.5, so that for heavy gluonis, the branching ratio is about 50\%. For $m_{\tilde g}\to0$, the decay via $\pi_8$ dominates, whereas it is kinematically forbidden in the top left.}
	\label{fig:Brpi8pi3}
\end{figure}

\begin{figure}
	\centering
	\captionsetup[subfigure]{justification=centering}
	\begin{subfigure}[]{0.31\linewidth}
		\centering
		\includegraphics[width=\textwidth]{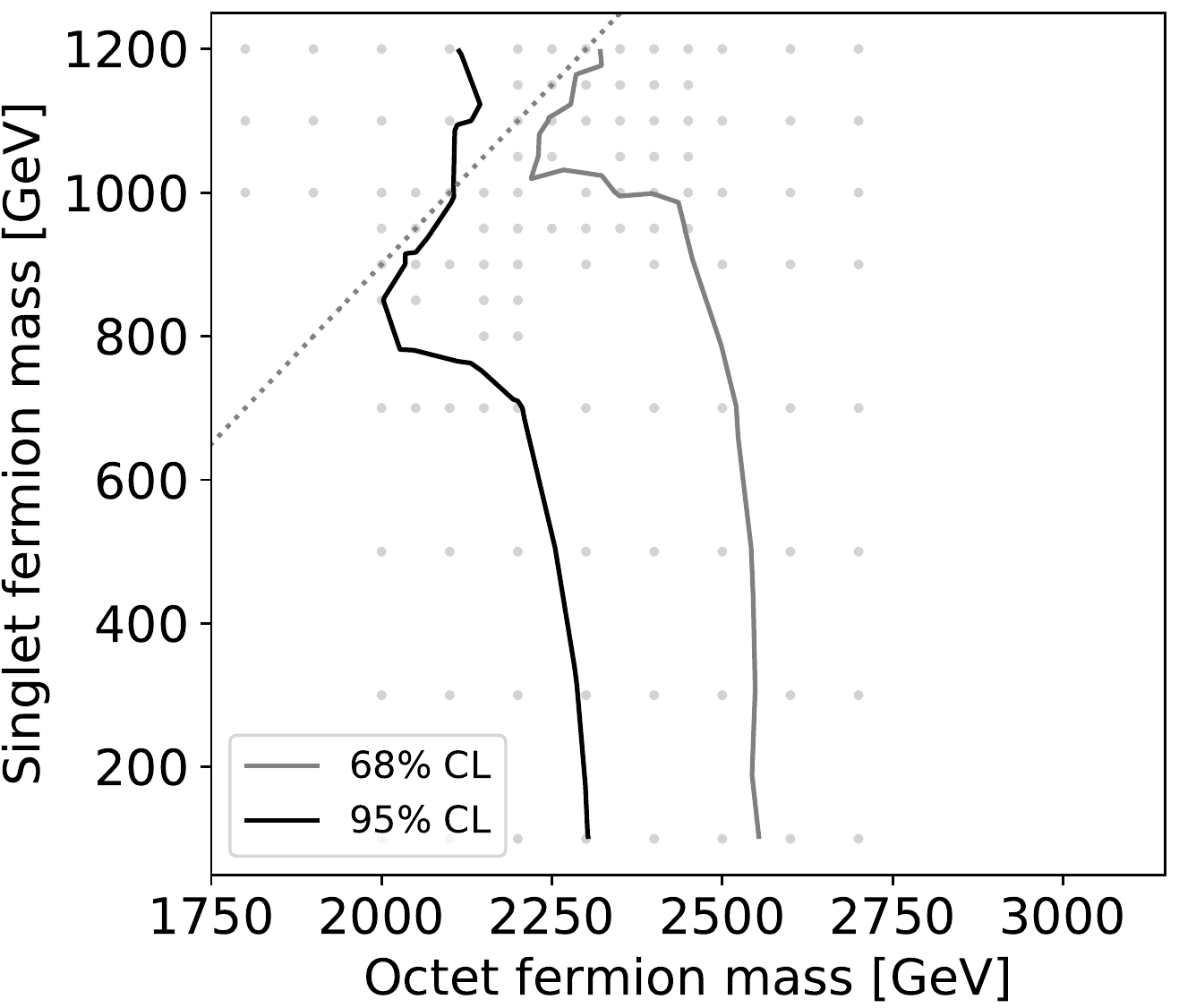} 
		\caption{$\tilde g\to \bar t\pi_3,t \pi_3^*$ or $\tilde g\to \pi_8 \tilde B$}\label{fig:fig13_gluoni_pi3pi8_MA}
	\end{subfigure}
	\begin{subfigure}[]{0.31\linewidth}
		\centering
		\includegraphics[width=\textwidth]{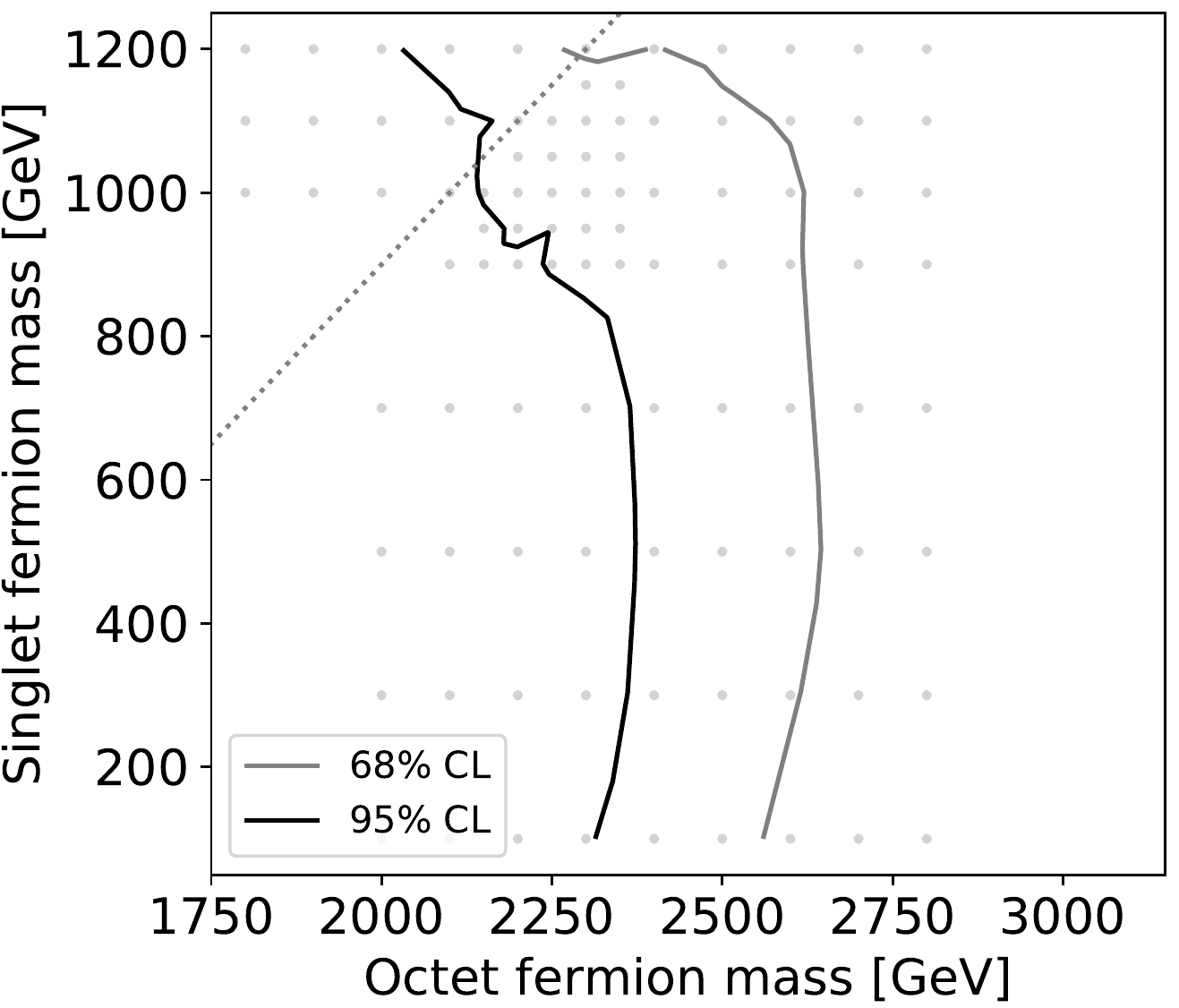} 
		\caption{$\tilde G^+\to \bar b\pi_3$ or $\tilde G^+\to \pi_8\tilde h^+$} \label{fig:fig14_Gplus_pi3pi8_MA}
	\end{subfigure}
	\begin{subfigure}[]{0.31\linewidth}
		\centering
		\includegraphics[width=\textwidth]{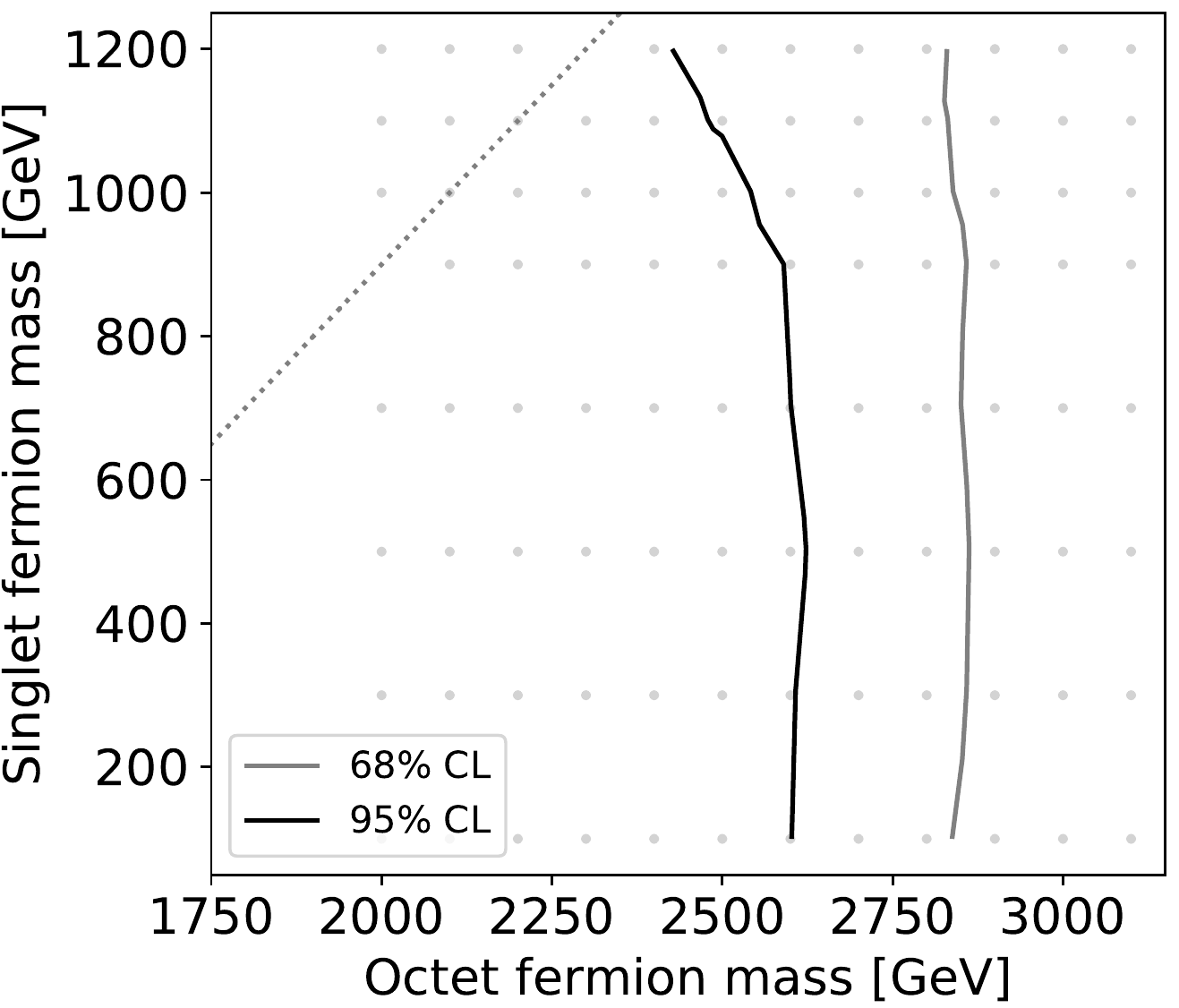} 
		\caption{$\tilde Q_8\to \bar q\pi_3$ or $\tilde Q_8\to \pi_8Q_1$}\label{fig:fig15_multiplet_pi3pi8_MA}
	\end{subfigure}
	\caption{Bounds on the fermion masses for QCD pair production of color octet top partners $Q_8$ with subsequent decay via $\pi_3$ and $\pi_8$ with comparable branching ratios. The scalar masses are fixed to $m_{\pi_3}=1.4$~TeV and $m_{\pi_8}=1.1$~TeV.
	Above the dotted line, $\mathrm{Br}(Q_8\to \pi_8 Q_1)=0$.
	In case of $\pi_8$ we take $\mathrm{Br}(\pi_8\to gg)=\mathrm{Br}(\pi_8\to t\bar t)=50\%$.}
	\label{fig:bounds_mixed}
\end{figure}

Finally, we consider a scenario in which color octet top partner decays to both pNGB triplets and octets are present. 
To study this case in more detail, we performed a scan in which we set the couplings of the color octet top partners to $\pi_3$ and $\pi_8$ to a fixed ratio of $1.5$ which yields comparable branching ratios for the decays under study.  We scan over the octet and singlet fermion masses, keeping $m_{\pi_8}= 1.1$~TeV, $m_{\pi_3}= 1.4$~TeV and the couplings fixed. Due to the decay phase space factors, the $Q_8$ branching ratios vary depending on the $Q_8$ and $Q_1$ masses, as is shown in  \cref{fig:Brpi8pi3}. For the $\pi_8$ decay, we assume a branching ratio of 50\% into $t\bar{t}$ and $gg$. 
We display the resulting bounds on the color octet top partner masses in \cref{fig:bounds_mixed} assuming $\tilde g$ pair production only (left), $\tilde G^+ \tilde G^-$ production only (middle), and pair production of all color octet top partner states (right).
As expected, the bounds are comparable to the cases of Br$(Q_8 \rightarrow \bar{q} \pi_3)=1$ and Br$(Q_8 \rightarrow \pi_8 Q_1)=1$.

\section{Conclusions and outlook} \label{sec:outlook}

This is the first of a series of papers where we explore the collider phenomenology of
composite Higgs models with a concrete UV completion. Here we have focused on the 
so-called M5 model class. We have worked out the details of the strongly interacting
pNGBs as well as of the possible hyper-baryon configurations.
A peculiar feature of this model is that the hyper-baryon spectrum contains color-octet fermionic states. They are predicted to be among the lightest top partners
and, therefore, play the leading role in the LHC phenomenology of this model. We have worked out
the generic phenomenological features of this model.  It turns out
that, in scenarios where both lepton and baryon number are conserved, the color singlet
top partners can be potentially dark matter candidates. This in turn implies that the color
octet top partners share several features of gluinos in SUSY models with conserved R-parity. Consequently, we have used existing recast tools to obtain mass bounds of up to
$2.7$~TeV  on these fermions due to existing LHC analyses. The usual color triplet top partners are expected to be in the same mass range or even  heavier
than the octet baryons, which would be an explanation of the null results in the direct
LHC searches.

The collider aspects can be explored in various ways: (i) How do the bounds change if
one allows for lepton or baryon number violating decay modes? (ii) What is the reach
of proposed future collider experiments? In previous studies on prospective pp colliders
with 33 TeV or 100 TeV  \cite{Golling:2016gvc} the reach for gluinos in supersymmetric
scenarios have been presented, focusing on same-sign leptons in the final states.
Performing a naive re-scaling of the corresponding cross section, in the case
of a 33 TeV collider we estimate a discovery (exclusion) reach up to $4.1$ ($4.8$) TeV. A 100 TeV
pp collider could find (exclude) such states up to masses of $8$ ($9.8$) TeV using this
signature. We note here that this particular channel has a rather small branching ratio while being
at the same time practically background free. However, the success of machine learning
techniques for hadronic final states, e.g.~in case of $t\,\bar{t}\,h$ production, indicates that
a significantly larger reach might be possible exploiting the fully hadronic channels. We will exploit this avenue in an up-coming work.

\section*{Acknowledgements}

This work has been supported by the ``DAAD, Frankreich'' and ``Partenariat Hubert Curien (PHC)'' PROCOPE 2021-2023, project number 57561441. 
TF is supported by a KIAS Individual Grant (AP083701) via the Center for AI and Natural Sciences at Korea Institute for Advanced Study.
We are grateful to the Mainz Institute for Theoretical Physics (MITP) of the DFG
Cluster of Excellence PRISMA+ (Project ID 39083149) for its hospitality and support
during the initial stages of this work.

\appendix

\section{Details on the electroweak embedding}
\label{app:ew_pngbs}
We briefly summarize here for completeness some main results of \cite{Agugliaro:2018vsu} for the electroweak (EW) Goldstone boson sector. It can be parameterized by a scalar field $\Sigma_\psi$ in the symmetric 2-tensor representation $\bf{15}$ of $\SU(5)$, transforming like $g \Sigma_\psi g^T$ where $g\in \SU(5)$.

The EW preserving vacuum which respects the $\SO(5)$ subgroup of $\SU(5)$ reads
\begin{equation}
\Sigma_{0,\psi}= \left(\begin{array} {cc|c}  & i \sigma_2 &   \\ - i \sigma_2 &  & \\\hline  & & 1 \end{array}\right).
\end{equation}
It has a slightly unusual form but this form helps to uniquely identify the
$\SO(4)$ part which at the level of the Lie algebras is isomorphic to 
$\SU(2)_L \times \SU(2)_R$. This facilitates also the identification of the quantum
numbers of the hyper-baryons.
The (unbroken) generators of $\SU(2)_{L,R}$ of the EW preserving vacuum
are:
\begin{eqnarray}
T^i_L=\frac {1}{2} \left(\begin{array} {c|c} \id_2 \otimes \sigma^i &  \\\hline   & 0 \end{array}\right) & \,\,\, , \,\,\,&
T^i_R=\frac {1}{2} \left(\begin{array} {c|c} \sigma^i  \otimes \id_2  & \\\hline   & 0 \end{array}\right) 
\end{eqnarray}

The Goldstone-boson matrix reads 
\begin{equation}
U_\psi = e^{i 2 \Pi^I_\psi X^I/f_\psi} \, , 
\end{equation}
with
\begin{equation}
\Pi^I_\psi X^I= \frac{1}{2}\begin{pmatrix}
\frac{\eta}{\sqrt{10}}\id_2+\pi_0 & \pi_+ & H \\
	\pi_- & \frac{\eta}{\sqrt{10}}\id_2-\pi_0  & -\tilde{H} \\
		H^\dagger & -\tilde{H}^\dagger & - \frac{4}{\sqrt{10}} \eta
\end{pmatrix}\, , 
\end{equation}
where $\eta$ is an electroweak singlet, $H$ is the Higgs doublet, $\tilde{H} = i \sigma_2 H^*$, and $\pi_0=\frac{1}{\sqrt{2}} \pi^i_0\sigma^i$, $\pi_-= \pi^i_0\sigma^i=\left(\pi_+\right)^\dagger$ form a $\SU(2)_L\times \SU(2)_R$ bi-triplet. Replacing as usual linear combinations of the real fields by complex fields, one gets more explicitly 
\begin{align}
\pi_0 = \begin{pmatrix} 
\frac{1}{\sqrt{2}} \pi^3_0 & \pi^+_0 \\ \pi^-_0 & -  \frac{1}{\sqrt{2}} \pi^3_0
\end{pmatrix}, \qquad  \pi_\pm = \begin{pmatrix} 
\pi^0_\pm & \sqrt{2}\pi^+_\pm \\ \sqrt{2} \pi^-_\pm & - \pi^0_\pm
\end{pmatrix} .
\end{align}
Moreover, we parameterize $H$ as
\begin{align}
H = \begin{pmatrix} \phi^+ \\ \frac{1}{\sqrt{2}} \left( h + i\, \phi^0 \right)\end{pmatrix}.
\end{align}

With respect to the vacuum $\Sigma_{0,\psi}$, $\Sigma_\psi$ is given by
\begin{equation}
\Sigma_\psi = U_\psi\Sigma_{0,\psi} U_\psi^T =  e^{2i\Pi^I_\psi X^I/f_\psi} \Sigma_{0,\psi}  e^{2i\Pi^I_\psi(X^I)^T/f_\psi} =e^{4i\Pi^I_\psi X^I/f_\psi} \Sigma_{0,\psi}  \, .
\end{equation}
However, the Higgs vacuum expectation value (\vev) $v$ breaks the electroweak symmetry, and we expand around the true (misaligned) \vev. The misaligned vacuum is given by 
\begin{equation}
\Sigma_\theta = \Omega_\theta \Sigma_{0,\psi} \Omega_\theta^T = 
\begin{pmatrix}
0 & 0 & 0 & 1 & 0 \\
0 & -s^2_\theta & -c^2_\theta & 0 & i s_{2\theta} / \sqrt{2} \\
0  & -c^2_\theta & -s^2_\theta &0 & -i s_{2\theta} / \sqrt{2} \\
1 & 0 & 0 & 0 & 0 \\
0 &  i s_{2\theta} / \sqrt{2}  & -i s_{2\theta} / \sqrt{2} & 0 & c_{2\theta}
\end{pmatrix} \, ,
\end{equation}
where $s_\theta = \sin (\theta) = v / f_\psi$ and 
\begin{equation}
\Omega_\theta = e^{4iX^{\hat{h}}\frac{\theta}{2}} =\begin{pmatrix}
1 & 0 & 0 & 0 & 0 \\
0 & {c^2_{\theta/2}} & {s^2_{\theta/2}} & 0 & i s_{\theta} / \sqrt{2} \\
0  & {s^2_{\theta/2}} & {c^2_{\theta/2}} &0 & -i s_{\theta} / \sqrt{2} \\
0 & 0 & 0 & 1 & 0 \\
0 &  i s_{\theta} / \sqrt{2}  & -i s_{\theta} / \sqrt{2} & 0 & c_{\theta}
\end{pmatrix} , \quad
X^{\hat h}= \frac{1}{2\sqrt{2}}
\begin{pmatrix}
	&&&&0\\&&&&1\\&&&&-1\\&&&&0\\0&1&-1&0&0
\end{pmatrix}
\end{equation}
is the misalignment of the vacuum along the Higgs direction due to the non-zero Higgs \vev.

The Goldstone bosons with respect to the misaligned vacuum ($\tilde{U}_\psi$) are
\begin{equation}
\tilde{U}_\psi = \Omega_\theta U_\psi \Omega^{-1}_\theta 
\label{eq:Utrue}
\end{equation}
and with respect to the misaligned vacuum, we have 
\begin{equation}
\Sigma_\psi = \tilde{U}_\psi\Sigma_{\theta,\psi} \tilde{U}_\psi^T = \Omega_\theta U_\psi \Omega^{-1}_\theta \Omega_\theta  \Sigma_{0,\psi} \Omega^T_\theta \Omega^{-1 T}_\theta U^T_\psi  \Omega^{T}_\theta = \Omega_\theta U_\psi \Sigma_{0,\psi} U^T_\psi  \Omega^{T}_\theta \, .
\end{equation}

The Higgs \vev\ breaks $\SU(2)_L\times \SU(2)_R \rightarrow \SU(2)_D$ and causes mixing between the $\SU(2)_L$ pion triplets $\pi_{0,+,-}$, which under $\SU(2)_D$ decompose as $\left(\bf{3},\bf{3}\right)\rightarrow \bf{5}\oplus\bf{3}\oplus \bf{1}$, with
\begin{eqnarray}
\pi^+_+=\eta_5^{++}  \,\, ,& \,\, \pi^0_+=\displaystyle\frac{i\eta_3^+-\eta_5^{+}}{\sqrt{2}} \,\, ,& \,\, \pi^+_0=\frac{-i\eta_3^+-\eta_5^{+}}{\sqrt{2}}\, , \\
\pi_0^3 =\frac{\eta_1^0-\sqrt{2}\eta_5^{0}}{\sqrt{3}} \,\, ,& &\,\,
\pi_+^- =\frac{\sqrt{2}\eta_1^0+\eta_5^{0}}{\sqrt{6}} +i\frac{\eta_3^0}{\sqrt{2}}\, ,
\end{eqnarray}
and $\pi_-^-=(\pi_+^+)^\dagger$, $\pi_-^0=(\pi_+^0)^\dagger$,  $\pi_-^+=(\pi_+^-)^\dagger$, $\pi_0^-=(\pi_0^+)^\dagger$.

The hyper-baryons come in $\Sp(6)\times \SO(5)$ representations. Both factors get
dressed by the respective Goldstone matrix, see \cref{eq:dressedoperators}, so that $\SU(6)\times \SU(5)$ invariant
interactions can be constructed with the spurions containing the SM fermions, in particular
the 3rd generations quarks. 
In \cref{sec:hyperbaryons}, we neglected the EW misalignment when calculating the partial compositeness interactions.
More correctly, the top partners have to be defined in the misaligned basis.
Taking into account the rotation by $\Omega_\theta$, we find
\begin{align}
    Q_3^c \zeta_L 
	&= \left( s_{\theta/2}^2 X_{2/3} + c_{\theta/2}^2 T^c_L - \frac{s_\theta}{\sqrt{2}} T^{ c}_R \right) t_L+B^c_L b_L  \\ 
	&= 	\frac{1}{\sqrt{2}} \left( -s_\theta T_1^c- c_\theta T_2^c + T_3^c \right) t_L+ B^c_L b_L,\\
	Q_3 \zeta_R &= \frac{s_\theta}{\sqrt{2}} (X_{2/3} -T_L) t_R^c + c_\theta T_R t_R^c \\ &= (c_\theta T_1 + s_\theta T_2) t_R^c,
\end{align}
where we introduced 
\begin{align}
T_1 = T_R ,\qquad T_2 = \frac{1}{\sqrt{2}} \left( X_{2/3} - T_L \right)
,\qquad T_3 = \frac{1}{\sqrt{2}} \left( X_{2/3} + T_L \right) 
\end{align}
to express the couplings in terms of $\SU(2)_D$ eigenstates:
a triplet $(X_{5/3},T_3,B)$  and two singlets $T_1,T_2$.
The remaining couplings are given by
\begin{align}
	Q_8 \zeta_L &= \left( -c_\theta^2 \tilde G_u^0- s_\theta^2 \tilde G_d^0 + \frac{s_\theta}{\sqrt{2}} \tilde g \right) t_L + \tilde G_u^+ b_L \\
	Q_8 \zeta_R &= - \frac{s_\theta}{\sqrt{2}} \left(\tilde G_u^0-\tilde G_d^0 \right) t_R^c + c_\theta \tilde g t_R^c \\
	Q_1 \zeta_L &= \left( -c_\theta^2 \tilde h_u^0- s_\theta^2 \tilde h_d^0 + \frac{s_\theta}{\sqrt{2}} \tilde B \right) t_L + \tilde h_u^+ b_L \\
	Q_1 \zeta_R &= - \frac{s_\theta}{\sqrt{2}} \left(\tilde h_u^0-\tilde h_d^0 \right) t_R^c + c_\theta \tilde B t_R^c
\end{align}
Finally, we also give the couplings of the top partners to a SM quark and a EW pNGB, \begingroup\allowdisplaybreaks
\begin{align}
    \mathcal L_\mathrm{mix} &\supset - \frac 12y_L  Q_3^c \zeta_L - \frac 12 y_R  Q_3 \zeta_R +\hc \\
    &\supset -\frac{y_L}{2f_\psi}\Bigg[ \sqrt 2i \, \eta_5^{++} X_{5/3}^c b_L - \frac{1}{\sqrt 2} (c_\theta \eta_3^+ +i \eta_5^+) X_{5/3}^c t_L \nonumber \\
	&\qquad\qquad\quad  + \left( \frac{i}{\sqrt{10}} \eta+ \frac{i}{\sqrt{6}} \eta_1^0 - \frac{i}{\sqrt 3} \eta_5^0 \right) B^c_L b_L + \frac{1}{\sqrt{2}} (c_\theta\eta_3^-+ i \eta_5^-) B^c_L t_L\nonumber \\
	&\qquad\qquad\quad + \left( -\frac{c_\theta}{\sqrt2} h+ \frac{2is_\theta}{\sqrt5} \eta \right) T_1^c t_L\nonumber \\
	&\qquad\qquad\quad + \left( \frac{s_\theta}{\sqrt 2} h -\frac{ic_\theta}{2\sqrt 5} \eta + \frac{\sqrt 3ic_\theta}{2} \eta_1^0 -\frac{1}{\sqrt 2}\eta_3^0 \right) T_2^c t_L + \eta_3^+ T_2^c b_L\nonumber \\
	&\qquad\qquad\quad + \left( \frac{i}{2\sqrt5} \eta + \frac{i}{2\sqrt3} \eta_1^0 - \frac{c_\theta}{\sqrt 2} \eta_3^0 + \frac{\sqrt 2i}{\sqrt 3} \eta_5^0 \right) T_3^c t_L + i\eta_5^+ T_3^c b_L +\hc\Bigg]\nonumber  \\
	&\phantom{=}-\frac {y_R}{2f_\psi} \Bigg[ -s_\theta \, \eta_3^- X_{5/3} t_R^c + s_\theta\, \eta_3^+ B_L t_R^c  -\left( s_\theta h +\frac{2\sqrt 2 i c_\theta}{\sqrt 5} \eta \right) T_1 t_R^c \nonumber \\
	&\qquad\qquad \quad +\left( c_\theta h + \frac{is_\theta}{\sqrt{10}} \eta- \frac{\sqrt 3 is_\theta}{\sqrt 2} \eta_1^0  \right) T_2 t_R^c - 	s_\theta\, \eta_3^0 T_3 t_R^c +\hc \Bigg] 
\end{align}
\endgroup
Here we have expanded the EW Goldstone matrix to linear order.

\section{Alternative hyper-baryon embedding}
\label{app:altebb}

In the main text, we presented in detail the case where the top partial compositeness involves operators transforming as the $({\bf 5}, {\bf 15})$ of the global symmetry $\SU(5) \times \SU(6)$. Here we recap the other two cases, corresponding to $({\bf 5}, {\bf \overline{15}})$ and $({\bf \bar 5}, {\bf 35})$.

For the former, $({\bf 5}, {\bf \overline{15}})$, the embedding is very similar, and one has
\begin{equation}
    \zeta_{L,A} = \begin{pmatrix}   0 & 0 \\ 0  & -\frac 12 \epsilon_{ijk} \zeta_{L,k} \end{pmatrix}\, \quad \mathcal{O}_{\bar{A},14} = U_\chi^\dagger (U_\psi \cdot \Psi_{14} ) U_\chi^\ast\,, \quad \mbox{etc.}
\end{equation}
This implies a simple change in sign of the colored pNGB couplings in Eqs.~\eqref{eq:mixingAL0} and \eqref{eq:mixingAR0}. Hence, the phenomenology of this case will be the same as that described in the main text.

For the latter, $({\bf \bar 5}, {\bf 35})$, one main difference is the presence of the $\bf 21$ of $\Sp(6)$ hyper-baryon:
\begin{equation} 
\Psi_{21} =  \begin{pmatrix}  -Q_6 &  \frac{1}{2\sqrt{2}} Q_8^{\prime a}\ \lambda^a  + Q_1^\prime \\  \frac{1}{2\sqrt{2}} Q^{\prime a}_8\ (\lambda^a)^T + Q_1^\prime & -Q_6^c  \end{pmatrix}\,.
\end{equation}
To write couplings to the top, stemming from the four-fermion interactions, analog to Eq.~\eqref{eq:topPC4F},
\begin{equation}
    \frac{\xi_L}{\Lambda_t^2} \bar{\psi} \bar{\chi} \chi q_{L,3} + \frac{\xi_R}{\Lambda_t^2} \bar{\psi} \bar{\chi} \chi t_{R}^c \,.
\end{equation}
The embeddings of the hyper-baryons in the adjoint of $\SU(6)$ and anti-fundamental of $\SU(5)$ read
\begin{equation}
    \mathcal O_{D,14} = U_\chi (\Psi_{14} \cdot U_\psi^\dagger) U_\chi^\dagger\,, \qquad 
    \mathcal O_{D,21}= U_\chi (\Psi_{21} \cdot U_\psi^\dagger) U_\chi^\dagger\,,
\end{equation}
where the singlet $Q_1$ does not appear as is does not enter in the four-fermion operators that couple to the top quark fields.
Instead, the elementary top fields are embedded in the $\SU(6)$ adjoint as
\begin{equation}
    \zeta_{L,D}   =  \begin{pmatrix}  0 & 0 \\ \frac 12 \epsilon_{ijk} \zeta_{L,k}^c  & 0 \end{pmatrix}\,, \qquad \zeta_{R,D}   =  \begin{pmatrix}  0 & -\frac 12 \epsilon_{ijk} \zeta^c_{R,k}  \\ 0 & 0 \end{pmatrix}\,,
\end{equation}
where the embedding in the fundamental of $\SU(5)$ reads:
\begin{equation}
    \zeta_L^c = (0, 0, t_L, b_L, 0)\,, \quad \zeta_R^c = (0,0,0,0,i t_R^c)\,.
\end{equation}
The main difference with the previous cases is that couplings of a single colored pNGB with two top partners or one top partner and one top field are absent. 
The lowest order couplings contain two colored pNGB, hence the phenomenology of this case differs enormously from the other cases. We leave this case for further exploration.

\section{Documentation of the \textsc{FeynRules} implementation}
\label{app:FRimp}

We document the \textsc{FeynRules} (FR) implementation used for simulations in \cref{sec:numerics}. The implementation is based on the publicly available FR implementation for vector-like quarks (VLQs) $X,T,B,Y$ with charges $5/3,2/3,-1/3,-4/3$, respectively \cite{urlFR,Buchkremer:2013bha,Fuks:2016ftf}.
Additional fields are added by including different independent modules, which include the interactions with the SM and the VLQs. This modular implementation for exotic decays of vector-like quarks has been initiated at the MITP workshop ''Fundamental Composite Dynamics: opportunities for future colliders and cosmology (2019)'' by G. Cacciapaglia, A. Deandrea, T. Flacke, B. Fuks, L. Panizzi, and W. Porod. Components of it have been / are used in other studies \cite{Cacciapaglia:2019zmj,Xie:2019gya,Benbrik:2019zdp,Corcella:2021mdl}. For this work, we extended the implementation by color-non-triplet fermions as well as BSM-BSM-SM interactions required for our simulations.
The following modules are implemented (and available upon request): 
\begin{itemize}
	\item A neutral color singlet scalar $S_1^0$. 
	\item A singly charged color singlet scalar $S_1^1$. 
	\item A doubly charged color singlet scalar $S_1^2$. 
	\item A neutral color octet scalar $S^0_8$.  
	\item A color triplet scalar with charge $Q=2/3$ ($S_3^{2/3}$).
	\item Color singlet fermions with charges $Q=1$ ($Q_1^1$), $Q=0$ ($Q^0_1$) and a Majorana ($Q_{1,M}^0$).
	\item Color octet fermions with charges $Q=1$ ($Q_8^1$), $Q=0$ ($Q^0_8$) and a Majorana ($Q_{8,M}^0$).
\end{itemize}
An overview of the notation for the fields is given in Tab.~\ref{tab:frfields}, where we also indicate the corresponding fields in M5.
For modules that correspond to multiple fields, e.g.\ \texttt{S10}, we add copies of the corresponding files with the replacements \texttt{S10}~$\to$~\texttt{S102,S103} etc.

In the following we present the Lagrangians for the new fields, excluding lepton and baryon number violating terms.
The Lagrangians are given in the mass eigenbasis, so the fields are eigenstates of $\SU(3)_c\times\U(1)_\mathrm{em}$.

\begin{table}	
	\centering
	\ra{1.0}
	\begin{tabular}{@{}ccccc@{}}\hline
		Field & Spin & $\SU(3)_c\times\U(1)_\mathrm{em}$ &In FR &  BSM fields in M5 \\ \hline 
		$\nu_\ell$ & 1/2 & $\mathbf{1}_{0}$ & \texttt{vl}  &\\
		$\ell$ & 1/2 & $\mathbf{1}_{-1}$ & \texttt{l} &\\
		$q_u$ & 1/2 & $\mathbf{3}_{+2/3}$ & \texttt{uq} & \\
		$q_d$ & 1/2 & $\mathbf{3}_{-1/3}$ & \texttt{dq} & \\
		$h$ & 0 & $\mathbf{1}_0$ & \texttt{H}&\\
		$A$ & 1 & $\mathbf{1}_0$ & \texttt{A}&\\
		$Z$ & 1 & $\mathbf{1}_0$ & \texttt{Z}&\\
		$W$ & 1 & $\mathbf{1}_{+1}$ & \texttt{W}&\\
		$g$ & 1 & $\mathbf{8}_0$ & \texttt{G}&\\				
		\hline
		$X$ & 1/2 & $\mathbf{3}_{+5/3}$ & \texttt{x}& $X_{5/3}$\\
		$T$ & 1/2 & $\mathbf{3}_{+2/3}$ & \texttt{tp}  & $T_1$, $T_2$, $T_3$ \\
		$B$ & 1/2 & $\mathbf{3}_{-1/3}$ & \texttt{bp}  & $B$ \\
		$Y$ & 1/2 & $\mathbf{3}_{-4/3}$ & \texttt{y}  & --\\ 
		\hline 
		$S_1^0$ & 0 & $\mathbf{1}_{0}$ & \texttt{S10} &  $a,\eta,\eta_1^0,\eta_3^0,\eta_5^0$\\
		
		$S_1^1$ & 0 & $\mathbf{1}_{+1}$ & \texttt{S11}& $\eta_3^+ $, $\eta_5^+$\\
		
		$S_1^2$ & 0 & $\mathbf{1}_{+2}$ & \texttt{S12} & $\eta_5^{++}$\\
		
		$S_8^0$ & 0 & $\mathbf{8}_{0}$ & \texttt{S80} & $\pi_8$\\
		
		$S_3^{2/3}$ & 0 & $\mathbf{3}_{+2/3}$ & \texttt{S323} & $\pi_3$\\
		
		$Q_1^1$ & 1/2 & $\mathbf{1}_{+1}$ & \texttt{Q11} & $\tilde h^+$\\
		$Q_{1}^0$ & 1/2 & $\mathbf{1}_{0}$ & \texttt{Q10}  & $\tilde h^0$\\
		$Q_{1,M}^0$ & 1/2 & $\mathbf{1}_{0}$ & \texttt{Q10M} & $\tilde B$\\
		
		$Q_8^1$ & 1/2 & $\mathbf{8}_{+1}$ & \texttt{Q81}  & $\tilde G^+$\\
		$Q_8^0$ & 1/2 & $\mathbf{8}_{0}$ & \texttt{Q80} & $\tilde G^0$\\
		$Q_{8,M}^0$ & 1/2 & $\mathbf{8}_{0}$ & \texttt{Q80M} & $\tilde g$\\
		\hline
	\end{tabular}
	\caption{\textsc{FeynRules} (FR) notation for SM and BSM fields. The upper two panels form the basis of the implementation. For the fields in the lower panel, the interactions with the upper two are implemented. The right most column shows the fields in M5 that can be described with the respective module.}
	\label{tab:frfields}
\end{table}

\noindent
General Lagrangian for $S_1^0\in \mathbf 1_0$:
\begin{align}
{\cal L}_{S_1^0} = &\
\frac12 \partial^\mu S_1^0 \partial_\mu S_1^0
- \frac12 m^2_{S_1^0} S_1^0 S_1^0
+ S_1^0 \Big[
\bar B\Big(\Gamma_{B}^{L}P_L+\Gamma_{B}^{R}P_R\Big)q_d
+ \bar T\Big(\Gamma_{T}^{L}P_L+\Gamma_{T}^{R}P_R\Big)q_u
+ {\rm h.c.} \Big]\nonumber\\
&\quad + \bigg[
S_1^0 \bar q_u      \big[ \Gamma_{u} + i\gamma_5\tilde\Gamma_{u}
\big] q_u
+ S_1^0 \bar q_d      \big[ \Gamma_{d} + i\gamma_5\tilde\Gamma_{d}
\big] q_d \Big.\nonumber \\
&\Big. \quad\quad + S_1^0 \bar \ell     \big[ \Gamma_{e} + i\gamma_5\tilde\Gamma_{e}
\big] \ell
+ S_1^0 \bar \nu_\ell \big[ \Gamma_{n} + i\gamma_5\tilde\Gamma_{n}
\big] \nu_\ell + {\rm h.c.} \bigg]\nonumber\\
&\quad  
+ S_1^0 \bar B      \big[ \Gamma_{BB} + i\gamma_5\tilde\Gamma_{BB}
\big] B
+ S_1^0 \bar T      \big[ \Gamma_{TT} + i\gamma_5\tilde\Gamma_{TT}
\big] T\nonumber \\
&\quad
+ S_1^0 \bar X      \big[ \Gamma_{XX} + i\gamma_5\tilde\Gamma_{XX}
\big] X
+ S_1^0 \bar Y      \big[ \Gamma_{YY} + i\gamma_5\tilde\Gamma_{YY}
\big] Y \nonumber\\
&\quad
+ \kappa_{G}^0\ \frac{g_s^2}{16\pi^2 v} S_1^0 G_{\mu\nu}^a G_a^{\mu\nu}
+ \tilde\kappa_{G}^0\ \frac{g_s^2}{16\pi^2 v} S_1^0 G_{\mu\nu}^a\tilde G_a^{\mu\nu}\nonumber\\
&\quad
+ \kappa_{W}^0\ \frac{g^2  }{16\pi^2 v} S_1^0 W^+_{\mu\nu} W^{-\,\mu\nu}
+ \kappa_{ZZ}^0\ \frac{e^2/s_W^2c_W^2 }{16\pi^2 v} S_1^0 Z_{\mu\nu}   Z^{\mu\nu}\nonumber\\
&\quad
+ \kappa_{Z\gamma}^0\ \frac{2 e^2/s_Wc_W  }{16\pi^2 v} S_1^0 Z_{\mu\nu}   A^{\mu\nu}
+ \kappa_{\gamma\gamma}^0\ \frac{e^2 }{16\pi^2 v} S_1^0 A_{\mu\nu}   A^{\mu\nu}\nonumber\\
&\quad
+ \tilde\kappa_{W}^0\ \frac{g^2  }{16\pi^2 v}
S_1^0 W^+_{\mu\nu}\tilde W^{-\,\mu\nu}
+ \tilde\kappa_{ZZ}^0\ \frac{e^2/s_W^2c_W^2  }{16\pi^2 v}
S_1^0 Z_{\mu\nu}  \tilde Z^{\mu\nu}\nonumber\\
&\quad
+ \tilde\kappa_{Z\gamma}^0\ \frac{2 e^2/s_W c_W  }{16\pi^2 v}
S_1^0 Z_{\mu\nu}  \tilde A^{\mu\nu}
+ \tilde\kappa_{ \gamma\gamma}^0\ \frac{e^2 }{16\pi^2 v}
S_1^0 A_{\mu\nu}  \tilde A^{\mu\nu}   + {\cal L}_{S_1^0}^h    
\end{align}
with
\begin{align}
{\cal L}_{S_1^0}^h = 
\frac{\kappa^{(1)}_h}{v} h \partial^\mu S_1^0 \partial_\mu S_1^0
+ \kappa^{(2)}_h  v h S_1^0 S_1^0
+ \kappa_{hh} v h^2 S_1^0
+ \kappa_{hz} h \partial_\mu S_1^0 Z^\mu \ .
\end{align}

\noindent
General Lagrangian for $S_1^1\in \mathbf 1_1$:
\begin{align}
    {\cal L}_{S_1^1} = &\
      \partial^\mu S_1^{-1} \partial_\mu S_1^1
    -  m^2_{S_1^1} S_1^{-1} S_1^1  + \bigg[
      S_1^1 \bar q_u      \big[ \Gamma_{u} + i\gamma_5\tilde\Gamma_{qq}
         \big] q_d
    + S_1^1 \bar \ell     \big[ \Gamma_{e} + i\gamma_5\tilde\Gamma_{ll}
         \big] \nu_\ell 
     + {\rm h.c.} \bigg]\nonumber\\
    & +  \Big[
     S_1^{-1} \bar B\Big(\Gamma_{Bu}^{L}P_L+  \Gamma_{Bu}^{R}P_R\Big)q_u  + S_1^{1}\bar T\Big(\Gamma_{Td}^{L}P_L+\Gamma_{Td}^{R}P_R\Big)q_d \Big.\nonumber\\
     &\quad \Big.
   + S_1^{1}\bar X\Big(\Gamma_{Xu}^{L}P_L+\Gamma_{Xu}^{R}P_R\Big)q_u 
   + S_1^{-1}\bar Y\Big(\Gamma_{Yd}^{L}P_L+\Gamma_{Yd}^{R}P_R\Big)q_d 
   + {\rm h.c.} \Big]\nonumber\\
   & +  \Big[
      S_1^{1}\bar X\Big(\Gamma_{XT}^{L}P_L+\Gamma_{XT}^{R}P_R\Big)T 
   + S_1^{1}\bar T\Big(\Gamma_{TB}^{L}P_L+\Gamma_{TB}^{R}P_R\Big)B\Big.\nonumber\\
     &\quad \Big.
   + S_1^{1}\bar B\Big(\Gamma_{BY}^{L}P_L+\Gamma_{BY}^{R}P_R\Big)Y 
   + {\rm h.c.} \Big]\nonumber\\
   &
      + \Big[ \kappa_{W\gamma}^1\ \frac{e^2/ s_W }{16\pi^2 v} S_1^1 A^+_{\mu\nu} W^{-\,\mu\nu}
      + \kappa_{WZ}^1\ \frac{e^2/ s_W c_W  }{16\pi^2 v} S_1^1 Z_{\mu\nu} W^{-\,\mu\nu} \Big.\nonumber\\
      &\quad \Big.
      + \tilde\kappa_{W\gamma}^1\ \frac{e^2/ s_W  }{16\pi^2 v} S_1^1 A^+_{\mu\nu} \tilde W^{-\,\mu\nu}
      + \tilde\kappa_{WZ}^1\ \frac{e^2/ s_W c_W }{16\pi^2 v} S_1^1 Z_{\mu\nu} \tilde W^{-\,\mu\nu} + {\rm h.c.} \Big] + {\cal L}_{S_1^1}^h  
\end{align}
with
\begin{align}
  {\cal L}_{S_1^1}^h = 
     \frac{\kappa^{(1,1)}_h}{v} h \partial^\mu S_1^{-1} \partial_\mu S_1^1
   + \kappa^{(1,2)}_h  v h S_1^{-1} S_1^1
   + \left[ \kappa^1_{hw} h \partial_\mu S_1^1 W^{-\mu} + {\rm h.c.} \right]\ .
\end{align}

\noindent
General Lagrangian for $S_1^2\in \mathbf 1_2$:
\begin{align}\label{eq:LS12}
	\mathcal L_{S_1^2} =~& (D_\mu S_1^2)^\dagger D^\mu S_1^2 - m^2_{S_1^2} (S_{1}^2)^\dagger S_1^2 \nonumber\\
	&+ \Big[ S_1^2 \adj X \left( \Gamma_{S_1^1}^{Xd,L} P_L + \Gamma_{S_1^1}^{Xd,R} P_R \right) q_d + (S_1^2)^\dagger \adj Y \left( \Gamma_{S_1^1}^{Yu,L} P_L + \Gamma_{S_1^1}^{Yu,R} P_R \right) q_u + \hc \Big] \nonumber\\
	&+ \Big[ S_1^2 \adj X \left( \Gamma_{S_1^1}^{XB,L} P_L + \Gamma_{S_1^1}^{XB,R} P_R \right) B + (S_1^2)^\dagger \adj Y \left( \Gamma_{S_1^1}^{YT,L} P_L + \Gamma_{S_1^1}^{YT,R} P_R \right) T + \hc \Big] \nonumber\\
	&+ \Big[ \kappa_{S_1^2}^W \frac{g^2}{16 \pi^2 v} S_1^2 W^{-,\mu\nu} W^-_{\mu\nu} + \tilde\kappa_{S_1^2}^W \frac{g^2}{16 \pi^2 v} S_1^2 W^{-,\mu\nu} \tilde W^-_{\mu\nu} + \hc \Big] \nonumber\\
	&+ \frac 1v \kappa_{S_1^2}^{h,(1)} h (D_\mu S_1^2)^\dagger D^\mu S_1^2 + \kappa_{S_1^2}^{h,(2)} vh (S_1^2)^\dagger S_1^2 + \kappa_{S_1^2} ^{hh} h^2 (S_1^2)^\dagger S_1^2 
\end{align}

\noindent
General Lagrangian for $S_3^{2/3}\in \mathbf 3_{2/3}$:
	\begin{align}\label{eq:LS323}
		\mathcal L_{S_3^{2/3}} =~& (D_\mu S_3^{2/3})^\dagger D^\mu S_3^{2/3} - m^2_{S_3^{2/3}} (S_3^{2/3})^\dagger S_3^{2/3}\nonumber \\
		&+ \frac 1v \kappa_{S_3^{2/3}}^{h,(1)} h (D_\mu S_3^{2/3})^\dagger D^\mu S_3^{2/3} + \kappa_{S_3^{2/3}}^{h,(2)} vh (S_3^{2/3})^\dagger S_3^{2/3} + \kappa_{S_3^{2/3}} ^{hh} h^2 (S_3^{2/3})^\dagger S_3^{2/3} 
	\end{align}

\noindent
General Lagrangian for $S_8^0=S_8^{0,a}T^a\in \mathbf 8_0$:
	\begin{align}
		\mathcal L_{S_8^0} = &~\frac 12 D_\mu S_8^{0,a} D^\mu S_8^{0,a} - \frac 12 m^2_{S_8^0} S_8^{0,a} S_8^{0,a} \nonumber \\
		& + \Big[ \adj B \left( \Gamma_{S_8^0}^{Bd,L} P_L + \Gamma_{S_8^0}^{Bd,R} P_R  \right) S_8^0 q_d + \adj T \left( \Gamma_{S_8^0}^{Tu,L} P_L + \Gamma_{S_8^0}^{Tu,R} P_R  \right) S_8^0 q_u + \hc \Big] \nonumber\\
		&+ \left[\adj q_u \left( \Gamma_{S_8^0}^u + \mathrm i \gamma_5 \tilde \Gamma_{S_8^0}^u \right) S_8^0 q_u +  \adj q_d \left( \Gamma_{S_8^0}^d + \mathrm i \gamma_5 \tilde \Gamma_{S_8^0}^d \right) S_8^0 q_d +\hc\right] \nonumber\\ 
		&+ \Big[\adj X \left( \Gamma_{S_8^0}^X + \mathrm i \gamma_5 \tilde \Gamma_{S_8^0}^X \right) S_8^0 X
		+ \adj T \left( \Gamma_{S_8^0}^T + \mathrm i \gamma_5 \tilde \Gamma_{S_8^0}^T \right) S_8^0 T \nonumber \nonumber\\
		&\quad+ \adj B \left( \Gamma_{S_8^0}^B + \mathrm i \gamma_5 \tilde \Gamma_{S_8^0}^B \right) S_8^0 B
		+ \adj Y \left( \Gamma_{S_8^0}^Y + \mathrm i \gamma_5 \tilde \Gamma_{S_8^0}^Y \right) S_8^0 Y +\hc\Big] \nonumber\\
		&+ \kappa_{S_8^0}^{G} \frac{g_s^2}{16\pi^2v} \Tr(S^0_8 G^{\mu\nu} G_{\mu\nu}) + \tilde\kappa_{S_8^0}^{G} \frac{g_s^2}{16\pi^2v} \Tr(S^0_8 G^{\mu\nu} \tilde G_{\mu\nu})\nonumber  \\
		&+ \kappa_{S_8^0}^{GZ} \frac{g_s e/s_Wc_W}{16\pi^2v} S^{0,a}_8 G^{a,\mu\nu} Z_{\mu\nu} + \kappa_{S_8^0}^{GA} \frac{g_s e}{16\pi^2v} S^{0,a}_8 G^{a,\mu\nu} A_{\mu\nu} \nonumber\\
		&+ \tilde\kappa_{S_8^0}^{GZ} \frac{g_s e/s_Wc_W}{16\pi^2v} S^{0,a}_8 G^{a,\mu\nu} \tilde Z_{\mu\nu} + \tilde\kappa_{S_8^0}^{GA} \frac{g_s e}{16\pi^2v} S^{0,a}_8 G^{a,\mu\nu} \tilde A_{\mu\nu}  \nonumber\\
		&+ \frac 1v \kappa_{S_8^0}^{h,(1)} h (D_\mu S_8^0)^a (D^\mu S_8^0)^a + \kappa_{S_8^0}^{h,(2)} vh S_8^{0,a} S_8^{0,a} + \kappa_{S_8^0} ^{hh} h^2 S_8^{0,a} S_8^{0,a}
	\end{align}

\noindent
General Lagrangian for Dirac fermion $Q_1^1\in \mathbf 1_{1}$: 
\begin{align}\label{eq:LQ11}
	\mathcal L_{Q_1^1} =\adj Q_1^1 \left(\mathrm i \slashed D - m_{Q_1^1}\right) Q_1^1 + \left(\frac{g}{2c_W} \adj Q_1^1 \slashed Z \left[ \kappa_{Q_1^1}^{Z} + \mathrm i \gamma_5 \tilde\kappa_{Q_1^1}^{Z} \right] Q_1^1 +\hc\right)
\end{align}
General Lagrangian for Dirac fermion $Q_1^0\in \mathbf 1_{0}$: 
\begin{align}
	\mathcal L_{Q_1^0} = \adj Q_1^0 \left(\mathrm i \slashed \partial - m_{Q_1^0}\right) Q_1^0 +\left( \frac{g}{2c_W} \adj Q_1^0 \,\slashed Z \left[ \kappa_{Q_1^0}^{Z} + \mathrm i \gamma_5 \tilde\kappa_{Q_1^0}^{Z} \right] Q_1^0  +\hc\right)
\end{align}
General Lagrangian for Majorana fermion $Q_{1,M}^0\in \mathbf 1_{0}$: 
\begin{align}
	\mathcal L_{Q_{1,M}^0} =~ &\frac 12 \adj Q_{1,M}^0 \left(\mathrm i \slashed \partial - m_{Q_{1,M}^0}\right) Q_{1,M}^0 + \left(\tilde\kappa_{Q_{1,M}^0}^Z\frac{g}{4c_W} \adj Q_{1,M}^0 \,\slashed Z\,  \mathrm i \gamma_5 Q_{1,M}^0  +\hc\right)
\end{align}

\noindent
General Lagrangian for Dirac fermion $Q_8^1\in \mathbf 8_{1}$: 
\begin{align}
	\mathcal L_{Q_8^1} =~ &\adj Q_8^{1,a} \left(\mathrm i \slashed D - m_{Q_8^{1}}\right) Q_8^{1,a} +\left( \frac{g}{2c_W} \adj Q_8^{1,a} \slashed Z \left[ \kappa_{Q_8^{1}}^{Z} + \mathrm i \gamma_5 \tilde\kappa_{Q_8^{1}}^{Z} \right] Q_8^{1,a}  +\hc\right)
\end{align}
General Lagrangian for Dirac fermion $Q_8^0\in \mathbf 8_{0}$: 
\begin{align}
	\mathcal L_{Q_8^0} =~ &\adj Q_8^{0,a} \left(\mathrm i \slashed D - m_{Q_8^{0}}\right) Q_8^{0,a}  +\left( \frac{g}{2c_W} \adj Q_8^{0,a} \slashed Z \left[ \kappa_{Q_8^{0}}^{Z} + \mathrm i \gamma_5 \tilde\kappa_{Q_8^{0}}^{Z} \right] Q_8^{0,a}  +\hc\right)
\end{align}
General Lagrangian for Majorana fermion $Q_{8,M}^0\in \mathbf 8_{0}$: 
\begin{align}\label{eq:LQ80M}
	\mathcal L_{Q_{8,M}^0} =~ &\frac 12 \adj Q_{8,M}^{0,a} \left(\mathrm i \slashed D - m_{Q_{8,M}^{0}}\right) Q_{8,M}^{0,a} + \left(\tilde\kappa_{Q_{8,M}^{0}}^{Z}  \frac{g}{4c_W} \adj Q_{8,M}^{0,a} \slashed Z \, \mathrm i \gamma_5  Q_{8,M}^{0,a}  +\hc\right)
\end{align}

We have also implemented selected vertices that mix different modules:
	\begin{align}
		\mathcal L_\mathrm{M5}= ~&C_{Q_{8,M}^0}^{S_3^{2/3}t} (S_3^{2/3})^\dagger  \adj Q^0_{8,M} P_R t + C_{Q_{8}^0}^{S_3^{2/3}t} (S_3^{2/3})^\dagger  \adj Q^{0,c}_{8,M} P_L t + C_{Q_{8}^1}^{S_3^{2/3}t} (S_3^{2/3})^\dagger  \adj Q^{1,c}_{8} P_L b \nonumber\\
		&+\frac 12 C_{Q_{8,M}^0}^{S_8^0 Q^0_{1,M}} S_8^{0,a}  \adj Q^{0,a}_{8,M} Q^0_{1,M} +C_{Q_{8}^0}^{S_8^0 Q^0_{1}} S_8^{0,a}  \adj Q^{0,a}_{8} Q^0_{1} + C_{Q_{8}^1}^{S_8^0 Q^1_{1}} S_8^{0,a}  \adj Q^{0,a}_{8} Q^1_{1} \\
		&+ \frac{g}{2 c_W} C_{Q_1^0}^{Q_{1,M}^0 Z} \adj Q_1^0 \slashed Z Q^0_{1,M} + \frac{g}{\sqrt{2}} C_{Q_1^1}^{Q_{1}^0 W} \adj Q_1^1 \slashed W^+ Q^0_{1}+ C_{S_3^{2/3}} ^{Q^0_{1,M} t} (S_3^{2/3})^\dagger \adj Q^0_{1,M} P_R t+ \hc\nonumber
	\end{align}
These are necessary for describing the decays studied in \cref{sec:numerics}.
Technically, the couplings in the second line should have a derivative of the scalar, since they originate from the derivative couplings in \cref{eq:Lder141}.
This does not majorly affect the results since we only work with two-body decays in the narrow-width approximation. The branching ratios are not affected as the momentum can be absorbed in the coupling constant. The effect on decay kinematics is minor as the color octet fermions are not produced with a high boost.

For simulations with \textsc{MADGRAPH5\_AMC@NLO}, we generate a model UFO file containing all M5 model fields and effective couplings, and then use MG restriction cards for simulations of the various scenarios in each of which we then perform scans.

\section{Comparison of the bounds from different searches}
\label{app:tools_comparison}

\begin{figure}
	\centering
	\begin{subfigure}[]{0.41\linewidth}
		\centering
		\includegraphics[width=\textwidth]{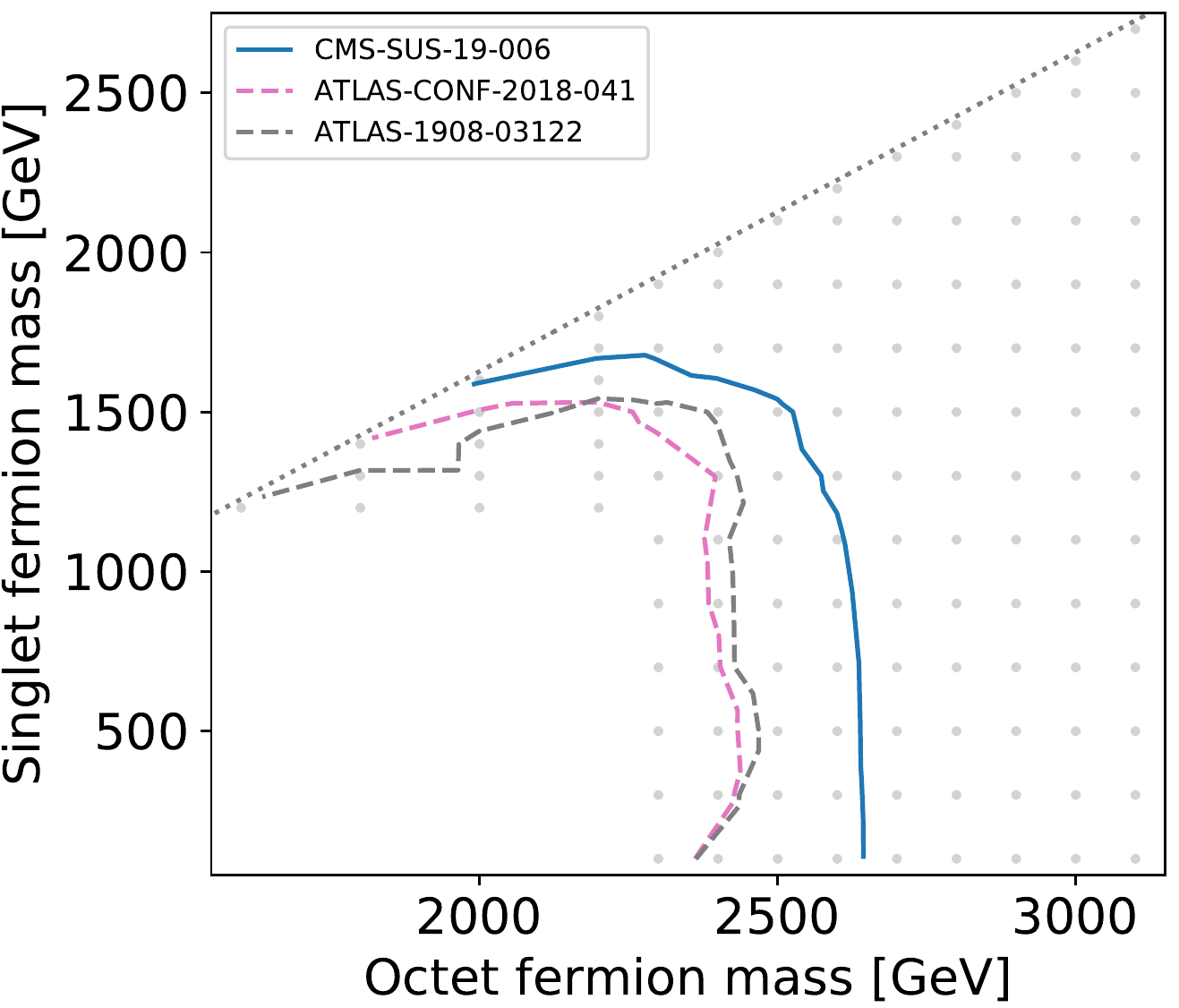} 
		\caption{Scenario 1 with $m_{Q_8}-m_{\pi_3}=200$~GeV} \label{fig:fig5_multiplet_stop_varyingmass_MAandCM}
	\end{subfigure}
	\hspace{2ex}
	\begin{subfigure}[]{0.41\linewidth}
		\centering
		\includegraphics[width=\textwidth]{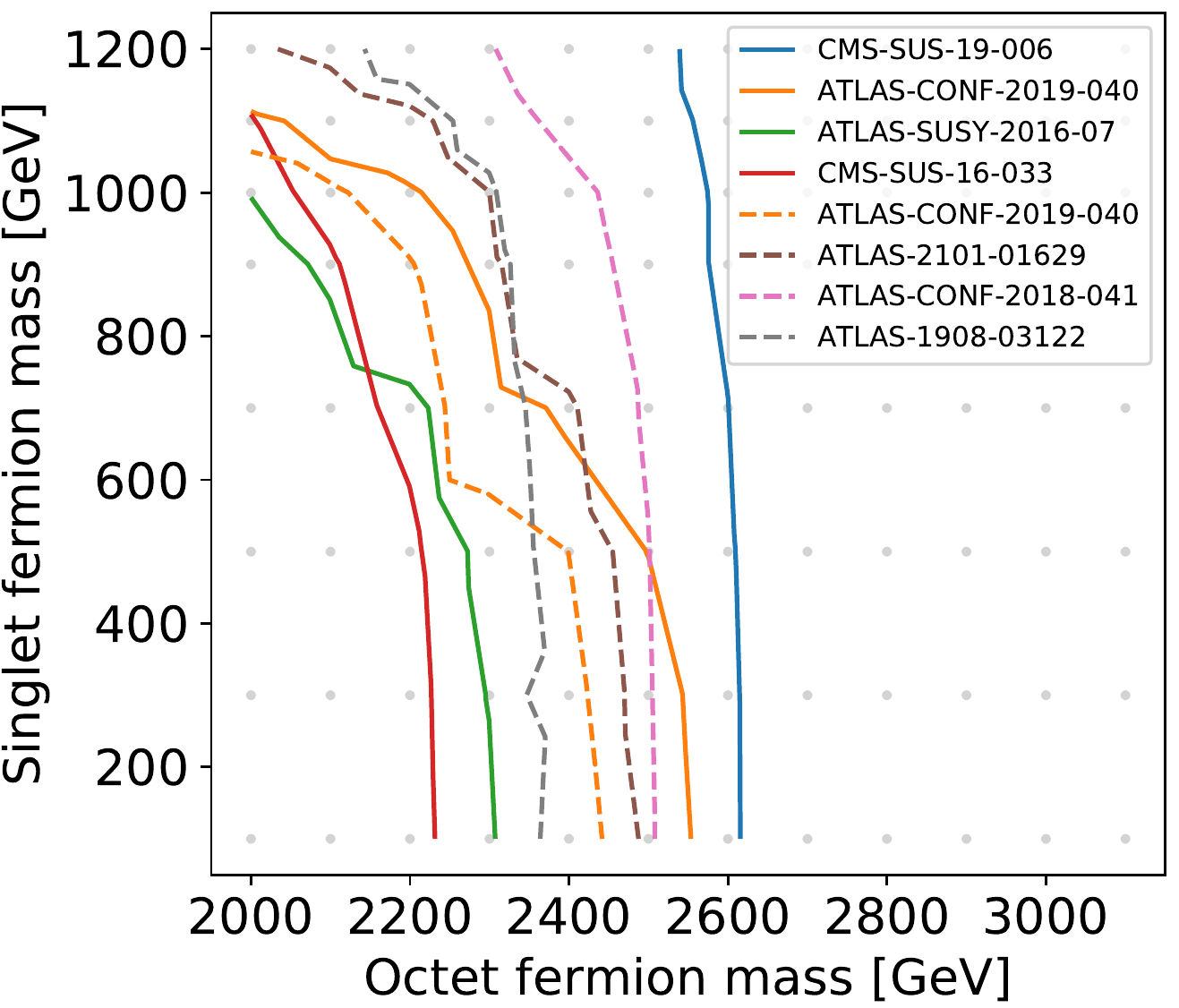} 
		\caption{Scenario 1 with $m_{\pi_3}=1.4$~TeV}
		\label{fig:fig6_multiplet_stop_fixedmass_MAandCM}
	\end{subfigure}
	\vspace{2ex}
	
	\begin{subfigure}[]{0.41\linewidth}
		\centering
		\includegraphics[width=\textwidth]{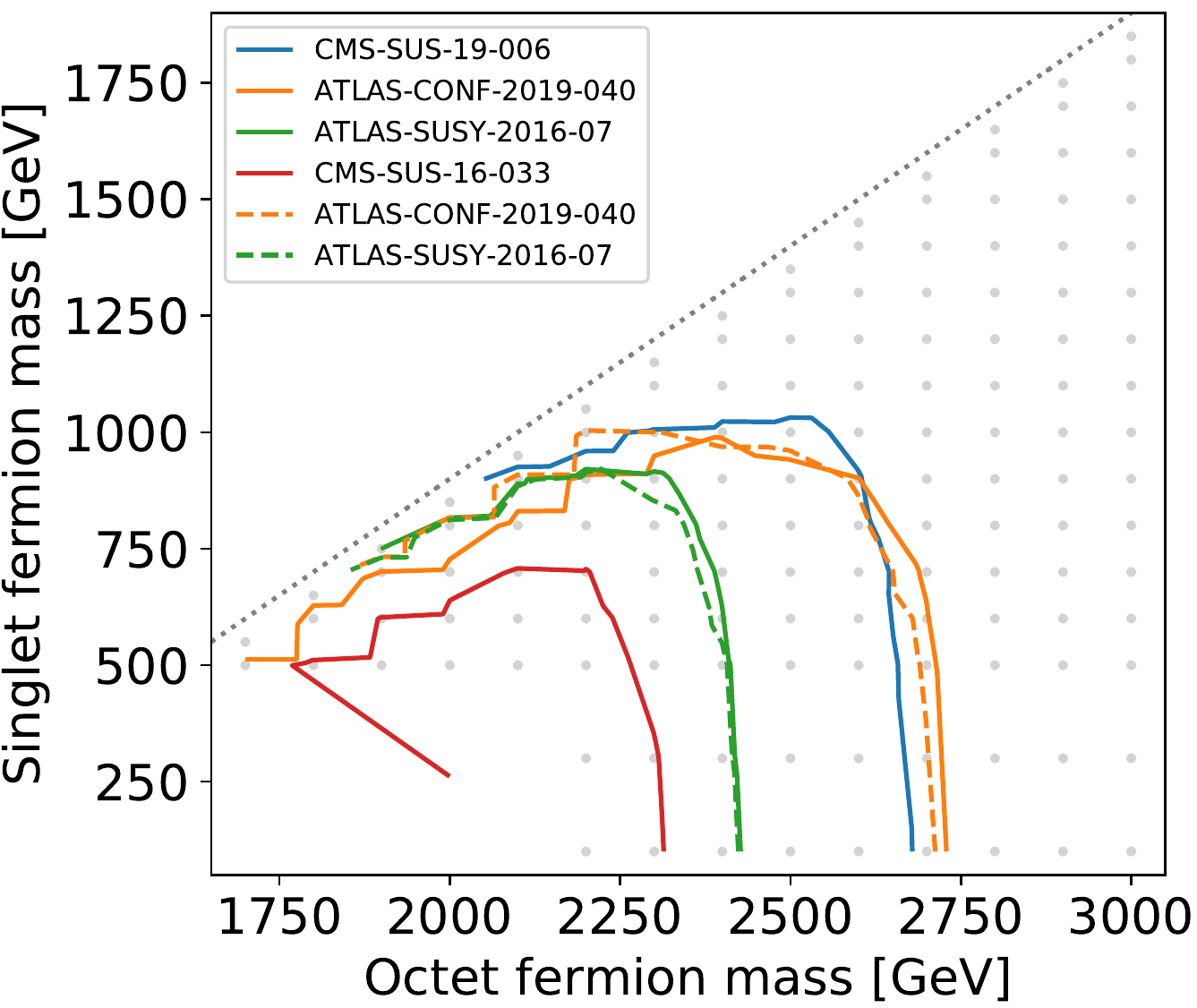} 
		\caption{Scenario 2 with $\pi_8\to gg$} \label{fig:fig10_multiplet_pi8_gg_fixedmass_MA_comparison}
	\end{subfigure}
	\hspace{2ex}
	\begin{subfigure}[]{0.41\linewidth}
		\centering
		\includegraphics[width=\textwidth]{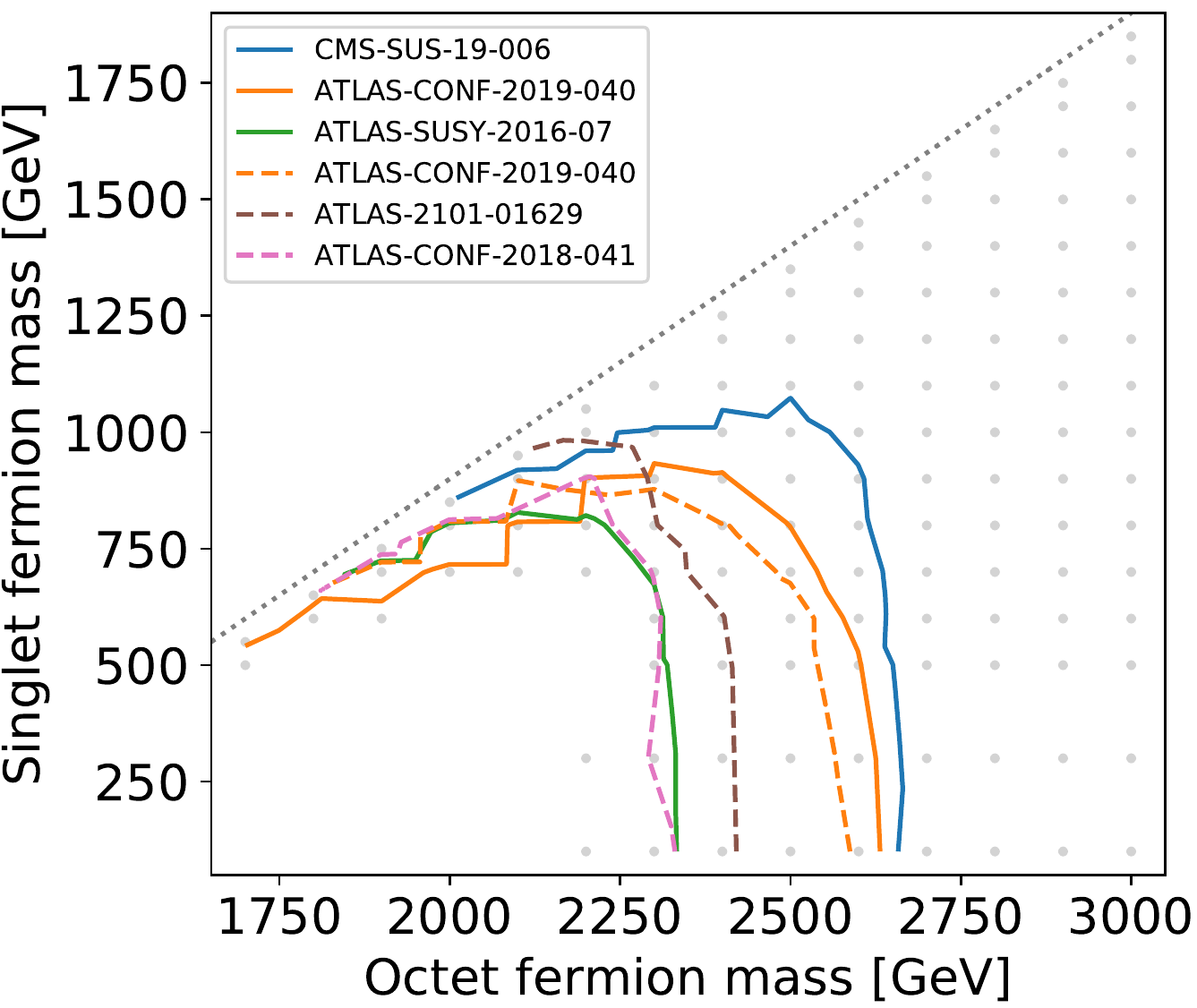} 
		\caption{Scenario 2 with $\pi_8\to gg,tt$}
		\label{fig:fig11_multiplet_pi8_ggtt_fixedmass_MA_comparison}
	\end{subfigure}
	\vspace{2ex}
	
	\begin{subfigure}[]{0.41\linewidth}
		\centering
		\includegraphics[width=\textwidth]{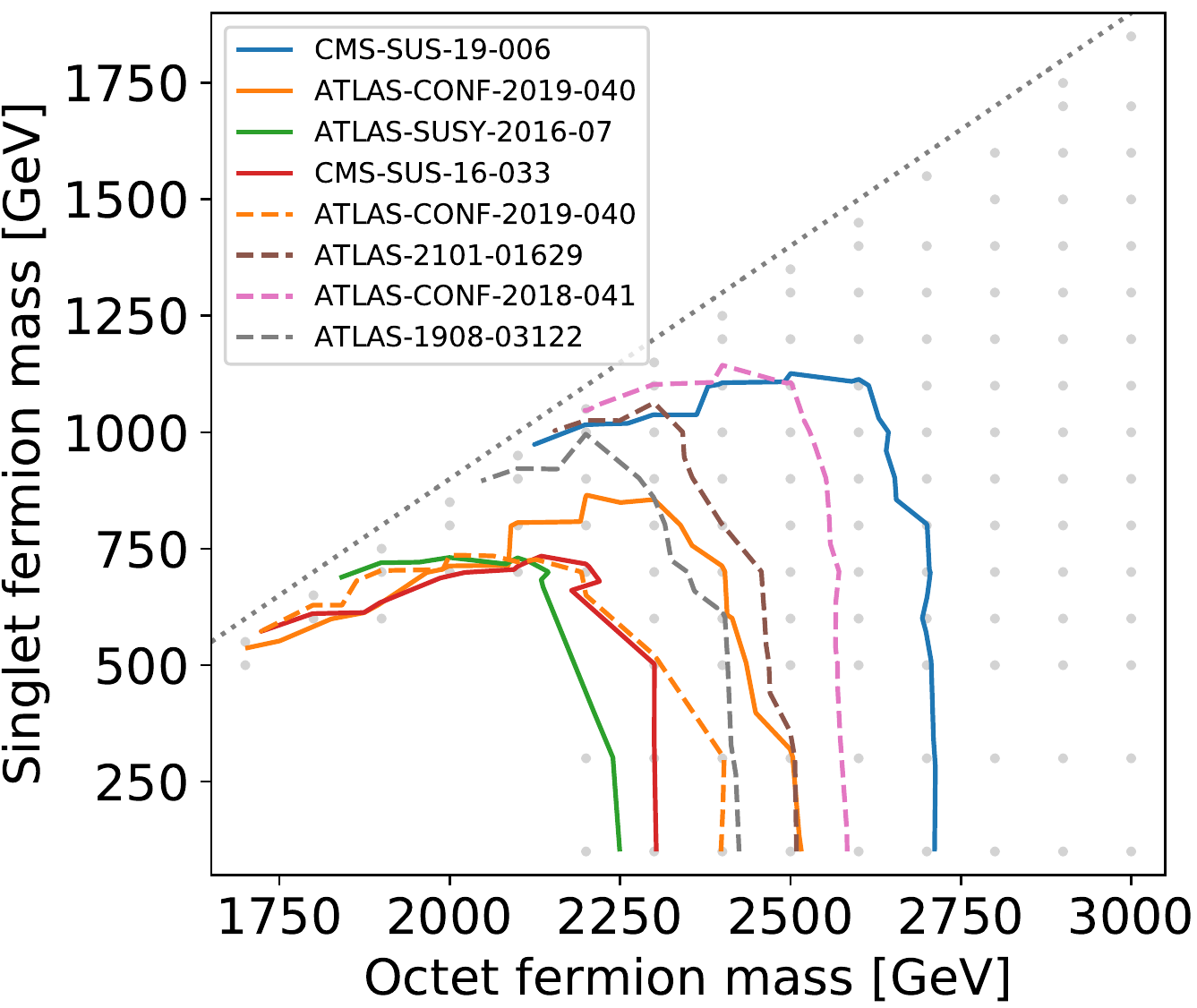} 
		\caption{Scenario 2 with $\pi_8\to tt$} \label{fig:fig12_multiplet_pi8_tt_fixedmass_MA_comparison}
	\end{subfigure}
	\caption{Comparison of the bounds at 95\% CL obtained from different searches \cite{CMS:2019zmd, ATLAS:2019vcq, ATLAS:2017mjy, CMS:2017abv, ATLAS:2021twp, ATLAS:2018yhd, ATLAS:2019gdh} implemented in MA (solid lines) and CM (dashed lines) for scenario (S1c) in (a)-(b) and scenario (S2b) with $m_{\pi_8}=1.1$~TeV in (c)-(e).}
	\label{fig:bounds_allsearches}
\end{figure}

In this appendix, we highlight the recasted searches that are most sensitive to the decays of the octet top partners.
To this end, we show the mass bounds for the scenarios 1 and 2 separately for each search in Fig.~\ref{fig:bounds_allsearches}.
Here, the gray dots are the simulated points, for which we generated 10000 events each.
The coarse structures in the contour lines are due to the limited resolution of the grid and could be improved if a more precise knowledge of the bounds is required.
Fig.~\ref{fig:bounds_allsearches} shows that the bounds are dominated by only a few searches.

\begin{itemize}
	\item \textbf{CMS-SUS-19-006} \cite{CMS:2019zmd}: This is a search for gluino and squark pair production with multiple jets and large MET in the final state using 137~fb$^{-1}$ of data.
	The results are interpreted within multiple simplified models, including the $4t+\ptmiss$, $4b+\ptmiss$, $4q+\ptmiss$ and $4q+2V+\ptmiss$ final states from gluinos, where $q=u,d,s,c$ are light quarks and $V=W,Z$.
	The signal candidates are divided into 174 orthogonal SRs, and covariance and correlation matrices for the SRs are provided.
	These are used by the recast implemented in MA \cite{Mrowietz:2020ztq} to perform a statistical combination of the SRs.
	This explains why this search gives the strongest bound for most scenarios.
	The optimization for both $4t$ and $4j$ final states makes the recast competitive both for $\pi_8\to t\bar t$ and $\pi_8\to gg$.
	
	\item \textbf{ATLAS-CONF-2019-040} \cite{ATLAS:2019vcq}: This search looks for gluinos and squarks in final states containing jets and MET but no charged leptons.
	It uses the full Run~2 dataset of $139$~fb$^{-1}$. 
	The simplified model for the gluinos assumes $\tilde g\to qq\tilde \chi^0_1$ or $\tilde g\to q'q W \tilde \chi_1^0$, where $q^{(\prime)}$ are light quarks.
	We therefore expect the recast to be very sensitive to final states with multiple light jets, such as those from $\pi_8\to gg$.
	This is confirmed by comparing Figs.~\ref{fig:fig10_multiplet_pi8_gg_fixedmass_MA_comparison}-e.
	For the final states dominated with top quarks, however, this search is subdominant.
	Note that it is implemented in both MA and CM.
\end{itemize}

Figure~\ref{fig:bounds_allsearches}  shows the mass bounds from several other searches, which however are less sensitive to our signatures. 
We briefly summarize those:
ATLAS-CONF-2018-041 \cite{ATLAS:2018yhd} presents a search for gluino pair production with decays to third generation quarks and neutralinos using 79.8~fb$^{-1}$ of data.
ATLAS-1908-03122  \cite{ATLAS:2019gdh} searches for bottom-squark production with Higgs bosons in the final state.
ATLAS-SUSY-2016-07 \cite{ATLAS:2017mjy} is a search for gluinos and squarks in final states with light quarks and no leptons.
It is implemented in both MA and CM.
CMS-SUS-16-033 \cite{CMS:2017abv} searches for pair production of gluinos and stops decaying to light or third-generation quarks, similarly to CMS-SUS-19-006 but using only $35.9$~fb$^{-1}$.
Finally, ATLAS-2101-01629 \cite{ATLAS:2021twp} searches for pair production and chain decays of gluinos $\tilde g\to q\bar q' \tilde \chi_1^\pm $ and squarks $\tilde q\to q'\tilde \chi_1^\pm$ with $\tilde \chi_1^\pm\to W^\pm \tilde \chi_1^0$.

\clearpage

\bibliographystyle{utphys}
\bibliography{CHlit}

\end{document}